\documentclass[iop,apj,twocolappendix,numberedappendix]{emulateapj}
\usepackage[usenames]{color}
\usepackage{hyperref}
\hypersetup{colorlinks,citecolor=Blue,linkcolor=Red,urlcolor=Blue}

\usepackage[varg]{txfonts}
\usepackage{bm}

\newcommand{\beq}{\begin{equation}}
\newcommand{\eeq}{\end{equation}}
\newcommand{\beqn}{\begin{eqnarray}}
\newcommand{\eeqn}{\end{eqnarray}}
\renewcommand{\leq}{\leqslant}

\newcommand{\eqref}[1]{(\ref{#1})}

\newcommand{\dfrac}[2]{ {\displaystyle\frac{#1}{#2}} }

\newcommand{\pfrac}[2]{ \biggl(\dfrac{#1}{#2}\biggr) }

\newcommand{\pd}{\partial}
\newcommand{\AU}{{\rm AU}} 
\newcommand{\yr}{{\rm yr}}
\newcommand{\Myr}{{\rm Myr}}
\newcommand{\kB}{k_{\rm B}}

\shorttitle{Sintering-induced Multiple Ring Formation in Protoplanetary Disks}
\shortauthors{Okuzumi et al.}

\begin{document}
\title{Sintering-induced Dust Ring Formation in Protoplanetary Disks: Application to the HL Tau Disk}
\author{Satoshi Okuzumi\altaffilmark{1,2}, Munetake Momose\altaffilmark{3}, Sin-iti Sirono\altaffilmark{4}, Hiroshi Kobayashi\altaffilmark{5}, and Hidekazu Tanaka\altaffilmark{6}}
\affil{$^1$Department of Earth and Planetary Sciences, 
Tokyo Institute of Technology, Meguro, Tokyo, 152-8551, Japan; okuzumi@geo.titech.ac.jp\\
$^2$Jet Propulsion Laboratory, California Institute of Technology, Pasadena, CA 91109, USA\\
$^3$College of Science, Ibaraki University, Mito, Ibaraki 310-8512, Japan\\
$^4$Department of Earth and Environmental Sciences, Nagoya University, Nagoya 464-8601, Japan\\
$^5$Department of Physics, Nagoya University, Nagoya, Aichi 464-8602, Japan\\
$^6$Institute of Low Temperature Science, Hokkaido University, Sapporo 060-0819, Japan
}

\begin{abstract}
The latest observation of HL Tau by ALMA revealed spectacular concentric dust rings in its circumstellar disk. We attempt to explain the multiple ring structure as a consequence of aggregate sintering. {Sintering is known to reduce the sticking efficiency of dust aggregates and occurs at temperatures slightly below the sublimation point of their constituent material. We here present a dust growth model incorporating sintering and use it to simulate global dust evolution due to sintering, coagulation, fragmentation, and radial inward drift in a modeled HL Tau disk. We show that aggregates consisting of multiple species of volatile ices experience sintering, collisionally disrupt, and pile up at multiple locations slightly outside the snow lines of the volatiles.} At wavelengths of 0.87--1.3 mm, these sintering zones appear as bright, optically thick rings with a spectral slope of $\approx 2$, whereas the non-sintering zones as darker, optically thinner rings of a spectral slope of $\approx 2.3$--$2.5$. {The observational features of the sintering and non-sintering zones are consistent with those of the major bright and dark rings found in the HL Tau disk, respectively.} Radial pileup and vertical settling occur simultaneously if disk turbulence is weak and if monomers constituting the aggregates are $\sim 1~\micron$ in radius. For the radial gas temperature profile of $T = 310(r/1~\AU)^{-0.57}~{\rm K}$, our model perfectly reproduces the brightness temperatures of the optically thick bright rings, and reproduces their orbital distances to an accuracy of $\la  30\%$. 
\end{abstract}
\maketitle

\section{Introduction}\label{sec:intro}
HL Tau is a flat spectrum T Tauri star with a circumstellar disk that is very luminous at millimeter wavelengths \citep{BSCG90}. Although the age of HL Tau has not been well constrained, its low bolometric temperature, high mass accretion rate \citep[e.g.,][]{WH04} and the presence of an optical jet \citep{MBR88} and an infalling envelope \citep{HOM93} suggest that the stellar age is likely less than 1 Myr. For these reasons, HL Tau is considered as an ideal observational target for studying the very initial stages of disk evolution and of planet formation.

The recent Long Baseline Campaign of the Atacama Large Millimeter/submillimeter Array (ALMA) has provided spectacular images of the HL Tau disk \citep{ALMA+15}. ALMA resolved the disk at three millimeter wavelengths with unprecedented spatial resolution of $\approx 3.5$ AU at 0.87 mm. The observations revealed a pattern of multiple bright and dark rings that are remarkably symmetric with respect to the central star. The spectral index at 1 mm is $\sim 2$ in the central emission peak and in some of the bright rings, and is $\sim 2.3$--$3$ in the dark rings \citep{ALMA+15,ZBB15}. The fact that the millimeter spectral index is $\la 3$ in the dark, presumably optically thin rings suggests that dust grains in the HL Tau disk have already grown into aggregates whose radius is larger than a few millimeters, assuming that the aggregates are compact \citep[][see \citealt{KOTN14} for how the aggregates' porosity alters this interpretation]{D06}. The observed continuum emission is best reproduced by models assuming substantial dust settling \citep{KLM11,KLMW15,PDM+16}, implying that the large aggregates are dominant in mass and that the turbulence in the gas disk is weak. Since dust growth and settling are key processes in planet formation, understanding the origin of this axisymmetric dust structure is greatly relevant to understanding how planets form in protoplanetary disks.

There are a variety of mechanisms that can produce axisymmetric dust rings and gaps in a protoplanetary disk. One of the most common mechanisms creating a dust ring is dust trapping at local gas pressure maxima under the action of gas drag \citep{W72}. In a protoplanetary disk, pressure bumps may be created by disk--planet interaction \citep{PM04,PM06,FGM10,ZNDEH12,PBB12,GPMMF12,DZW15,DPL+15}, magnetorotational instability \citep{JYK09,UKFH11}, and/or steep radial variation of the disk viscosity \citep{KL07,DFTKH10,FRD+15}. {Axisymmetric dust rings may also be produced by secular gravitational instability \citep{Y11,TI14}, by baroclinic instability arising due to dust settling \citep{LB15}, or by a combined effect of dust coagulation and radial drift \citep{L14,GLM+15}. Planet-carved gaps may explain the observed features of the HL Tau disk even if dust trapping at the pressure maxima is ineffective \citep{KMT+15}.
}
 
Another intriguing possibility is that the multiple ring patterns of the HL Tau disk are related to the snow lines of various solid materials. Recently, \citet{ZBB15} used a temperature profile based on a previous study \citep{MHF99} and showed that the major dark rings seen in the ALMA images lie close to the sublimation fronts of some main cometary volatiles such as ${\rm H_2O}$ and ${\rm NH_3}$. They interpreted this as the evidence of rapid particle growth by condensation as recently predicted by \citet{RJ13} for ${\rm H_2O}$ ice particles. However, it is unclear at present whether relatively minor volatiles such as ${\rm NH_3}$ and clathrates indeed accelerate dust growth.

In this study, we focus on another important mechanism that can affect dust growth near volatile snow lines: sintering. Sintering is the process of fusing grains together at a temperature slightly below the sublimation point. Sintered aggregates are characterized by thick joints, called necks, that connect the constituent grains (e.g., see Figures~3 and 4 of \citealt*{P03}; Figure~1 of \citealt*{B07}). A familiar example of a sintered aggregate is a ceramic material (i.e., pottery), which is an agglomerate of micron-sized clay particles fused together by sintering. Sintered aggregates are less sticky than unsintered ones, because the necks prevent collision energy from being dissipated through plastic deformation. For example, unsintered dust aggregates are known to absorb much collision energy through rolling friction among constituent grains \citep{DT97}. However, sintered aggregates are unable to lose their collision energy in this way, and therefore their collision tends to end up with bouncing, fragmentation, or erosion rather than sticking \citep{S99,SU14}. Thus, sintering suppresses dust growth in regions slightly outside the snow lines.

The importance of sintering in the context of protoplanetary dust growth was first pointed out by \citet{S99}, and has been studied in more detail by \citet{S11a,S11b} and  \citet{SU14}. {\citet{S99} simulated collisions of two-dimensional sintered and unsintered aggregates (both made of $0.1~\micron$-sized icy grains) with a wall taking into account the high mechanical strength of the sintered necks. At collision velocities lower than $10~{\rm m~s^{-1}}$, the sintered aggregates are found to bounce off the wall whereas the unsintered ones stick to the wall. By using the same sintered neck model, \citet{SU14} simulated collisions between two identical sintered aggregates each of which consists of up to $10^4$ icy grains (again of $0.1~\micron$ radius) and has a porosity of 30--80\%.} {They} found that the aggregates erode each other rather than stick if the collision velocity is above $20~{\rm m~s^{-1}}$. This threshold is considerably lower than that for unsintered aggregates, which is around $50~{\rm m~s^{-1}}$ when the constituent grains are ice and $0.1~\micron$ in radius \citep{DT97,WTS+09,WTO+13}.

{Another important fact about sintering is that it can occur at multiple locations in a protoplanetary disk as already pointed out by \citet{S99,S11b}.} In contrast to condensational growth as envisioned by \citet{ZBB15}, sintering requires only a small amount of volatiles because the volume of a neck is generally a small fraction of the grain volume. For example, the volume fraction is only $0.2\%$ even if the neck radius is as large as $30\%$ of the grain radius \citep{S11b}. Therefore, inclusion of ${\rm NH_3}$ ice at a standard cometary abundance ($\sim 0.2$--$1.4\%$; \citealt{MC11}) is enough to sinter the grains near the ${\rm NH_3}$ snow line. 

In this study, we investigate how this ``sintering barrier'' against dust coagulation affects the global evolution of dust in a protoplanetary disk. We present a simple recipe to account for the change in the mechanical strength of dust aggregates due to sintering, and apply it to global simulations of dust evolution in a disk that take into account coagulation, fragmentation, and radial inward drift induced by gas drag \citep{AHN76,W77a}. Our simulations for the first time show that sintering-induced fragmentation leads to a pileup of dust materials in the vicinity of each volatile snow line. We will demonstrate that at millimeter wavelengths, these pileups can be seen as multiple bright dust rings as observed in the HL Tau disk.

The structure of this paper is as follows. We begin by modeling the HL Tau gas disk in Section~\ref{sec:disk}. Section~\ref{sec:subsint} introduces our model for aggregate sintering and sublimation. Section~\ref{sec:method} describes our simulation method, Section~\ref{sec:results} presents the results from our fiducial simulation run, and Section~\ref{sec:param} presents a parameter study. The validity and possible limitations of our model are discussed in Section~\ref{sec:discussion}. A summary is presented in Section~\ref{sec:summary}.

\section{Disk Model}\label{sec:disk}
We model the HL Tau protoplanetary disk as a static, axisymmetric, and vertically isothermal disk. The radial
profiles of the temperature and gas density are presented in Sections~\ref{sec:temp} and \ref{sec:density},
respectively.

\subsection{Temperature Profile}\label{sec:temp}
\begin{figure}[t]
\centering
\includegraphics[width=8.5cm]{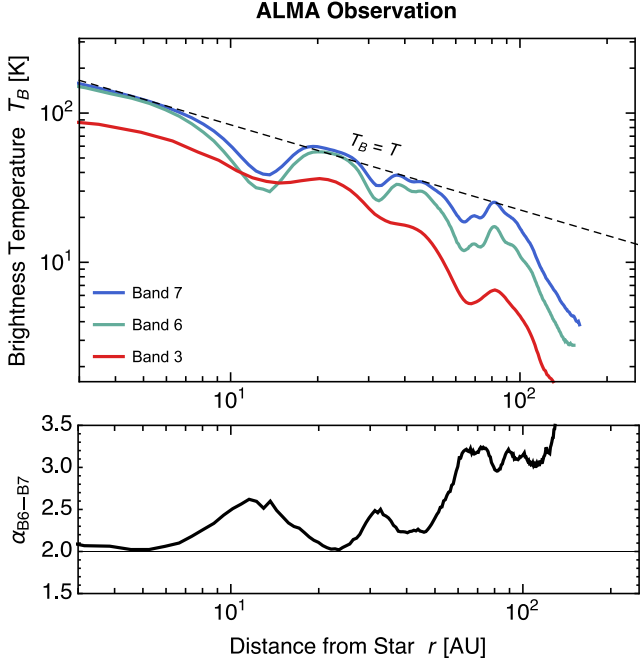}
\caption{Upper panel: radial profiles of the brightness temperature $T_B$ of the HL Tau disk 
retrieved from the ALMA data (\citealt{ALMA+15}; see text for details). 
The red, green, and blue curve are for ALMA Bands 3, 6, and 7, respectively. 
The dashed line shows the gas temperature profile $T(r)$ adopted in this study (Equation~\eqref{eq:T}). 
Lower panel: spectral index between Bands 6 and 7, $\alpha_{\rm B6\textrm{--}B7}$, versus radial distance $r$. 
}
\label{fig:T}
\end{figure}

We construct a radial  temperature profile $T(r)$ of the HL Tau disk based on the data of the surface brightness profiles provided by \citet{ALMA+15}. We deproject the intensity maps of the disk's dust continuum at ALMA Bands 3, 6, and 7 into circularly symmetric views assuming the disk inclination angle of $46.7^{\circ}$ and the position angle of $138^{\circ}$ \citep{ALMA+15}. We then obtain the radial profiles of the intensities $I_\nu$ by azimuthally averaging the deprojected images. The upper panel of Figure~\ref{fig:T} shows the derived radial emission profiles. Here, the intensities are expressed in terms of the Planck brightness temperature $T_{\rm B}$. Shown in the lower panel is the spectral index between Bands 6 and 7, $\alpha_{\rm B6\textrm{--}B7} \equiv \ln(I_{\rm B7}/I_{\rm B6})/\ln(\nu_{\rm B7}/\nu_{\rm B6})$, where $\nu_{\rm B6} = 233.0~{\rm GHz}$ and $\nu_{\rm B7} = 343.5~{\rm GHz}$ are the frequencies at Bands 6 and 7, respectively.

As already pointed out by \citet{ALMA+15}, the HL Tau disk has a pronounced central emission peak at $\la 10$ AU, and three major bright rings at $\sim 20$, 40, and 80 AU. The central emission peak and two innermost bright rings have a spectral index of $\alpha_{\rm B6\textrm{--}B7} \approx 2$. In general, this indicates either that the emission at these wavelengths is optically thick,
or that the emission is optically thin but is from dust particles larger than millimeters. 
The brightness temperature is equal to the gas temperature in the former case, and is lower in the latter case. 
While \citet{ZBB15} adopted the latter interpretation, we here pursue the former interpretation.
Specifically, we assume that the Band 7 emission is optically thick and hence $T(r) = T_{\rm B}$ 
at the center, 20 AU, and 40 AU.
The simplest profile  satisfying this assumption is the single power law
\beq
T(r) = 310\pfrac{r}{1~\AU}^{-0.57}~{\rm K},
\label{eq:T} 
\eeq
which is shown by the dashed line in the upper panel of Figure~\ref{fig:T}. 
We will adopt this temperature profile in this study.

\subsection{Density Structure}\label{sec:density}
\begin{figure}[t]
\epsscale{0.6} 
\centering
\includegraphics[width=8.5cm]{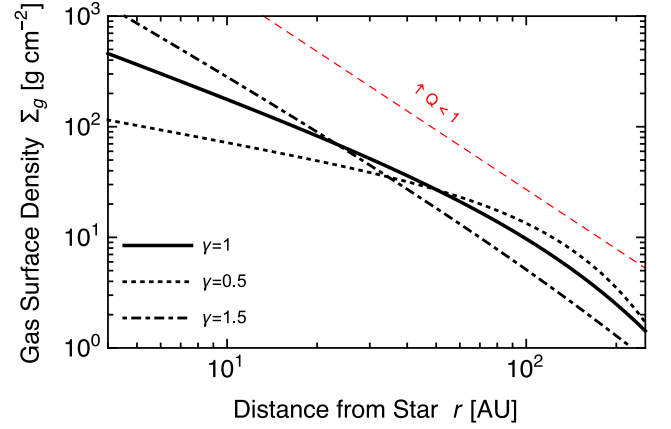}
\caption{Radial profiles of the gas surface density $\Sigma_{\rm g}$ adopted 
in this study (Equation~\eqref{eq:Sigma_g}). 
The solid, dotted, and dot-dashed lines are for $\gamma = 1$ (fiducial), 0.5, and 1.5, respectively.
The dashed line shows $Q \equiv c_s\Omega/\pi G \Sigma_{\rm g} = 1$.
}
\label{fig:Sigma_g}
\end{figure}

The density structure of the HL Tauri gas disk is unknown. 
Therefore, we simply assume that the gas surface density $\Sigma_{\rm d}(r)$
obeys a power law with an exponential taper \citep{HCGD98,KMY+02,AWH+09},
\beq
\Sigma_{\rm g}(r) =\frac{(2-\gamma)M_{\rm disk}}{2\pi r_c^2}\pfrac{r}{r_{\rm c}}^{-\gamma} 
\exp\left[-\pfrac{r}{r_c}^{2-\gamma} \right],
\label{eq:Sigma_g}
\eeq
where $r_{\rm c}$ and $M_{\rm disk}$ are a characteristic radius and the total mass 
of the gas disk, respectively,
and $\gamma$ ($0<\gamma <2$) is the negative slope of $\Sigma_{\rm g}$ at $r \ll r_{\rm c}$.
We take $\gamma = 1$ as the fiducial value but also consider $\gamma = 0.5$ and 1.5.
The dependence of our simulation results on $\gamma$ will be studied in Section~\ref{sec:gamma}.
The values of $M_{\rm disk}$ and $r_{\rm c}$ are fixed to $0.2M_\sun$ and $150~\AU$, respectively.
The adopted disk mass is about twice the upper end of the previous mass estimates for the HL Tau disk \citep{GDPB11,KLM11,KLMW15}.
We assume such a massive disk because the dust mass in the disk decreases with time 
due to the radial drift of dust particles (see Section~\ref{sec:time}). 
Figure~\ref{fig:Sigma_g} shows the surface density profiles for $\gamma = 1$, 0.5, and 1.5. 

Since the disk is assumed to be vertically isothermal, 
the vertical distribution of the gas density obeys a Gaussian with the midplane value 
$\rho_{\rm g} = \Sigma_{\rm g}/(\sqrt{2\pi}H_{\rm g})$, 
where $H_{\rm g} = c_s/\Omega$ is the gas scale height, $c_{\rm s}$ is the sound speed, 
and $\Omega$ is the Keplerian frequency. 
The isothermal sound speed is given by $c_{\rm s} = \sqrt{\kB T/m_\mu}$,
where $\kB$ is the Boltzmann constant and $m_\mu$ is the mean molecular mass 
of the disk gas assumed to be $m_\mu = 2.3~{\rm amu}$.  
The Keplerian frequency is given by $\Omega = \sqrt{GM_*/r^3}$, 
where $G$ is the gravitational constant and $M_*$ is the stellar mass. 
We adopt $M_* = 1M_\odot$ so that the sum of the stellar and disk masses in the fiducial model,
$M_* + M_{\rm disk} = 1.2M_\sun$, 
is within the range of the previous estimates for the HL Tau system \citep{BSCG90,SB91,ALMA+15}. 

We note here that the assumed disk model is marginally gravitationally stable: 
the Toomre stability parameter $Q \equiv c_{\rm s}\Omega/(\pi G\Sigma_{\rm g})$ satisfies $Q \ga 1$ 
at all radii for all choices of $\gamma$. This can be seen in Figure~\ref{fig:Sigma_g}, 
where the surface density corresponding to $Q = 1$ is shown by the dashed line.  

\section{Sublimation and Sintering of Icy Dust}\label{sec:subsint}
We model dust in the HL Tau disk as aggregates of (sub)micron-sized grains.
Each constituent grain, which we call a monomer, is assumed to be 
coated by an ice mantle composed of various volatile molecules (Section~\ref{sec:volatiles}). 
The composition of the mantle at a given distance from the central star
is determined by using the equilibrium vapor pressure curves for the volatile species 
(Sections~\ref{sec:Pev} and \ref{sec:snow}). 
The equilibrium vapor pressures also determine the rate at which the sintering of aggregates
proceeds at each orbital distance (Section~\ref{sec:sint}).
The sintering rate will be used to determine the sticking efficiency of the aggregates
in our dust coagulation simulations (see Section~\ref{sec:Dm}).

\subsection{Volatile Composition}\label{sec:volatiles}
\begin{deluxetable}{lcc}
\tablecaption{Abundances of Major Cometary Volatiles Relative to ${\rm H_2O}$ (in Percent) }
\tablecolumns{3}
\tablewidth{0pt}
\tablehead{
\colhead{Species} &  \colhead{Cometary Value\tablenotemark{a}} & \colhead{Adopted Value $f_{j}$} 
}
\startdata
${\rm H_2O}$ & 100 & 100 \\
${\rm NH_3}$ & 0.2--1.4 & 1 \\
${\rm CO_2}$ & 2--30 &   10 \\
${\rm H_2S}$ & 0.12--1.4 & 1 \\
${\rm C_2H_6}$ & 0.1--2 & 1 \\
${\rm CH_4}$ &  0.4--1.6 & 1 \\
CO                 &  0.4--30 & 10
\enddata
\tablenotetext{a}{\citet{MC11}}
\label{tab:abundance}
\end{deluxetable}
We assume that the volatile composition of the HL Tau disk is similar 
to that of comets in our solar system.  
We select six major cometary volatiles in addition to ${\rm H_2O}$
and take their abundances relative to ${\rm H_2O}$ 
to be consistent with cometary values \citep{MC11}.
The volatiles we select are ammonia (${\rm NH_3}$), carbon dioxide (${\rm CO_2}$), 
hydrogen sulfide (${\rm H_2S}$), ethane (${\rm C_2H_6}$), methane (${\rm CH_4}$), 
and carbon monoxide (CO).
We neglect another equally abundant species, methanol (${\rm CH_3OH}$), 
because the snow line of ${\rm CH_3OH}$ is very close to that of more abundant ${\rm H_2O}$ \citep{S11b}.
Table~\ref{tab:abundance} lists the abundances we adopt and  
the observed ranges of cometary abundances taken from \citet{MC11}.

\subsection{Equilibrium Vapor Pressures}\label{sec:Pev}
\begin{deluxetable}{lcccc}
\tablecaption{Vapor Pressure Parameters}
\tablecolumns{4}
\tablewidth{0pt}
\tablehead{
\colhead{Species} &  \colhead{$L_j$~(K)} & \colhead{$A_j$}  & Ref. & \colhead{$L_{j,\rm tuned}$~(K)}
}
\startdata
${\rm H_2O}$ & 6070 & 30.86    & 1  & 5463   \\
${\rm NH_3}$ & 3754 & 30.21     & 2 & 3379   \\
${\rm CO_2}$ & 3148 & 30.01    & 2  &   \nodata         \\
${\rm H_2S}$ & 2860 & 27.70     & 3  &  \nodata          \\
${\rm C_2H_6}$ & 2498 & 30.24    & 4 & 2248 \\
${\rm CH_4}$ &  1190 & 24.81   & 2 &  \nodata   \\
CO                 &  981.8 & 26.41  & 2 & \nodata 
\enddata
\tablerefs{(1) \citet{BFDGS97}, (2) \citet{YNF83}, (3) \citet{CRC14}, (4) \citet{MAY92}.}
\label{tab:Pev}
\end{deluxetable}
The equilibrium vapor pressures of volatiles
determine the temperatures at which sublimation and sintering occurs. 
In this study, we approximate the equilibrium vapor pressure 
for each volatile species $j$ by the Arrhenius form
\beq
P_{{\rm ev},j} = \exp\left(-\frac{L_j}{T} + A_j \right) ~{\rm dyn~cm^{-2}},
\label{eq:Pev}
\eeq
where $L_j$ is the heat of sublimation in Kelvin and $A_j$ is a dimensionless constant.  
Table~\ref{tab:Pev} summarizes the values of $L_j$ and $A_j$ for the seven volatile species
considered in this study.
For ${\rm CO}$, ${\rm CH_4}$, ${\rm CO_2}$, and ${\rm NH_3}$, 
we follow \citet{S11b} and take the values from Table 2 of \citet{YNF83}. 
The values for ${\rm C_2H_6}$ are derived from the analytic expression of the vapor pressure  
by \citet[][see their Table III; note that we here neglect the small offset in $T$ 
in their original expression]{MAY92}.
For ${\rm H_2S}$, we determined $L_j$ and $A_j$ by fitting Equation~\eqref{eq:Pev} 
to the vapor pressure data provided by \citet[][page 6-92]{CRC14}. 
Figure~\ref{fig:Pev} shows $P_{{\rm ev},j}$ of the seven volatile species 
as a function of $r$ for the temperature distribution given by Equation~\eqref{eq:T}. 

Strictly speaking, the vapor pressure data given in Table~\ref{tab:Pev} only apply to pure ices.
In protoplanetary disks, volatiles  
may be trapped inside the ${\rm H_2O}$ mantle of dust grains instead 
of being present as pure ices.
If this is the case, the volatiles would sublimate not only at the sublimation temperatures for pure ices
but also at higher temperatures where monolayer desorption from the ${\rm H_2O}$ substrate,
phase transition of ${\rm H_2O}$ ice, 
or co-desorption with the ${\rm H_2O}$ ice takes place \citep{CDF+03,CAC+04,MMBG14}.
However, all these high-temperature desorption processes are irrelevant to neck formation (sintering) 
because the desorbed molecules are unable to recondense onto grain surfaces at such high temperatures. 
By using the vapor pressure data for pure ices, we effectively neglect all these desorption processes.

When estimating the locations of the snow lines, 
it is important to note that the vapor pressure data in the literature are subject to 
small but still non-negligible uncertainties. 
For example, the values of the sublimation energies we use for 
${\rm H_2O}$, ${\rm NH_3}$, ${\rm CO_2}$, and ${\rm CO}$ are 10--20\% higher 
than those derived from the very recent temperature programmed desorption experiments 
by \citet[][see their Table 4]{MMBG14}.
\citet{LSSD14} compiled the sublimation energies of major cometary volatiles
from different experimental methods, and showed that the published sublimation energies
have standard deviation of 14, 8, 11, and 8\% 
for ${\rm NH_3}$, ${\rm CO_2}$, ${\rm CH_4}$ and ${\rm CO}$, 
respectively (see their Table~2 and Figures~4 and 5).
Such uncertainties might be present in the vapor pressure data for other volatiles species. 
As we will demonstrate in Section~\ref{sec:L}, 
even a 10\% uncertainty in $L_j$ can lead to 
a 20--30\% uncertainty in the location of its snow line,
and a better match between our simulation results and 
the ALMA observation can be achieved if 
the sublimation energies of ${\rm H_2O}$, ${\rm NH_3}$, and ${\rm C_2H_6}$ 
are taken to be 10\% lower than the fiducial values. 
We will denote these tuned sublimation energies by $L_{j, \rm tuned}$ (see Table~\ref{tab:Pev}).
For ${\rm H_2O}$ and ${\rm NH_3}$, the tuned sublimation energies are 
more consistent with the results of \citet{MMBG14}.

\subsection{Snow Lines}\label{sec:snow}
For each volatile species $j$, we define the snow line as the orbit inside which 
the equilibrium pressure $P_{{\rm ev},j}$ exceeds the partial pressure $P_j$.
Assuming that the disk gas is well mixed in the vertical direction, 
$P_j$ is related to the surface number density of $j$-molecules in the gas phase, $N_j$, as 
\beq
P_j = \frac{N_j \kB T}{\sqrt{2\pi}H_{\rm g}}.
\label{eq:Pj}
\eeq
In this study, we do not directly treat the evolution of $N_j$
but instead estimate them by assuming that the ratio between $N_j$ 
and the surface number density of ${\rm H_2O}$ molecules in the solid phase
is equal to the cometary abundance $f_j$  given in Table~\ref{tab:abundance}.
We also assume that the mass fraction of ${\rm H_2O}$ ice inside the aggregates is 50\%. 
Under these assumptions, $N_j$ can be expressed as 
\beq
N_j = \frac{0.5 f_j \Sigma_{\rm d}}{m_{\rm mol,H_2O}}.
\label{eq:Sigmaj}
\eeq
The adopted relative abundances give the relations 
$P_j = 0.1P_{\rm H_2O}$ for $j = {\rm CO}$ and ${\rm CO}_2$,
and $P_j = 0.01P_{\rm H_2O}$ for $j = {\rm NH_3}$, ${\rm H_2S}$, ${\rm C_2H_6}$, and ${\rm CH_4}$.
For the purpose of calculating the locations of the snow lines, 
the simplification made here is acceptable as a first-order approximation,
because the locations of the snow lines are predominantly determined 
by the strong dependence of $P_{{\rm ev},j}$ on $T(r)$ 
and are much less sensitive to a change in $P_j$.

\begin{figure}[t]
\centering
\includegraphics[width=8.5cm]{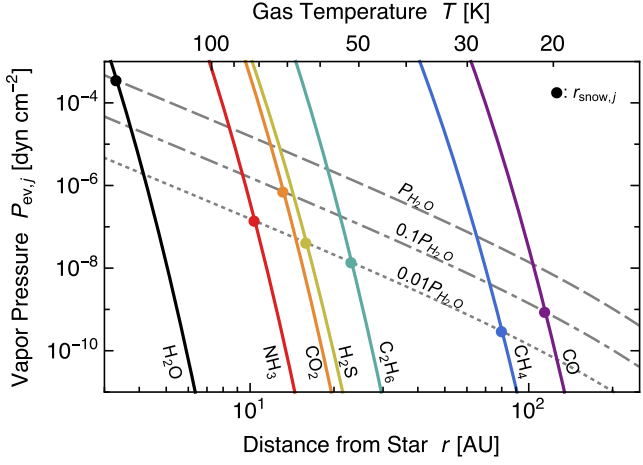}
\caption{
Equilibrium vapor pressures $P_{{\rm ev},j}$ (Equation~\eqref{eq:Pev}) of major cometary volatiles 
as a function of orbital distance $r$ for the temperature profile given by Equation~\eqref{eq:T}.
From left to right:  ${\rm H_2O}$ (black), ${\rm NH_3}$ (red), ${\rm CO_2}$ (orange),
${\rm H_2S}$ (yellow), ${\rm C_2H_6}$ (green), ${\rm CH_4}$ (blue), and CO (purple).
The gray curves show the partial pressures of the volatiles (dashed curve for ${\rm H_2O}$, 
dot-dashed curve for ${\rm CO}$ and ${\rm CO_2}$, dotted curve for the other species) for the gas density profile given by Equation~\eqref{eq:Sigma_g} with $\gamma = 1$ and $\Sigma_{\rm d} = 0.01\Sigma_{\rm g}$.
The filled circles indicate the locations of the snow lines, $r = r_{{\rm snow},j}$.
}
\label{fig:Pev}
\end{figure}

To see the approximate locations of the snow lines in our HL Tau disk model,
we shall now temporarily assume the standard dust-to-gas mass ratio of 
$\Sigma_{\rm d} = 0.01\Sigma_{\rm g}$ throughout the disk.
The dashed, dot-dashed, and dotted lines in Figure~\ref{fig:Pev} show 
the partial pressure curves $P_j = \{1, 0.1, 0.01\} P_{\rm H_2O}$ 
for the fiducial disk model ($\gamma = 1$).
For each volatile species, the location of the snow line is given by the intersection of 
$P_{{\rm ev},j}$ and $P_j$, which is indicated by a filled circle in Figure~\ref{fig:Pev}. 
In this example, the ${\rm H_2O}$ snow line lies at a radial distance of 3.5 AU,
which is well interior to the innermost dark ring of the HL Tau disk lying at $\sim$ 13~AU (see Figure~\ref{fig:T}). 
\citet{ZBB15} suggested that the ${\rm H_2O}$ snow line lies on the 13 AU dark ring
assuming  a higher gas temperature than ours.
The snow lines of ${\rm NH_3}$, ${\rm CO_2}$, ${\rm H_2S}$, and ${\rm C_2H_6}$ are 
narrowly distributed over the intermediate region of 10--30 AU, 
and those of ${\rm CH_4}$ and ${\rm CO}$ are located in the outermost region of 100--150 AU.

\subsection{Sintering Zones}\label{sec:sint}
Sintering is the process of neck growth, and its timescale is inversely proportional to 
the rate at which the neck radius increases \citep[e.g.,][]{SA81}. 
The timescale depends on the size of monomers, with larger monomers  generally 
leading to slower sintering.
In this study, we simply assume monodisperse monomers
and treat their radius $a_0$ as a free parameter (see Section~\ref{sec:a0} for parameter study).
We only consider $a_0 \leq 4~\micron$ because sintering is 
too slow to affect dust evolution in a protoplanetary disk beyond this size range (see below).

When neck growth is driven by vapor transport of volatile $j$, 
its timescale is given by \citep{S11b}
\beq
t_{{\rm sint}, j} = 4.7\times 10^{-3} \frac{a_0^2(2\pi m_{{\rm mol},j})^{1/2} (\kB T)^{3/2}}
{P_{{\rm ev},j}(T) V_{{\rm mol},j}^2 \gamma_j},
\label{eq:tsint}
\eeq
where $m_{{\rm mol},j}$, $V_{{\rm mol},j}$ and $\gamma_j$ 
are the molecular mass, molecular  volume,  and surface energy of the species, respectively.
{The small prefactor $4.7\times 10^{-3}$ comes from the fact that 
the neck radius is much smaller than $a_0$.} 
In general, $t_{{\rm sint}, j}$ rapidly decreases with increasing $T$
because $P_{{\rm ev},j}$ strongly depends on $T$.
For species other than ${\rm H_2S}$ and ${\rm C_2H_6}$, 
we use the same set of $V_{{\rm mol},j}$ and $\gamma_j$ adopted by \citet{S11b}.  
The molecular volumes of ${\rm H_2S}$ and ${\rm C_2H_6}$ 
are estimated as $V_{\rm mol, H_2S} = 5.7\times 10^{-23}~{\rm cm^3}$ and
 $V_{\rm mol, C_2H_6} = 7.1\times 10^{-23}~{\rm cm^3}$ 
assuming that the densities of ${\rm H_2S}$ and ${\rm C_2H_6}$ solids 
are $1~{\rm g~cm^{-3}}$ and $0.7~{\rm g~cm^{-3}}$ \citep{MAY92}, respectively. 
We assume that the surface energy of ${\rm H_2S}$ ice is equal to that of ${\rm H_2S}$ liquid, 
$30~{\rm erg~cm^{-2}}$ \citep{M77}.
For ${\rm C_2H_6}$, we use $\gamma_{\rm C_2H_6} =40~{\rm erg~cm^{-2}}$,
which is the value at $\sim 50~{\rm K}$ \citep{MAY92}.
The locations of the sintering lines discusses below are insensitive to 
the values of $V_{{\rm mol},j}$ and $\gamma_j$
because of the strong temperature dependence of $P_{{\rm ev},j}$.

The necks do not only grow but also are destroyed by at least two processes. 
\begin{enumerate}
\item 
The necks evaporate when the ambient gas temperature exceeds the sublimation temperature
of volatile $j$ hat constitutes the necks. This occurs at $r < r_{{\rm snow},j}$,
where $r_{{\rm snow},j}$ is the orbital radius of the snow line of the volatile.
\item 
The necks break when the aggregate is plastically deformed by another aggregate 
upon collision. 
Unsintered aggregates are known to experience substantial plastic deformation 
even if the collision velocity is much below the fragmentation threshold \citep[][]{DT97,WTS+08}.
Therefore, fully sintered aggregates form only if they never collide with each other 
until the sintering is completed, i.e., only if the sintering timescale is shorter than 
their collision timescale, which we will denote by $t_{\rm coll}$. 
Since $t_{{\rm sint}, j}$ falls off rapidly toward the central star, 
there exists a location $r = r_{{\rm sint},j}$ inside which $t_{{\rm sint}, j} < t_{\rm coll}$
for a given volatile species $j$. 
In this study, we call such locations the sintering lines.
\end{enumerate}
Taken together, each volatile species $j$ causes aggregate sintering 
only inside the annulus defined by $r_{{\rm snow},j} < r < r_{{\rm sint},j}$.
We call such regions the sintering zones.
Our sintering zones are essentially equivalent to the ``sintering regions'' of \citet{S11b}.

In order to know the locations of sintering zones, one needs to estimate $t_{\rm coll}$.
In principle, the collision time can be calculated from the size, number density, 
and relative speed of aggregates as we will do in our simulations (see Equation~\eqref{eq:tcoll}).
In this subsection, we shall avoid such detailed calculations and 
instead use a useful formula $t_{\rm coll} = 100\Omega^{-1}$, 
which is an approximate expression for the collision timescale of macroscopic aggregates 
in a turbulent disk with the dust-to-gas mass ratio of 0.01 \citep{TL05,BDH08}.
This is a rough estimate \citep[see][]{SOI16}, but still provides a reasonable estimate
for $r_{{\rm sint},j}$ because $t_{{\rm sint}, j}$ is a steep function of $r$.   

\begin{figure}
\centering
\includegraphics[width=8.5cm]{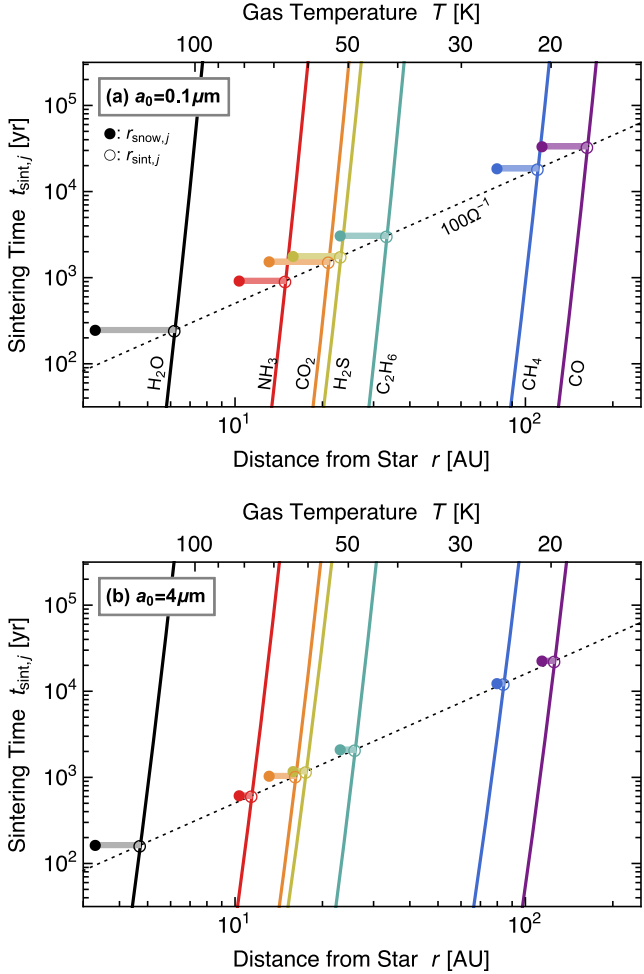}
\caption{
Sintering timescales $t_{\rm sint}$ (Equation~\eqref{eq:tsint}) for major cometary volatiles 
as a function of orbital distance $r$ for our HL Tau disk model.
From left to right:  ${\rm H_2O}$ (black), ${\rm NH_3}$ (red), ${\rm CO_2}$ (orange),
${\rm H_2S}$ (yellow), ${\rm C_2H_6}$ (green), ${\rm CH_4}$ (blue), and CO (purple).
Panels (a) and (b) are for monomer sizes $a_0 = 0.1~\micron$ and $4~\micron$, respectively. 
The dashed gray curve indicates $100\Omega^{-1}$, 
the typical timescale for particle collision in a disk \citep{TL05,BDH08}. 
The filled circles indicate the locations of the snow lines, $r = r_{{\rm snow},j}$ (see also Figure~\ref{fig:Pev}),
while the open circles indicate the locations of the sintering lines, $r = r_{{\rm sint},j}$.
Icy aggregates experience sintering in the sintering zones defined by $r_{{\rm snow},j} < r < r_{{\rm sint},j}$ (horizontal bars). 
}
\label{fig:tsint}
\end{figure}

In Figure~\ref{fig:tsint}, we plot the sintering timescales of the seven volatile species 
as a function of $r$ for our HL Tau disk model with $\gamma = 1$. 
We also indicate $t_{\rm coll} = 100\Omega^{-1}$ by the dotted lines, the location of the sintering lines ($r_{{\rm sint},j}$) by the open circles, and the locations of the sintering zones
($r_{{\rm snow},j} < r < r_{{\rm sint},j}$) by the horizontal bars. 
We here consider two different values of $a_0$, $0.1~\micron$ and $4~\micron$ (panels a and b, respectively), to highlight the importance of this parameter in our sintering model.
For $a_0 = 0.1~\micron$, the width of each sintering zone is $30$--$50\%$ of $r_{{\rm sint},j}$,
and the sintering zones of ${\rm NH_3}$, ${\rm CO_2}$, and ${\rm H_2S}$ significantly overlap with each other.
A larger $a_0$ leads to longer sintering timescales and hence to narrower sintering zones.
For $a_0 = 4~\micron$, the sintering zones for species other than ${\rm H_2O}$ almost disappear.
For this reason, we will restrict ourselves to $a_0 \leq 4~\micron$ in the following sections. 

\section{Simulation Method}\label{sec:method}
As introduced in Section~\ref{sec:intro}, sintering is expected to
reduce the sticking efficiency of dust aggregates.
In protoplanetary disks, this can occur in the sintering zones defined in Section~\ref{sec:sint}. 
To study how the presence of the sintering zones affects 
the radial distribution of dust in a disk, 
we conduct global simulations of dust evolution including sintering-induced fragmentation.
In our simulations, we calculate the evolution of the surface density and representative size 
of icy aggregates due to coagulation and radial drift 
using the single-size approach (Section~\ref{sec:global}).
The radial drift is due to the aerodynamical drag by the gas disk \citep{AHN76,W77a},
and its velocity depends on the size of the aggregates and on the gas surface density (Section~\ref{sec:vr}).  
We also consider turbulence in the gas disk to compute the vertical scale height and collision velocity 
of the aggregates (Section~\ref{sec:tcoll}). 
The sticking efficiency of the aggregates is given 
as a function of their sintering timescale (Section~\ref{sec:Dm}).
Based on the work of \citet{S99} and \citet{SU14}, 
we assume that sintered aggregates have a lower sticking efficiency than unsintered aggregates.
{We do not consider spontaneous (noncollisional) breakup of icy aggregates 
due to sintering \citep{S11a} and sublimation 
\citep{SS11} near the snow lines. 
These effects might be important in the vicinity of the ${\rm H_2O}$ snow line
where grains constituting the aggregates would lose a significant fraction of 
their volume (see Section~\ref{sec:ripening}). 
The aggregate internal density is fixed to be $0.26~{\rm g~cm^{-3}}$
assuming a material (monomer) density of $1.3~{\rm g~cm^{-3}}$ 
and a constant aggregate porosity of 80\%. 
Possible effects of porosity evolution will be discussed in Section~\ref{sec:porosity}. 
}

The output of the simulations is then used to generate the radial profiles of dust thermal emission
(Sections~\ref{sec:opacity} and \ref{sec:emission}), 
which we will compare with the ALMA observation of the HL Tau disk in 
Sections~\ref{sec:results} and \ref{sec:param}.

\subsection{The Single-size Approach for Global Dust Evolution}\label{sec:global}
We simulate the global evolution of particles in a gas disk using the single-size approximation.  
We assume that the total solid mass at each orbital radius $r$
is dominated by particles having mass $m=m_*(r)$.
We then follow the evolution of the solid surface density $\Sigma_{\rm d}(r)$ and
``representative'' (mass-dominating) particle mass $m_*(r)$ 
taking into account aggregate collision and radial drift 
(see Equations~\eqref{eq:evol_Sigmad} and \eqref{eq:evol_mstar} below).
The single-size approach (or mathematically speaking, moment approach) 
has often been used in the modeling of particle growth 
in planetary atmospheres \citep[e.g.,][]{F94,O14} as well as in protoplanetary disks
\citep[e.g.][{see also Appendix of \citealt{SOI16} for the mathematical background 
of the single-size approximation and a comparison between single-size and full-size simulations}]{KSR01,G07,BKE12,ECM16,KODT16}.
This approach allows us to track the evolution of the mass budget of dust in a disk
without using computationally expensive Smoluchowski's coagulation equation.
{A drawback of this approach is that
one has to assume aggregates' size distribution at each orbit whenever it is needed. 
In this study, we will assume a power-law size distribution when we predict dust 
emission from the disk (see Section~\ref{sec:opacity}).}

Under the single-size approximation, the evolution of 
 $\Sigma_{\rm d}$ and $m_*$ is described by \citep{O14,SOI16}
\beq
\frac{\pd \Sigma_{\rm d}}{\pd t} + \frac{1}{r}\frac{\pd}{\pd r}(r v_{\rm r}\Sigma_{\rm d}) = 0,
\label{eq:evol_Sigmad}
\eeq
\beq
\frac{\pd m_*}{\pd t} + v_{\rm r}\frac{\pd m_*}{\pd r}  = \frac{\Delta m_*}{t_{\rm coll}},
\label{eq:evol_mstar}
\eeq
where $v_{\rm r}$ and $t_{\rm coll}$ are the radial drift velocity and mean collision time 
of the representative particles, respectively, and $\Delta m_*$ is the change of $m_*$ 
upon a single aggregate collision.   
Equation~\eqref{eq:evol_Sigmad} expresses the mass conservation for solids, 
while Equation~\eqref{eq:evol_mstar} states that the growth rate of representative particles 
along their trajectory, ${\rm D}m_*/{\rm D}t \equiv \pd m_*/\pd t + v_{\rm r} \pd m_*/\pd r$, 
is equal to $\Delta m_*/t_{\rm coll}$.
The mass change $\Delta m_*$ is equal to $m_*$ when the collision results in pure sticking,
while $\Delta m_* < m_*$ when fragmentation or erosion occurs. 
Our Equations~\eqref{eq:evol_Sigmad} and \eqref{eq:evol_mstar} correspond 
to Equations~(3) and (8) of \citet{O14}, respectively, although Equation~(8) of \citet{O14} 
assumes $\Delta m_* = m_*$.
The expressions for $v_{\rm r}$, $t_{\rm coll}$, and $\Delta m_*$ will be given in
Section~\ref{sec:vr}, \ref{sec:tcoll}, \ref{sec:Dm}, respectively. 

We solve Equations~\eqref{eq:evol_Sigmad} and \eqref{eq:evol_mstar} 
by discretizing the radial direction into 300 logarithmically spaced binds spanning 
from $r_{\rm in} = 1~\AU$ and $r_{\rm out} = 1000~\AU$.
The parameter sets and initial conditions used in the simulations 
will be described in Section~\ref{sec:runs}.

\subsection{Radial Drift}\label{sec:vr}
The motion of a particle in a gas disk is characterized
 by the dimensionless Stokes number ${\rm St} \equiv \Omega t_{\rm s}$, 
 where $\Omega$ is the Keplerian frequency and $t_{\rm s}$ is the particle's stopping time. 
When the particle radius is much smaller than the mean free path of the gas molecules in the disk, 
as is true for particles treated in our simulations, the stopping time is given by Epstein's drag law.
The Stokes number of a representative aggregate at the midplane can then be written as \citep{BDB10}
\beq
{\rm St} = \frac{\pi}{2}\frac{\rho_{\rm int} a_*}{\Sigma_{\rm g}},
\label{eq:St}
\eeq
where $\rho_{\rm int} = 0.26~{\rm g~cm^{-3}}$ and $a_* \equiv (3m_*/4\pi\rho_{\rm int})^{1/3}$ 
are the internal density and radius of the aggregate, respectively. 
We use Equation~\eqref{eq:St} whenever we calculate ${\rm St}$.

The drift velocity $v_{\rm r}$ is given by \citep{AHN76,W77a}
\beq
v_{\rm r} = -2\eta v_{\rm K} \frac{\rm St}{1+{\rm St^2}},
\label{eq:vr}
\eeq
where 
\beq
\eta = -\frac{1}{2}\pfrac{c_{\rm s}}{v_{\rm K}}^2 \frac{d\ln P_{\rm g}}{d\ln r}
\label{eq:eta}
\eeq
is the parameter characterizing the sub-Keplerian motion of the gas disk,  
$v_{\rm K} = r\Omega$ is the Keplerian velocity, 
and $P_{\rm g} = \rho_{\rm g} c_{\rm s}^2$ is the midplane gas pressure. 
In our disk model, $dP_{\rm g}/dr < 0$ and therefore $v_{\rm r} < 0$ everywhere.
At $r \ll r_{\rm c}$, we approximately have 
$\eta \approx 1.8\times 10^{-3}(r/1~{\rm AU})^{0.43}$
and $\eta v_{\rm K} \approx 52(r/1~{\rm AU})^{-0.07}~{\rm m~s^{-1}}$.

\subsection{Collision Time}\label{sec:tcoll}
We evaluate the collision term $\Delta m_*/t_{\rm coll}$ 
assuming that collisions between representative aggregates dominate the evolution of $m_*$.
Under this assumption, the collision time is approximately given by 
\beq
t_{\rm coll} = \frac{1}{4\pi a_*^2 n_* \Delta v},
\label{eq:tcoll}
\eeq
where $4\pi a_*^2$, $n_*$ and $\Delta v$ are the collisional cross section, 
number density, and collision velocity of the representative aggregates, respectively. 
We use the midplane values for $n_*$ and $\Delta v$. 
{We do not consider erosion of representative aggregates by a number of small grains \citep{SKK13,KODT15} because there remain uncertainties in the threshold velocity 
for erosive collisions as a function of the projectile mass \citep[see Section 2.3.2 of][]{KODT15}. }

To evaluate $n_*$, we consider disk turbulence 
and assume that vertical settling balances with turbulent diffusion for the representative aggregates.
We parameterize the strength of disk turbulence with the dimensionless parameter 
$\alpha_{\rm t} = D/c_{\rm s}H_{\rm g}$, 
where $D$ is the particle diffusion coefficient in the turbulence.
For simplicity, $\alpha_{\rm t}$ is assumed to be independent of time and distance from 
the midplane, but we allow $\alpha_{\rm t}$ to depend on $r$ (see Section~\ref{sec:runs}). 
Under this assumption, $n_*$ at the midplane can be written as 
\beq
n_* = \frac{\Sigma_{\rm d}}{\sqrt{2\pi} H_{\rm d}m_*},
\label{eq:nstar}
\eeq
where 
\beq
H_{\rm d} = \left( 1+\frac{{\rm St}}{\alpha_{\rm t}}\frac{1+2{\rm St}}{1+{\rm St}} \right)^{-1/2}H_{\rm g}.
\label{eq:Hd}
\eeq 
is the scale height of the representative aggregates  \citep{DMS95,YL07}.

The collision velocity $\Delta v$ is given by the root sum square of 
the contributions from Brownian motion, gas turbulence \citep{OC07}, 
and size-dependent drift relative to the gas disk \citep{AHN76,W77a}.
The expressions for these contributions can be found in, e.g., Section 2.3.2 of \citet{OTKW12}.  
The contributions from turbulence and drift motion are functions 
of the Stokes numbers of two colliding aggregates, ${\rm St}_1$ and ${\rm St}_2$. 
In this study, we set ${\rm St}_1 = {\rm St}$ and ${\rm St}_2 = 0.5{\rm St}$
because we consider collisions between aggregates similar in size.
\citet{SOI16} and \citet{KODT16} have recently shown that such a choice best reproduces the results 
of coagulation simulations that resolve the full size distribution of the aggregates.
As long as ${\rm St} \ll 1$, the collision velocity is an increasing function of ${\rm St}$.

For macroscopic aggregates satisfying $10^{-4} \la {\rm St} \ll 1$, 
either  turbulence or radial drift mostly dominates their collision velocity. 
For this range of ${\rm St}$, the collision velocities driven by turbulence and radial drift 
are approximately given by $\Delta v_{\rm t} \approx \sqrt{2.3\alpha_{\rm t}{\rm St}}c_{\rm s}$ 
and $\Delta v_{\rm r} \approx 2\eta v_{\rm K}|{\rm St}_1-{\rm St}_2| \approx \eta v_{\rm K}{\rm St}$, respectively, 
where we have used ${\rm St}_1 = {\rm St}$ and ${\rm St}_2 = 0.5{\rm St}$
(for the expression of $\Delta v_{\rm t}$, see Equation~(28) of \citealt{OC07}).

\subsection{Collisional Mass Gain/Loss}\label{sec:Dm}
\begin{figure}[t]
\centering
\includegraphics[width=8.5cm]{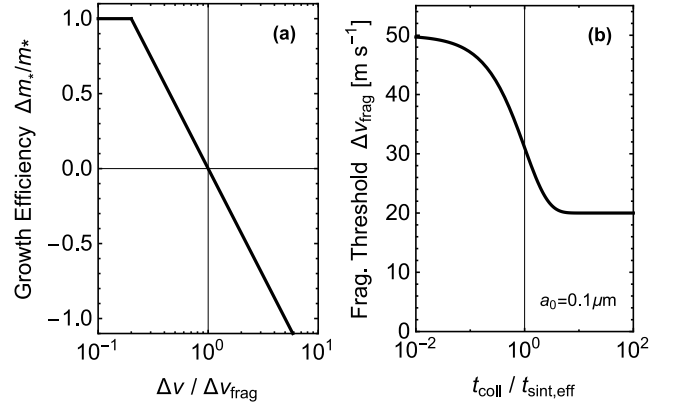}
\caption{Panel (a): growth efficiency $\Delta m_*/m_*$  (Equation~\eqref{eq:Dm})
versus the scaled collision velocity $\Delta v/\Delta v_{\rm frag}$.
Panel (b): catastrophic fragmentation threshold $\Delta v_{\rm frag}$  (Equation~\eqref{eq:Dvfrag})
versus the ratio between the collision and effective sintering timescales $t_{\rm coll}/t_{\rm sint, eff}$
for $a_0 = 0.1~\micron$. 
}
\label{fig:coll}
\end{figure}

{We denote the change in $m_*$ due to a single collision between two mass-dominating aggregates
by $\Delta m_*$. 
Following \citet{OH12}, we model $\Delta m_*$ as }
\beq
\Delta m_* =  \min\left\{1, -\frac{\ln\left(\Delta v/\Delta v_{\rm frag}\right)}{\ln 5} \right\} m_*,
\label{eq:Dm}
\eeq
where the fragmentation threshold $\Delta v_{\rm frag}$ 
characterizes the sticking efficiency of the colliding aggregates.  
The mass change is positive for $\Delta v<\Delta v_{\rm frag}$
and negative for $\Delta v > \Delta v_{\rm frag}$ as shown in Figure~\ref{fig:coll}(a).
{Equation~\eqref{eq:Dm} is a fit to the data of the collision simulations 
for unsintered aggregates by \citet[][their Figure~11]{WTS+09}.}
{For sintered aggregates, Equation~\eqref{eq:Dm} overestimates the sticking efficiency 
at low collision velocities where the aggregates bounce rather than stick \citep{S99}.
However, we will show in Section~\ref{sec:bouncing} that the bouncing hardly alters 
the evolution of $m_*$ in the sintering zone. For simplicity, we use Equation~\eqref{eq:Dm} 
for both unsintered and sintered aggregates.
}

To account for the effect of sintering on the aggregate sticking efficiency, 
we model the fragmentation threshold $\Delta v_{\rm frag}$ as
\beq
\Delta v_{\rm frag} 
=  \Delta v_{\rm frag, NS}\exp\left(-\frac{t_{\rm coll}}{t_{\rm sint,eff}}\right)
+ \Delta v_{\rm frag, S}\left[1- \exp\left(-\frac{t_{\rm coll}}{t_{\rm sint,eff}}\right)\right],
\label{eq:Dvfrag}
\eeq
where 
\beq
t_{\rm sint,eff} \equiv \left(\sum_j t_{{\rm sint},j}^{-1}\right)^{-1}
\label{eq:t_sint_eff}
\eeq
is the effective sintering timescale (see Equation~\eqref{eq:tsint} for the definition of 
the individual sintering timescale $t_{{\rm sint},j}$), 
and $\Delta v_{\rm frag, NS}$ and $\Delta v_{\rm frag, S} (< \Delta v_{\rm frag, NS})$ are 
the thresholds for unsintered and sintered aggregates, respectively.
In Equation~\eqref{eq:t_sint_eff}, the summation is taken over all solid-phase volatiles,
i.e., volatiles satisfying $r > r_{{\rm snow},j}$.
As described in Section~\ref{sec:snow}, 
the snow line location $r_{{\rm snow},j}$ for each species is calculated 
from the relation $P_{{\rm ev},j}(r_{{\rm snow},j}) = P_j(r_{{\rm snow},j})$ 
with Equations~\eqref{eq:Pj} and \eqref{eq:Sigmaj}.
At $r < r_{{\rm snow,H_2O}}$, where all volatile ices sublimate, 
we exceptionally set $\Delta v_{\rm frag} = \Delta v_{\rm frag,S}$
to mimic the low sticking efficiency of bare silicate grains compared to
(unsintered) ice-coated grains \citep[e.g.,][]{CTH93}. 

Equation~\eqref{eq:Dvfrag} is constructed so that 
$\Delta v_{\rm frag} \approx \Delta v_{\rm frag, NS}$ 
in the non-sintering zones ($t_{\rm coll} \ll t_{\rm sint,eff}$)
and $\Delta v_{\rm frag} \approx \Delta v_{\rm frag, S}$ in the sintering zone 
($t_{\rm coll} \gg t_{\rm sint,eff}$).
Simulations of aggregate collisions suggest that $\Delta v_{\rm frag, NS} \sim 50~{\rm m~s^{-1}}$ 
\citep{WTS+09} and $\Delta v_{\rm frag, S} \sim 20~{\rm m~s^{-1}}$ \citep{SU14}
if the colliding aggregates are identical and made of $0.1~\micron$-sized ice monomers.
The theory of particle sticking \citep{JKR71}, on which the simulations by \citet{WTS+09} are based,
indicates that $\Delta v_{\rm frag,NS}$ scales with the monomer size as $a_0^{-5/6}$ \citep{CTH93,DT97}.
For $\Delta v_{\rm frag,S}$, the scaling is yet to be studied, 
so we simply assume the same scaling as for $\Delta v_{\rm frag,NS}$. 
We thus model $\Delta v_{\rm frag,NS}$ and $\Delta v_{\rm frag,S}$ as 
\beq
\Delta v_{\rm frag, NS} = 50 \pfrac{a_0}{0.1~\micron}^{-5/6}~{\rm m~s^{-1}},
\label{eq:Dvfrag_NS}
\eeq
\beq
\Delta v_{\rm frag, S} = 
20 \pfrac{a_0}{0.1~\micron}^{-5/6}~{\rm m~s^{-1}}.
\label{eq:Dvfrag_S}
\eeq
Figure~\ref{fig:coll}(b) shows $\Delta v_{\rm frag}$ versus $t_{\rm coll}/t_{\rm sint,eff}$ 
for $a_0 = 0.1~\micron$.

\subsection{Aggregate Opacity}\label{sec:opacity}
We calculate the absorption cross section of porous aggregates 
using the analytic expression by \citet[][their Equation~(18)]{KOTN14},
which is based on Mie calculations with effective medium theory. 
Monomers are treated as composite spherical grains made of 
astronomical silicates, carbonaceous materials, and water ice 
with the mass abundance ratio of 2.64:3.53:5.55 \citep{PHB+94}.
We calculate the effective refractive index of the monomers using the Bruggeman mixing rule.
The optical constants of silicates, carbons, and water ice are 
taken from \citet{D03b}, \citet[][data for ACH2 samples]{ZMCB96}, 
and \citet[][data for $T=-60~{\rm^\circ C}$]{W84}, respectively.
We neglect the contribution of volatiles other than ${\rm H_2O}$ 
to the monomer optical properties.
The effective refractive index of porous aggregates are computed 
using the Maxwell--Garnett rule in which the monomers 
are regarded as inclusions in vacuum.  

{Since we adopt the single-size approach, we only track 
the evolution of aggregates dominating the dust surface density.
However, these aggregates do not necessarily dominate millimeter dust emission from a disk. 
While the mass-dominating aggregates are generally the largest aggregates in the population 
\citep[e.g.,][]{OTKW12,BKE12}, smaller ones can dominate the millimeter opacity of the population
when the largest aggregates are significantly larger than a millimeter in radius. 
In order to take into account this effect, we assume a size distribution 
only when we calculate dust opacities. 
Specifically, we assume a power law distribution 
\beq
N_{\rm d}'(a) = \left\{ \begin{array}{ll}
C a^{-3.5}, & a_{\rm min} < a < a_* \\
0, & {\rm otherwise},
\end{array}\right.
\label{eq:dNda}
\eeq
 where $N_{\rm d}'(a)$ is the column number density of aggregates per unit aggregate radius 
 $a$ ($<a_*$), $a_{\rm min}$ is the minimum aggregate radius, and $C$ is 
the normalization constant determined by the condition 
$\int_0^\infty m N_{\rm d}'(a) {\rm d}a = \Sigma_{\rm d}$. 
We fix $a_{\rm min}$ to be $0.1~\micron$ with the understanding that millimeter opacities
are insensitive to the choice of $a_{\rm min}$ as long as $a_{\rm min} \ll 1~{\rm mm}$. 
The slope of $-3.5$ is based on the classical theory of fragmentation cascades 
\citep[][see \citealt{BOD11} for how the coagulation of the fragments modifies this value]{D69,TIN96}.
Therefore, Equation~\eqref{eq:dNda} would overestimate the amount of fragments 
when the collisions between the largest aggregates (which are the source of the fragments)
do not lead to their catastrophic disruption. 
Possible effects of this simplification will be discussed in Section~\ref{sec:distr}. 
}

\begin{figure}[t]
\centering
\includegraphics[width=8.5cm]{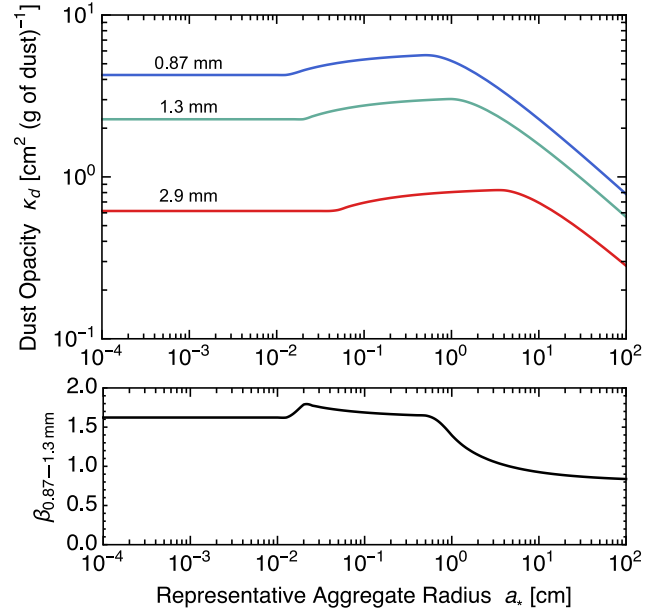}
\caption{Upper panel: 
absorption opacities $\kappa_{\rm d,\nu}$ (in units of ${\rm cm}^2$ per gram of dust) of dust at wavelengths 
$\lambda =$ 0.87 mm, 1.3 mm, and 2.9 mm as a function of the representative aggregates radius $a$.
The opacities are calculated by assuming that the aggregate size distribution (not resolved in our simulation)
obeys a power law $N_{\rm d}'(a) \propto a^{-3.5}$ {(Equation~\eqref{eq:dNda})}.
Lower panel: opacity slope at $\lambda = 0.87$--1.3 mm versus $a_*$.
}
\label{fig:kappa}
\end{figure}

The upper panel of Figure~\ref{fig:kappa} shows our dust opacities $\kappa_{{\rm d},\nu}$ 
at wavelengths $\lambda = 0.87$, 1.3, and 2.9 mm (corresponding to 
ALMA Bands 7, 6, and 3, respectively) as a function of $a_*$.
Note that the opacities are expressed in units of ${\rm cm^2}$ per gram of dust.
In the lower panel of  Figure~\ref{fig:kappa}, we plot the opacity slope 
measured at $\lambda = 0.87$--1.3 mm, 
$\beta_{\rm 0.87\textrm{--}1.3~mm} \equiv  
{\ln(\kappa_{\rm 0.87~mm}/\kappa_{\rm 1.3~mm})}/{\ln(\nu_{\rm 0.87~mm}/\nu_{\rm 1.3~mm})}$.
Our model gives $\beta_{\rm 0.87\textrm{--}1.3~mm} \approx 1.7$ at $a_* \ll 1~{\rm cm}$ 
and $\beta_{\rm 0.87\textrm{--}1.3~mm} \approx 0.8$ in the opposite limit. 

\subsection{Dust Thermal Emission}\label{sec:emission}
We calculate the intensities $I_\nu$ of dust thermal emission at each orbital radius $r$ as 
\beq
I_\nu(r) = \left[1-\exp\left( - \tau_\nu(r) \right) \right] B_\nu\left(T(r)\right),
\label{eq:Inu}
\eeq
where $B_\nu(T)$ is the Planck function,
\beq
\tau_\nu(r) = \frac{\kappa_{\rm d,\nu}(r)\Sigma_{\rm d}(r)}{\cos i} 
\label{eq:tau}
\eeq
is the line-of-sight optical depth, and $i$ is the disk inclination.
We use $i = 46.7^\circ$ for the HL Tau disk \citep{ALMA+15}.
The Planck brightness temperature $T_{\rm B}$ is computed by solving 
the equation $I_\nu = B_\nu(T_{\rm B})$ for $T_{\rm B}$.
Equation~\eqref{eq:tau} assumes that the dust disk is geometrically thin, 
i.e., the radial distance over which $\kappa_{\rm d,\nu}(r)$ and $\Sigma_{\rm d}(r)$ vary 
is longer than the dust scale height. This assumption is, however, not always satisfied 
in our simulations as we discuss in detail in Section~\ref{sec:a0}. 

We will also use the flux density 
\beq
F_\nu = \frac{2\pi \cos i}{d^2} \int_{r_{\rm in}}^{r_{\rm out}} I_\nu(r) r dr,
\label{eq:Fnu}
\eeq
where $d$ is the distance to HL Tau, 
$r_{\rm in} = 1~\AU$ and $r_{\rm out} = 1000~\AU$ 
are the boundaries of our computational domain (see Section~\ref{sec:global}),
and the factor $\cos i$ accounts for the ellipticity of the disk image.   
In accordance with \citet{ALMA+15}, we set $d = 140~{\rm pc}$, the standard mean distance to Taurus.

When comparing our simulation results with the ALMA observation, 
it is useful to smooth the simulated radial emission profiles at the spatial resolution of ALMA. 
In this study, we do this in the following two steps. 
First, we generate projected images of our simulation snapshots 
assuming the disk inclination of $46.7^\circ$. 
For simplicity, the geometrical thickness of the disks is neglected in this process.
Second, we smooth the ``raw'' images along their major axis using a circular Gaussian
with the FWHM angular resolutions of $\sqrt{85.3\times61.1}$, $\sqrt{35.1\times21.8}$, 
and $\sqrt{29.9\times19.0}$ mas for $\lambda = 2.9$, 1.3, and 0.87 mm, 
in accordance with the ALMA observation of HL Tau at Bands 3, 6, and 7, respectively \citep{ALMA+15}. 
Since we assume $d = 140~{\rm  pc}$, 
these angular resolutions translate into the spatial resolutions of 
$\approx$ 10, 3.9, and 3.3 AU at Bands 3, 6, and 7, respectively.

\subsection{Parameter Sets and Initial Conditions}\label{sec:runs}
We conduct ten simulation runs with different sets of model parameters.
Columns 1 through 4 of Table~\ref{tab:runs} list the run names and parameter choices
($\gamma$, $a_0$, $\alpha_{\rm t}$) for the simulation runs. 
Run Sa0 is our fiducial model and assumes 
$\gamma = 1$, $a_0 = 0.1~\micron$, and $\alpha_{\rm t} = 0.03(r/10~\AU)^{1/2}$.
Model Sa0-NoSint is the same as model Sa0 but neglects sintering.  
Runs Sa0-Lgam and Sa0-Hgam are designed to study the dependence of the results 
on the gas surface density slope $\gamma$. 
Runs Sa0-Lalp and Sa0-Halp will be used to study how the sintering-induced ring formation 
scenario constrains the radial distribution of $\alpha_{\rm t}$ in the HL Tau disk.
Runs La0 and LLa0 assume more fragile aggregates (i.e., larger $a_0$) and weaker turbulence
than in the fiducial run. As we will see, the set of $a_0$ and $\alpha_{\rm t}$ 
controls the degree of dust sedimentation, i.e., the geometrical thickness of the dust disk.
Runs Sa0-tuned and La0-tuned are the same as runs Sa0 and La0, respectively, 
except that they adopt slightly lower $L_j$ for ${\rm H_2O}$, ${\rm NH_3}$, and ${\rm C_2H_6}$
(see $L_{j,\rm tuned}$ in Table~\ref{tab:Pev}).
These runs will be used to quantify possible uncertainties of our results 
that might arise from the uncertainties in the vapor pressure data. 

\begin{deluxetable*}{lcccccccc}
\tablecaption{List of Simulation Runs}
\tablecolumns{9}
\tablewidth{0pt}
\tablehead{
\colhead{Run} &  \colhead{$\gamma$} & \colhead{$a_0$} & \colhead{$\alpha_{\rm t}$}
&  \colhead{$t_{\rm snap}$}
& \multicolumn{3}{c}{$F_\nu$ (Jy) at $t = t_{\rm snap}$} & Section \\
\cline{6-8} \\[-2mm]
\colhead{} & \colhead{} & \colhead{(\micron)} & \colhead{} & \colhead{(Myr)} & 2.9 mm & 1.3 mm & 0.87 mm
}
\startdata
Sa0  & $1$    & $0.1$ & $0.03(r/10~\AU)^{1/2}$ & 0.26 & 0.070 & 0.79 & 2.2 & 
\ref{sec:results} \\
Sa0-NoSint\tablenotemark{a} & $1$    & 0.1 & $0.03(r/10~\AU)^{1/2}$ & 0.12 & 0.063 & 0.79 & 2.3 & 
\ref{sec:ALMA} \\
Sa0-Lgam  & $0.5$ & $0.1$ & $0.03(r/10~\AU)^{1/2}$ & 0.29 & 0.076 & 0.76 & 2.1 &
\ref{sec:gamma} \\
Sa0-Hgam  & $1.5$ & $0.1$ & $0.03(r/10~\AU)^{1/2}$ & 0.05 & 0.072 & 0.80 & 2.3 &
\ref{sec:gamma} \\
Sa0-Lalp  & $1$ & $0.1$ & $0.03$ & 0.18 & 0.066 & 0.77 & 2.2  &
\ref{sec:alpha} \\
Sa0-Halp  & $1$ & $0.1$ & $0.1$ & 0.29 & 0.075 & 0.75 & 2.1 &
\ref{sec:alpha} \\
La0  & $1$    & 1& $10^{-3}(r/10~\AU)^{1/2}$ & 0.41 & 0.064 & 0.78 & 2.3 &
\ref{sec:a0} \\
LLa0  & $1$    & 4 & $10^{-4}(r/10~\AU)^{1/2}$ & 0.45 & 0.062 & 0.80 & 2.4 &
\ref{sec:a0} \\
Sa0-tuned\tablenotemark{b} & $1$    & 0.1 & $0.03(r/10~\AU)^{1/2}$ & 0.27 & 0.068 & 0.77 & 2.2 &
\ref{sec:L} \\
La0-tuned\tablenotemark{b} & $1$    & 1 & $10^{-3}(r/10~\AU)^{1/2}$ & 0.41 & 0.063 & 0.79 & 2.3 &
\ref{sec:L} 
\enddata
\tablenotetext{a}{No sintering}
\tablenotetext{b}{Uses $L_{j, {\rm tuned}}$ instead of $L_j$ for ${\rm H_2O}$, ${\rm NH_3}$, and ${\rm C_2H_6}$ (see Table~\ref{tab:Pev})}
\label{tab:runs}
\end{deluxetable*}

The initial conditions are given by $\Sigma_{\rm d}(t=0,r) = 0.01 \Sigma_{\rm g}(r)$ 
and $a_*(t=0,r) = a_0$, where we have assumed that the dust-to-gas mass ratio 
of the initial disk is 0.01.

\section{Results from the Fiducial Simulation}\label{sec:results}
We now present the results of our simulations in the following two sections. 
In this section, we particularly focus on the fiducial run Sa0 and analyze its results in detail. 
The dependence on model parameters will be discussed in Section~\ref{sec:param}.

\subsection{Evolution of the Total Dust Mass and Flux Densities}\label{sec:time}
In our simulations, the observational appearance of the disk changes 
with time  because dust particles grow and drift inward. 
In particular, the millimeter emission of the disk diminishes as the particles drain onto the central star. 
For this reason, we select from each simulation run one snapshot that best reproduces 
the millimeter flux densities of the HL Tau disk reported by \citet{ALMA+15}. 
Specifically, we calculate the relative errors between the simulated and observed 
flux densities at $\lambda = 0.87$, 1.3, and 2.9 mm as a function of time,
and search for the time $t =t_{\rm snap}$ at which the sum of the relative errors is minimized. 
For reference, the flux densities reported by the ALMA observation 
are 0.0743, 0.744, and 2.14 Jy at $\lambda = 0.87$, 1.3, and 2.9 mm (Bands 7, 6, and 3), respectively 
\citep{ALMA+15}.

\begin{figure}[t]
\centering
\includegraphics[width=8.5cm]{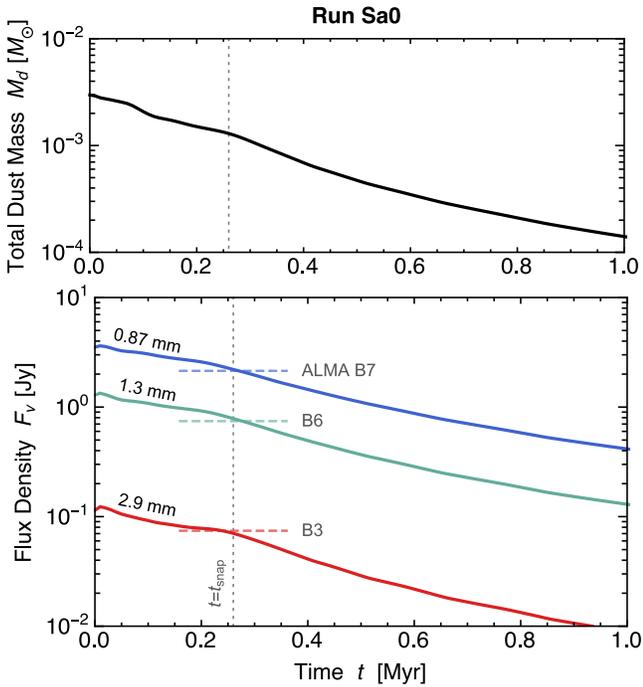}
\caption{Simulated time evolution of the total dust mass $M_{\rm d}$ (upper panel) 
and flux densities $F_\nu$ (lower panel) of the HL Tau disk from simulation run Sa0.
The blue, green, and red solid curves in the lower panel are $F_\nu$ 
at wavelengths $\lambda = 0.87$ mm, 1.3 mm, and 2.9 mm, respectively. 
The dashed horizontal line segments indicate the ALMA measurements
of the flux densities at these wavelengths (Bands 7, 6, and 3, respectively).
The vertical dotted line indicates the time $t = t_{\rm snap}$
at which the simulated flux densities best reproduce the ALMA measurements 
($t_{\rm snap} = 0.26~{\rm Myr}$ for run Sa0; see Section~\ref{sec:time}).
}
\label{fig:MFnu}
\end{figure}

Figure~\ref{fig:MFnu} illustrates such an analysis for our fiducial simulation run Sa0. 
This figure shows the simulated time evolution of the flux densities 
at the three wavelengths as well as the evolution of the total dust mass 
$M_{\rm d}$ within the computational domain.
The dust mass and flux densities decrease on a timescale of $\sim 1~\Myr$,
which reflects the timescale on which the dust near the disk's outer edge  (which dominates $M_{\rm d}$) 
grows into rapidly drifting pebbles \citep[see, e.g.,][]{SOI16}.
Comparing the flux densities from the simulation with those from the ALMA observations
(shown by the dashed horizontal line segments in the lower panel of Figure~\ref{fig:MFnu}),
we find that the sum of the relative errors in the flux densities is minimized when $t = 0.26~\Myr$. 
At this time, the flux densities in the simulation are 
0.070, 0.79, and 2.2 Jy at $\lambda = $ 2.9, 1.3, and 0.87 mm, respectively,
in agreement with the ALMA measurements to an accuracy of less than 6\%. 

Columns 5 through 8 of Table~\ref{tab:runs} list the values of 
$t_{\rm snap}$ and $F_\nu(t=t_{\rm snap})$ for all simulation runs.
We find that $t_{\rm snap}$ falls within the range $0.1$--$0.5~\Myr$.
Since $t_{\rm snap}$ may be regarded as the time after disk formation, 
our results are consistent with the idea that HL Tau is younger than 1 Myr. 
In fact, the age predicted from our simulations depends on the disk mass $M_{\rm disk}$ assumed: 
a higher $M_{\rm disk}$ leads to a larger $t_{\rm snap}$
because it takes longer for dust emission to decay to the observed level
when the initial dust mass is larger.
However, a disk mass much in excess of $0.2~M_\sun$ seems to be unrealistic
because the disk would then be gravitationally unstable at outer radii (see Section~\ref{sec:density}).

\subsection{Aggregate Size and Dust Surface Density}\label{sec:aSigma}
\begin{figure*}
\centering
\includegraphics[width=18cm]{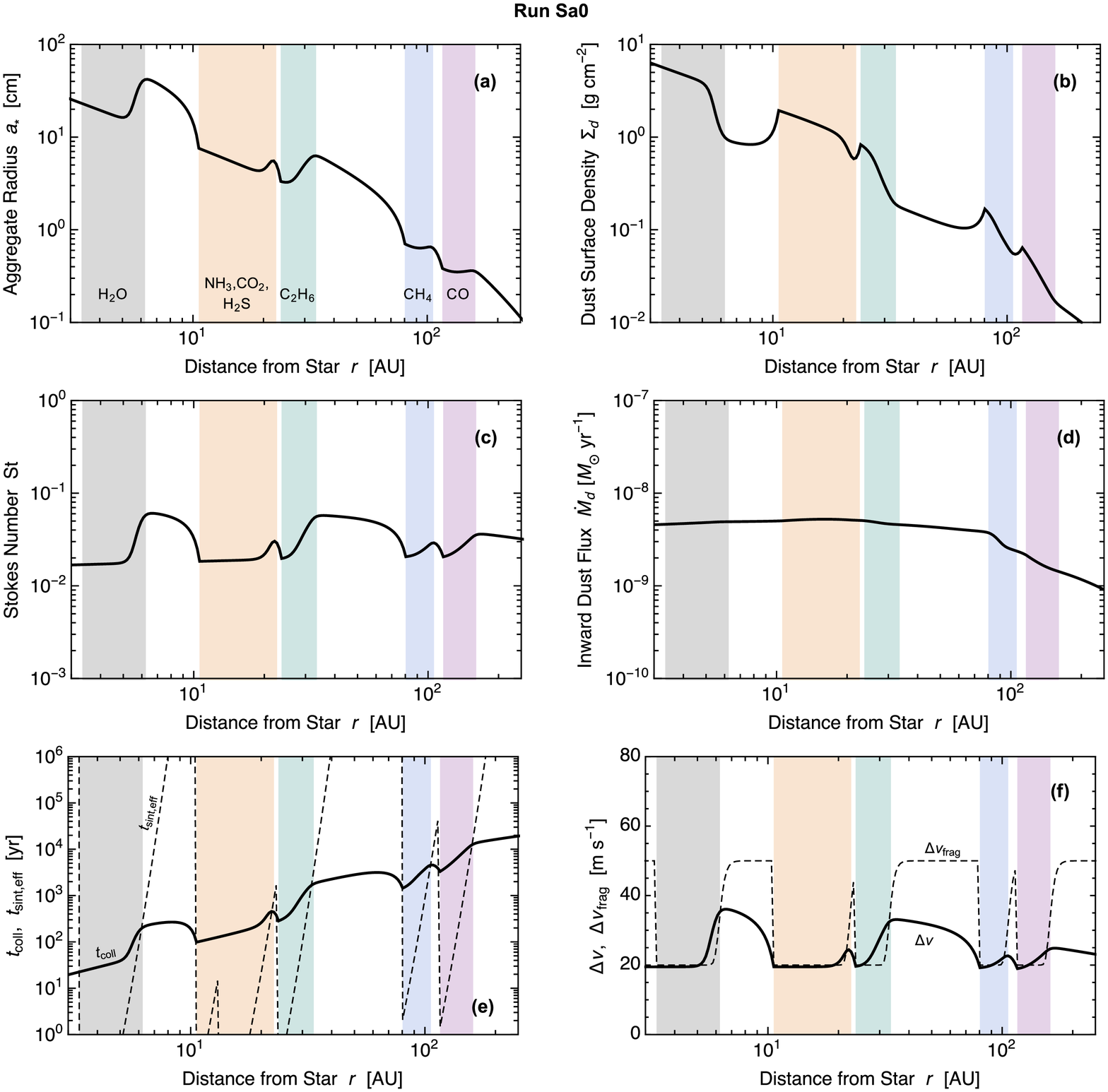}
\caption{Radial distribution of the representative aggregate radius $a_*$ (panel a), 
dust surface density $\Sigma_{\rm d}$ (panel b), aggregate Stokes number ${\rm St}$ (panel c), 
and radial inward dust flux $\dot{M}_{\rm d}$ (panel d)
at time $t = t_{\rm snap} = 0.26~\Myr$ from the fiducial simulation. 
The shaded areas mark the sintering zones defined by $r_{{\rm subl},j} < r < r_{{\rm subl},j}$ 
(see Section~\ref{sec:sint} for details).
Panel e shows the collision timescale $t_{\rm coll}$ (Equation~\eqref{eq:tcoll}; solid line) 
and effective sintering timescale $t_{\rm sint,eff}$ (Equation~\eqref{eq:t_sint_eff}; dashed line), 
while panel f plots the collision velocity $\Delta v$ (solid line) and 
fragmentation threshold $\Delta v_{\rm frag}$ (Equation~\eqref{eq:Dvfrag}; dashed line) 
for the representative aggregates.
}
\label{fig:aSigma}
\end{figure*}
Below we fix $t$ to be $t_{\rm snap} = 0.26~\Myr$ and look at 
the radial distribution of dust in detail. 
Figures~\ref{fig:aSigma}(a) and (b) show the radial distribution 
of the representative aggregate radius $a_*$ and dust surface density $\Sigma_{\rm d}$ at this time.
We also show in Figure~\ref{fig:aSigma}(c) the Stokes number ${\rm St}$ of the representative aggregates, 
which is more directly related to their dynamics than $a_*$. 
In these figures, the vertical stripes indicate the locations of the sintering zones.
Here, the sintering zone of each volatile $j$ is defined 
by the locations where $r > r_{{\rm snow},j}$ and $r < r_{{\rm sint},j}$, with the latter 
being equivalent to $t_{\rm coll} < t_{{\rm sint},j}$ 
(see Figures~\ref{fig:aSigma}(e) for the radial distribution of $t_{\rm coll}$ and $t_{\rm sint,eff}$).
The sintering zones of ${\rm NH_3}$, ${\rm CO_2}$, and ${\rm H_2S}$ 
partially overlap with each other and form a single sintering zone.  
The exact locations of the sintering zones are 
3--6 AU (${\rm H_2O}$), 11--23 AU (${\rm NH_3}$--${\rm CO_2}$--${\rm H_2S}$), 24--33 AU (${\rm C_2H_6}$),
80--106 AU (${\rm CH_4}$), and 116--160 AU (${\rm CO}$).
Strictly speaking, the locations of the sintering zones are time-dependent 
because the volatile partial pressures $P_j$ and aggregate collision timescale $t_{\rm coll}$
evolve with $\Sigma_{\rm d}$ (see Equations~\eqref{eq:Pj}, \eqref{eq:Sigmaj}, and \eqref{eq:tcoll}).
However, comparison between Figures~\ref{fig:tsint} and \ref{fig:aSigma} shows that 
the sintering zones little migrate during this $0.26~\Myr$. 
This is because the locations of the sintering zones depend
on the radial distribution of the gas temperature $T$
(which is taken to be time-independent) 
much more strongly than on the distribution of $\Sigma_{\rm g}$.

Figures~\ref{fig:aSigma}(a) and (b) show that sintering produces 
a clear pattern in the radial distribution of the dust component.
We see that dust aggregates in the sintering zones tend to have a high surface density 
and a small radius compared to those in the adjacent non-sintering zones. 
The small aggregate size is a direct consequence of fragmentation induced by sintering.
To see this, we plot in Figure~\ref{fig:aSigma}(f) the collision velocity $\Delta v$ and 
fragmentation threshold $\Delta v_{\rm frag}$ as a function of $r$. 
In the non-sintering zones, we find $\Delta v_{\rm frag} \approx 50~{\rm m~s^{-1}}$ 
and $\Delta v \approx 25$--$35~{\rm m~s^{-1}}$,
implying that no disruptive collisions occur for the unsintered aggregates
(as we will see below, the maximum size of the unsintered 
aggregates is determined by radial drift rather than by fragmentation).
In the sintering zones, $\Delta v_{\rm frag}$ is decreased to $20~{\rm m~s^{-1}}$, 
and $\Delta v$ is also suppressed down to the same value.
Since $\Delta v$ is an increasing function of $a_*$ (as long as ${\rm St} \ll 1$), 
this indicates that the sintered aggregates disrupt each other
so that $\Delta v$ never exceeds $\Delta v_{\rm frag}$.
The disrupted aggregates pile up there because the inward drift 
speed $|v_{\rm r}|$ decreases with decreasing $a_*$. 
These pileups provide the high surface densities in the sintering zones.

\begin{figure*}[t]
\centering
\includegraphics[width=18cm]{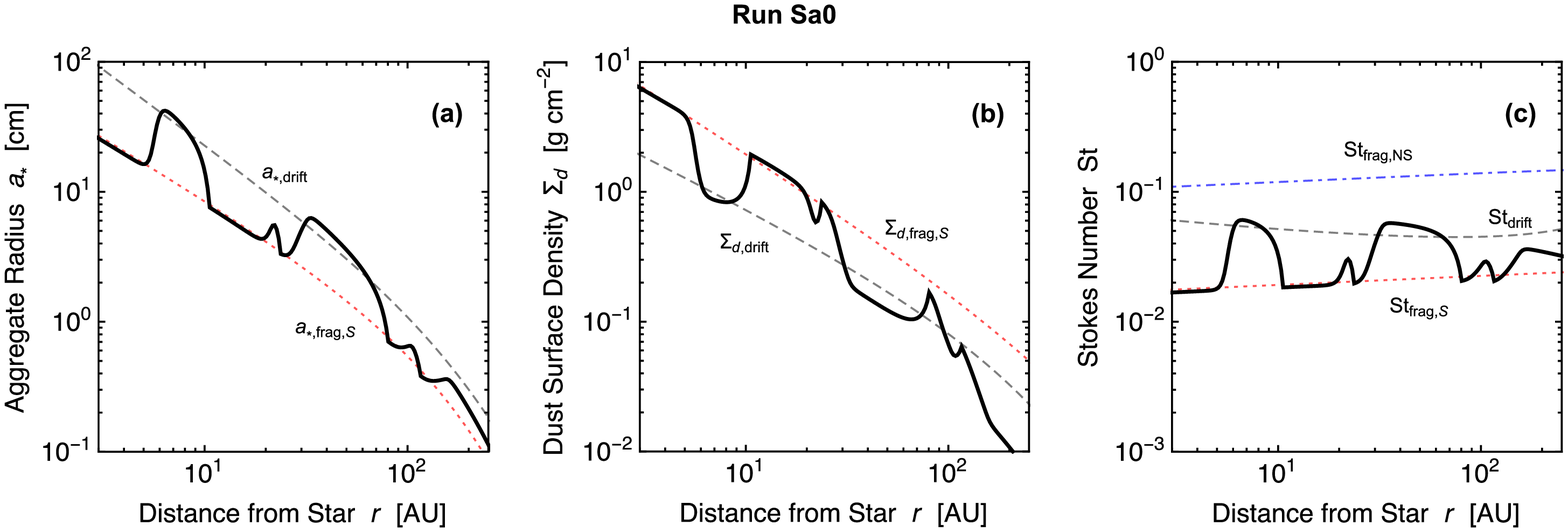}
\caption{Same as panels (a), (b), and (c) of Figure~\ref{fig:aSigma} 
but compared with the analytic estimates assuming steady inward dust flow 
($\dot{M}_{\rm d}=5\times 10^{-9}~{M}_\odot~\yr^{-1}$). 
The dashed curves show the drift-limited solution for unsintered aggregates 
(labeled by ``NS''), while the dotted curves the fragmentation-limited solution
of sintered aggregates (labeled by ``S''). 
}
\label{fig:comp}
\end{figure*}
To understand the radial distribution of $\Sigma_{\rm d}$ and $a_*$ more quantitatively, 
we now look at the radial inward mass flux of drifting aggregates,
\beq
\dot{M}_{\rm d} \equiv 2\pi r |v_{\rm r}|\Sigma_{\rm d}.
\label{eq:Mddot}
\eeq
The radial distribution of $\dot{M}_{\rm d}$ is shown in Figure~\ref{fig:aSigma}(d).
We can see that the mass flux is radially constant 
($\approx 5\times 10^{-9}~M_\sun~\yr^{-1}$) at $\la 100~\AU$.
This indicates that the radial dust flow in this region can be approximated by a {steady} flow. 
Such a quasi-steady dust flow is commonly realized when 
some mechanism like radial drift or fragmentation limits dust growth \citep[see, e.g.,][]{BKE12,LJ14}.
Substituting $|v_{\rm r}| \approx 2\eta v_{\rm K}{\rm St}$  (${\rm St} \ll 1$) 
into Equation~\eqref{eq:Mddot}, we obtain the relation between $\Sigma_{\rm d}$ and ${\rm St}$, 
\beq
\Sigma_{\rm d} = \frac{\dot{M}_{\rm d}}{2\pi r |v_{\rm r}|} 
\approx \frac{\dot{M}_{\rm d}}{4\pi \eta r v_{\rm K}{\rm St}}.
\label{eq:Sigmad_approx}
\eeq
Since ${\rm St} \propto a_*$, we have $\Sigma_{\rm d} \propto 1/a_*$ for constant $\dot{M}_{\rm d}$.

Using Equation~\eqref{eq:Sigmad_approx} together with the assumption 
that either radial drift or fragmentation limits dust growth, 
one can estimate the radial distribution of $\Sigma_{\rm d}$ and $a_*$ 
in the non-sintering and sintering zones in an analytic way.
When radial drift limits dust growth, 
the maximum aggregate size is determined by the condition \citep[][]{OTKW12}
\beq
\frac{t_{\rm coll}}{t_{\rm drift}} \approx \frac{1}{30},
\label{eq:cond_drift}
\eeq
where $t_{\rm coll}$ is the collision timescale already given by Equation~\eqref{eq:tcoll}
and 
\beq
t_{\rm drift} \equiv \frac{r}{|v_{\rm r}|} \approx \frac{1}{2\eta \Omega{\rm St}}
\label{eq:tdrift}
\eeq
is the timescale of radial drift.
Substituting 
$H_{\rm d} \approx (1+{\rm St}/\alpha_{\rm t})^{-1/2}H_{\rm g}$ (${\rm St} \ll 1$)
and $m_*/\pi a_*^2 = (4/3)\rho_{\rm int} a_* = (8/3\pi)\Sigma_{\rm g}{\rm St}$
into Equation~\eqref{eq:nstar}, the collision timescale can be evaluated as 
\beq
t_{\rm coll} \approx 
0.53\frac{\Sigma_{\rm g}}{\Sigma_{\rm d}}
\frac{H_{\rm g}{\rm St}}{\Delta v \sqrt{1+{\rm St}/\alpha_{\rm t}}}
\label{eq:tcoll_approx}
\eeq
Furthermore, the collision velocity can be approximated as the root square sum 
of the turbulence-driven velocity $\Delta v_{\rm t}$ 
and differential radial drift velocity $\Delta v_{\rm r}$,
\beqn
\Delta v \approx \sqrt{ (\Delta v_{\rm t})^2 + (\Delta v_{\rm r})^2 }
\approx \sqrt{ 2.3\alpha_{\rm t}c_{\rm s}^2{\rm St}  + (\eta v_{\rm K}{\rm St})^2},
\label{eq:Dv_approx}
\eeqn
where we have used the approximate expressions for $\Delta v_{\rm t}$ and 
and $\Delta v_{\rm r}$ already given in Section~\ref{sec:tcoll}. 
Substituting Equations~\eqref{eq:Sigmad_approx} and 
\eqref{eq:tdrift}--\eqref{eq:Dv_approx}
into Equation~\eqref{eq:cond_drift}, we obtain the equation 
for the maximum Stokes number in the drift-limited growth,
\beq
\frac{{\rm St}_{\rm drift}^{5/2}}{\sqrt{(2.3\alpha_{\rm t}c_{\rm s}^2  
+ \eta^2 v_{\rm K}^2 {\rm St}_{\rm drift})
(1+{\rm St}_{\rm drift}/\alpha_{\rm t})}} 
= \frac{0.0025\dot{M}_{\rm d}}{\eta^2 r v_{\rm K} c_{\rm s} \Sigma_{\rm g}},
\label{eq:St_drift}
\eeq
where we have labeled ${\rm St}$ by the subscript ``drift'' to emphasize drift-limited growth. 
In Figure~\ref{fig:comp}(c), we compare ${\rm St}_{\rm drift}$ for 
$\dot{M}_{\rm d} = 5\times 10^{-9}~M_\sun~{\rm yr}^{-1}$
with the Stokes number directly obtained from run Sa0 at $t=t_{\rm snap}$.
We find that ${\rm St}_{\rm drift}$ reproduces ${\rm St}$ in the non-sintering zones, 
implying that radial drift limits the growth of unsintered aggregates.
We are also able to estimate $a_*$ and $\Sigma_{\rm d}$ in the non-sintering zones 
by substituting ${\rm St} = {\rm St}_{\rm drift}$ into  
$a_* = (2/\pi)\Sigma_{\rm g}{\rm St}/\rho_{\rm int}$
and Equation~\eqref{eq:Sigmad_approx}, respectively. 
These are shown by the dashed lines in Figures~\ref{fig:comp}(a) and (b).

If fragmentation limits dust growth, 
the maximum Stokes number is simply determined by the balance
\beq
\Delta v({\rm St}) = \Delta v_{\rm frag}.
\label{eq:cond_frag}
\eeq
We will denote the solution to this equation by ${\rm St}_{\rm frag}$. 
If we approximate $\Delta v$ by Equation~\eqref{eq:Dv_approx}, 
Equation~\eqref{eq:cond_frag} can be rewritten as a quadratic equation for ${\rm St}$, 
and its positive root gives
\beq
{\rm St}_{\rm frag} = \frac{-2.3\alpha_{\rm t}c_{\rm s}^2 
+ \sqrt{(2.3\alpha_{\rm t}c_{\rm s}^2)^2+4(\eta v_{\rm K}\Delta v_{\rm frag})^2}}{2(\eta v_{\rm K})^2}.
\label{eq:St_frag}
\eeq
The dot-dashed and dotted lines in Figure~\ref{fig:comp}(c) show 
${\rm St}_{\rm frag}$ for $\Delta v_{\rm frag} = \Delta v_{\rm frag,NS}$ 
and $\Delta v_{\rm frag,S}$ (denoted by ${\rm St}_{\rm frag,NS}$
and ${\rm St}_{\rm frag,S}$), respectively. 
We can see that ${\rm St}_{\rm frag,S}$ reproduces ${\rm St}$ in the sintering zones,
which confirms that fragmentation limits the growth of sintered aggregates.
One can estimate the values of $a_*$ and $\Sigma_{\rm d}$ in the sintering zones 
by substituting ${\rm St} = {\rm St}_{\rm frag,S}$ 
into  $a_* = 2\Sigma_{\rm g}{\rm St}/(\pi \rho_{\rm int})$ and Equation~\eqref{eq:Sigmad_approx}. 
These estimates are in excellent agreement with the simulation results 
as shown by the dotted lines of Figures~\ref{fig:comp}(a) and (b). 

\subsection{Lifetime of the Ring Patterns}\label{sec:lifetime}
\begin{figure}[t]
\centering
\includegraphics[width=8.5cm]{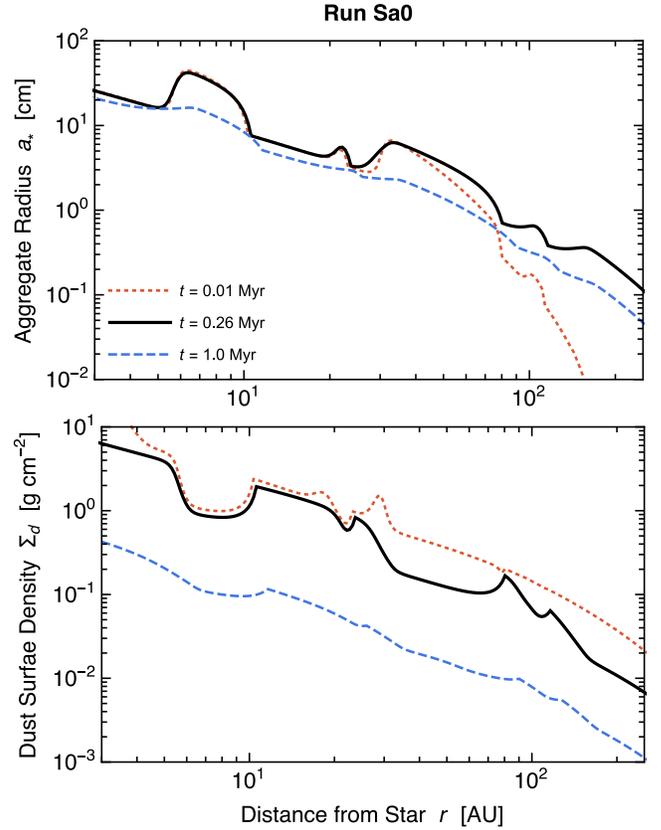}
\caption{Radial distribution of the representative aggregate radius 
$a$ (upper panel) and dust surface density $\Sigma_{\rm d}$ (lower panel) 
at $t = 0.01~\Myr$ (dotted lines), $0.26~\Myr$ (solid lines), and $1~\Myr$ (dashed line) 
from simulation run Sa0.
}
\label{fig:evol}
\end{figure}
It is worth mentioning at this point that 
the radial pattern of dust as shown in Figure~\ref{fig:aSigma}
fades out as the disk becomes depleted of dust.
As the dust-to-gas mass ratio $\Sigma_{\rm d}/\Sigma_{\rm g}$ decreases, 
the collision timescale $t_{\rm coll}$ of the aggregates increases, 
and consequently the maximum aggregate size set by radial drift (Equation~\eqref{eq:cond_drift}) decreases.
the radial pattern disappears when the radial drift barrier dominates over the fragmentation barrier at all $r$
because sintering has no effect on radial drift.
This is illustrated in Figure~\ref{fig:evol}, where we plot 
the radial distribution of $a_*$ and $\Sigma_{\rm d}$ for model Sa0
at different values of $t$.  
We find that the radial pattern that was present at $t = t_{\rm snap} =0.26~\Myr$ 
has disappeared by $t = 1~\Myr$.
In this particular example, ${\rm St}_{\rm drift}$ falls below ${\rm St}_{\rm frag, S}$ 
at all $r$ when ${M}_{\rm d}<10^{-11}~M_\sun~\yr^{-1}$.
In models La0 and LLa0, which assume weaker turbulence,  
dust evolution is slower than in model Sa0 owing to the lower turbulence-driven 
collision velocity (see $t_{\rm snap}$ in Table~\ref{tab:runs}).
However, even in these cases, the sintering-induced ring patterns are found to decay in 2~Myr.
We note that the lifetime of the pattern would be longer 
for radially more extended ($r_{\rm c} > 150~\AU$) disks, 
because the lifetime of dust flux in a disk generally 
scales with the orbital period at the disk's outer edge \citep{SOI16}.

\subsection{Optical Depths and Brightness Temperatures}\label{sec:tauTB}
\begin{figure*}[t]
\centering
\includegraphics[width=18cm]{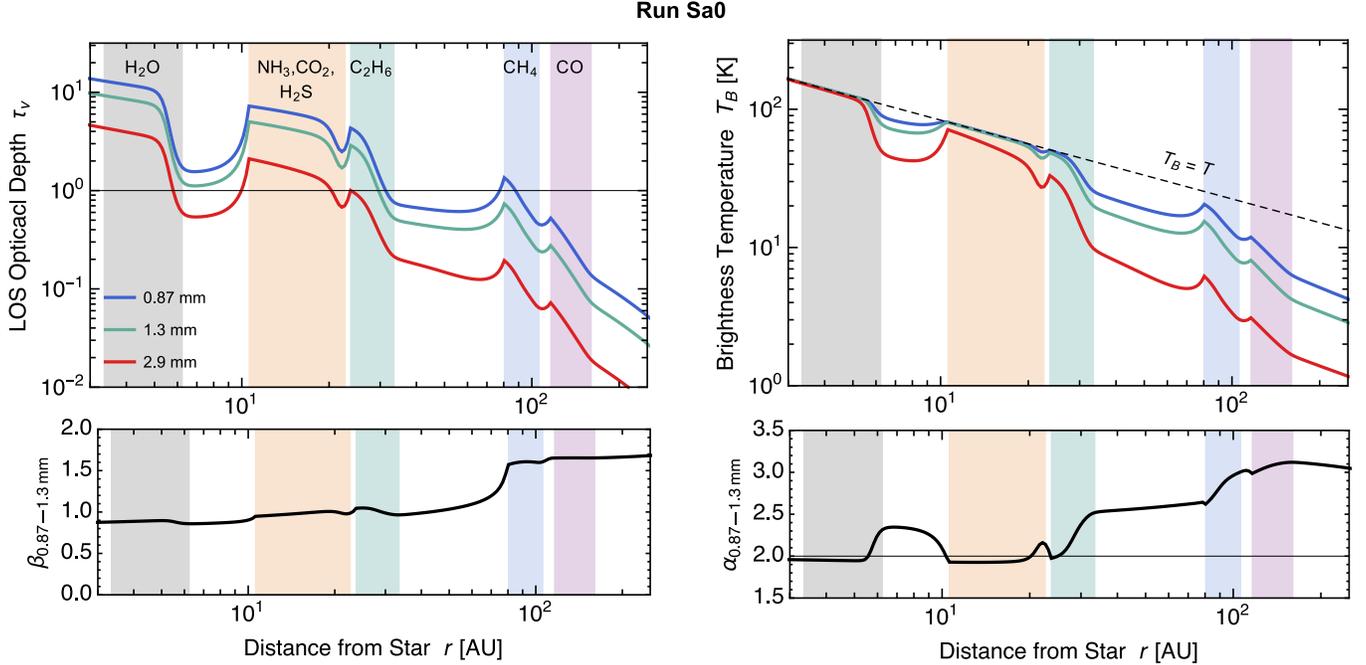}
\caption{Radial distribution of the line-of-sight optical depths $\tau_{\nu}$
(upper left panel), brightness temperatures $T_{\rm B}$ (upper right panel), 
opacity slope $\beta_{\rm 0.87\textrm{--}1.3~mm}$ (lower left panel), 
and spectral slope $\alpha_{\rm 0.87\textrm{--}1.3~mm}$ (lower right panel)
at $t = t_{\rm snap} = 0.26~\Myr$ from run Sa0.
The blue, green, and red curves in the upper panels correspond to wavelengths 
$\lambda = $ 0.87, 1.3, and 2.9 mm (ALMA Bands 7, 6, and 3), respectively. 
The dashed line in the upper right panel shows the gas temperature profile $T$ 
in our disk model (Equation~\eqref{eq:T}). 
The shaded areas mark the sintering zones.
}
\label{fig:tauTB}
\end{figure*}
We move on to the observational appearance of the sintering-induced dust rings
at millimeter wavelengths. 
The upper left panel of Figure~\ref{fig:tauTB} shows the radial distribution 
of the line-of-sight optical depths $\tau_{\nu}$ (Equation~\eqref{eq:tau}) 
from the snapshot of run Sa0 at $t=t_{\rm snap}$.
We here present the optical depths at three wavelengths $\lambda =$ 0.87, 1.3, and 2.9 mm,
which correspond to ALMA Bands 7, 6, and 3, respectively.

Overall, the optical depths in the sintering zones are higher than in the non-sintering zones.
This mainly reflects the higher dust surface density in the sintering zones (see Figure~\ref{fig:aSigma}(b)). 
The radial variation of  $\kappa_{{\rm d},\nu}$ represents only a minor contribution 
to the radial variation of $\tau_\nu$, in particular in outer regions 
where the representative aggregates are smaller than $\sim 1~{\rm cm}$ in radius. 
At  $\lambda =$ 0.87 and 1.3 mm, the three inner 
sintering zones (of ${\rm H_2O}$, ${\rm NH_3}$--${\rm CO_2}$--${\rm H_2S}$, and ${\rm C_2H_6}$)
are optically thick, while the two outer sintering zones (of ${\rm CH_4}$ and ${\rm CO}$)
are optically thin or marginally thick.
The ${\rm CO}$ sintering zone is much darker than the other sintering zones
because the disk surface density drops at $r > r_{\rm c} = 150~\AU$.
The non-sintering zones are optically thin or marginally thick at all three wavelengths.
The opacity index at $\lambda =$ 0.87--1.3 mm, $\beta_{\rm 0.87\textrm{--}1.3~mm} \equiv  
{\ln(\kappa_{\rm 0.87~mm}/\kappa_{\rm 1.3~mm})}/{\ln(\nu_{\rm 0.87~mm}/\nu_{\rm 1.3~mm})}$
is shown in the lower left panel of Figure~\ref{fig:tauTB}.
We see that  $\beta_{\rm 0.87\textrm{--}1.3~mm} \sim 1$ at $\la 70~\AU$ and 
approaches the interstellar value $\sim 1.7$ beyond 80 AU.   

The upper right panel of Figure~\ref{fig:tauTB} shows the distribution of the 
brightness temperatures $T_{\rm B}$ for the same snapshot.
For comparison, we also plot the gas temperature $T$ of our disk model given by Equation~\eqref{eq:T}.
We find that the three innermost sintering zones are
optically thick ($T_{\rm B} \approx T$) at $\lambda = $ 0.87 mm and 1.3 mm.

An interesting observational signature of the sintering-induced rings appears in 
the radial variation of the spectral slope.
In the lower right panel of Figure~\ref{fig:tauTB}, we plot the spectral index at 0.87--1.3 mm,
$\alpha_{\rm 0.87\textrm{--}1.3~mm} \equiv \ln(I_{\rm 0.87~mm}/I_{\rm 1.3~mm})/\ln(\nu_{\rm 0.87~mm}/\nu_{\rm 1.3~mm})$, as a function of $r$. 
In the three innermost sintering zones, we have $\alpha_{\rm 0.87\textrm{--}1.3~mm} \approx 2$
since these zones are  optically thick at these wavelengths. 
If these regions were optically thin, we would have $\alpha_{\rm 0.87\textrm{--}1.3~mm} \approx 3$
because $\beta_{\rm 0.87\textrm{--}1.3~mm} \approx 1$  
(see the lower left panel of Figure~\ref{fig:tauTB}).
In the non-sintering zones lying at $\sim$ 6--11 AU and $\sim$ 33--80 AU, 
we obtain $\alpha_{\rm 0.87\textrm{--}1.3~mm} \approx 2.3$--2.5,
which is in between the values in the optically thick and thin limits.
This reflects the fact that the non-sintering zones are marginally thick at 0.87--1.3 mm.

\subsection{Comparison with the ALMA Observation}\label{sec:ALMA}
\begin{figure*}[t]
\centering
\includegraphics[width=18cm]{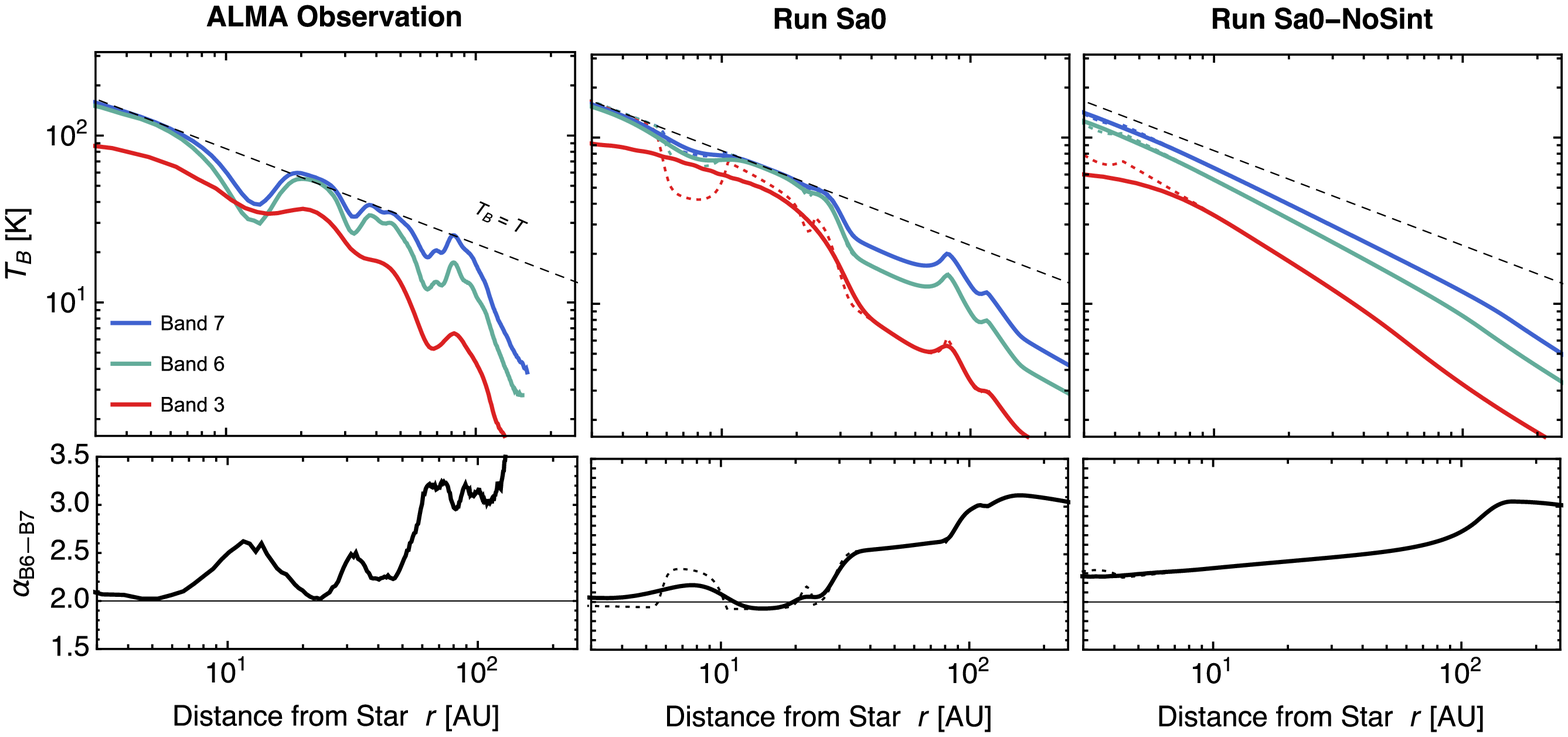}
\caption{Comparison between the ALMA observation and our fiducial model calculation Sa0.
The left panels show the radial profiles of $T_{\rm B}$ at Bands 3, 6, and 7
and $\alpha_{\rm B6\textrm{--}B7}$ of the HL Tau disk obtained from the ALMA observation 
(same as Figure~\ref{fig:T}). 
The center and right panels are from simulation runs Sa0 and Sa0-NoSint. 
The simulated profiles are obtained by smoothing the raw simulation data (dotted curves)
at the ALMA resolutions (see text for details).
The dashes lines show $T_{\rm B} = T$, where $T$ is the gas temperature given by Equation~\eqref{eq:T}. 
}
\label{fig:TBcomp_Sa0}
\end{figure*}
Now we make more detailed comparisons between the simulation results 
and the ALMA observation of the HL Tau disk. 
We here smooth the radial profiles of the intensities $I_\nu$ from run Sa0 ($t=t_{\rm snap}$)
at the ALMA resolutions as described in Section~\ref{sec:emission}.
In the center panels of Figure~\ref{fig:TBcomp_Sa0}, 
the solid lines show the radial profiles of the brightness temperatures $T_{\rm B}$ 
and spectral slope $\alpha_{\rm B6\textrm{--}B7}$ obtained from the smoothed $I_\nu$.
For comparison, $T_{\rm B}$ and $\alpha_{\rm B6\textrm{--}B7}$ from the raw $I_\nu$ 
are shown by the dotted lines. 
The left panels show the profiles from the ALMA observations (same as Figure~\ref{fig:T}).
We also show in the right panels the results from the no-sintering run Sa0-NoSint 
to clarify the features of the sintering-indued structures. 

After smoothing, the innermost emission dip lying at 6--11 AU 
has been partially smeared at 0.87 mm (Band 7) and 1.3 mm (Band 6).
{The emission dip in the smoothed images has a lower spectral slope than 
that in the raw images directly obtained from the simulation. 
This is a consequence of the frequency-dependent angular resolution: 
since Band 6 has a coarser resolution than Band 7, 
the emission dip seen at Band 6 is more significantly buried than that seen at Band 7, 
resulting in a decrease in the spectral slope after smoothing.}
At 2.9 mm (Band 3), the radial structure of $T_{\rm B}$ at $\la 10$ AU has been significantly smoothed out.

We find that our simulation reproduces 
many observational features of  the HL Tau disk.
First, the simulation predicts a central emission peak that closely resembles the observed one.
This central emission peak is associated with the ${\rm H_2O}$ sintering zone as shown in Figure~\ref{fig:tauTB}.
The radial extent of the simulated central peak is $\approx 6~\AU$, 
which is comparable to $\approx 8~\AU$ of the observed peak.
The simulation perfectly reproduces the emission features of the central region: 
the magnitudes and radial slopes of $T_{\rm B}$ at all three ALMA Bands, and 
the millimeter spectral slope of $\alpha_{\rm B6\textrm{--}B7} \approx 2$.
In our simulation, the spectral slope simply reflects the high optical thickness of the central region, 
having nothing to do with the optical properties of the aggregates in the region.
Sintering plays an essential role in the buildup of the optically thick region; 
without sintering, the disk would be entirely optically thin at these bands as shown 
by run Sa0-NoSint (see the right panels of Figure~\ref{fig:TBcomp_Sa0}). 
The fact that the central emission peak has a lower intensity at Band 3 than at Bands 6 and 7 
is explained as a consequence of the lower spatial resolution ($\approx 10~\AU$) at Band 3, 
i.e., this compact emission is underresolved at this band. 

The simulation also predicts two bright rings in the region of $\sim$ 10--30 AU
that might be identified with the two innermost bright rings of HL Tau observed at $\sim 20 ~\AU$ and 40 AU. 
In our simulation, the two emission rings are associated with  
the sintering zones of ${\rm NH_3}$--${\rm CO_2}$--${\rm H_2S}$ (11--23 AU) and 
${\rm C_2H_6}$ (24--33 AU).
These rings are optically thick at Bands 6 and 7 (and therefore $\alpha_{\rm B6\textrm{--}B7} \approx 2$)
and are optically thin at Band 3. These features are consistent 
with those of the two innermost bright rings of HL Tau.
However, the separation of the predicted rings are much smaller than in the observed rings
as we discuss below.

Farther out in the disk, the simulation predicts an optically thin emission peak 
at 80 AU associated with ${\rm CH_4}$ sintering. 
Its location coincides with the 80 AU bright ring in the ALMA image of HL Tau, 
and its brightness temperatures agree with those of the observed ring at all three wavelengths
to within a factor of two.
The simulation also predicts a less pronounced peak at 120 AU associated with ${\rm CO}$ sintering.
As mentioned in Section~\ref{sec:tauTB}, this 120 AU peak is much less pronounced than 
other inner peaks because this location is close to the outer edge of our modeled gas disk. 
Interestingly, the observed HL Tau disk also has one minor emission peak exterior to the 80 AU
ring ($r\sim 97~\AU$; see \citealt{ALMA+15}). 

The innermost dark ring seen at 6--11 AU in the simulated image  
is optically marginally thick and has $\alpha_{\rm B6\textrm{--}B7} >2$,
consistent with the observed innermost dark ring at $\sim 13~\AU$. 
Our simulation also explains why this innermost emission dip is much shallower at Band 3 than at 
Bands 6 and 7: as mentioned above, this is simply because the spatial resolution at Band 3 is
no high enough to distinguish the dark ring from the central emission peak. 

However, there are some discrepancies between the prediction 
from the fiducial model and the observation of the HL Tau disk.  
For example, the second innermost ring predicted by the model,
which is associated with the ${\rm C_2H_6}$ sintering zone,
is about 10 AU interior to that observed by ALMA.
For this reason, the dark ring just inside the ${\rm C_2H_6}$ sintering zone
is much narrower than the second innermost dark ring of HL Tau extending from 30 AU to 40 AU.
In addition, as we explain in detail in Section~\ref{sec:a0}, 
the fiducial model predicts that dust particles are vertically well mixed in the gas disk,  
which seems to be inconsistent with the observations of HL Tau suggesting 
that large dust particles settle to the disk midplane  \citep{KLM11,PDM+16}. 
In the following section, we will examine if these discrepancies 
can be removed or alleviated by tuning the parameters in our model. 

\section{Parameter Study}\label{sec:param}
The previous section has mainly focused on our fiducial simulation (run Sa0). 
We here study how the simulation results depend on the model parameters.

\subsection{Gas Surface Density Slope}\label{sec:gamma}
Since the gas density distribution of the HL Tau disk is unknown, 
it is important to quantify how strongly our results depend on the assumption about the gas density profile. 
We here present the results of two simulation runs Sa0-Lgam and Sa0-Hgam, 
in which we vary $\gamma$ in Equation~\eqref{eq:Sigma_g} to 0.5 and 1.5, respectively. 
The other parameters are unchanged form the fiducial run.

\begin{figure}[t]
\centering
\includegraphics[width=8.5cm]{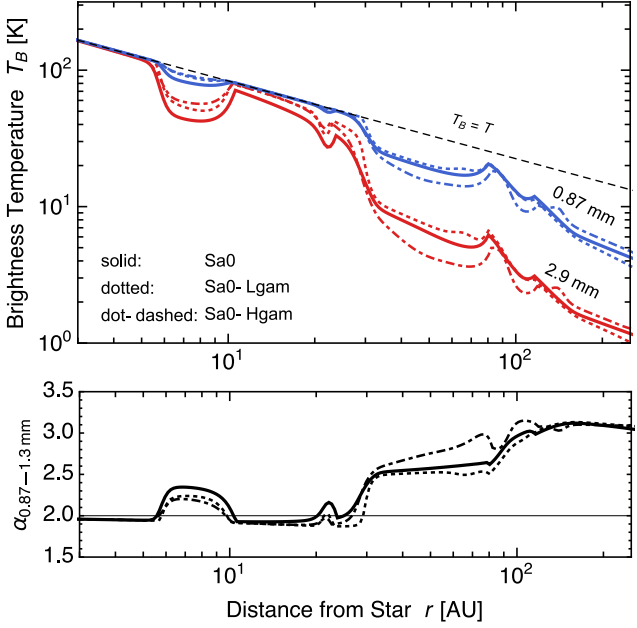}
\caption{Comparison between runs Sa0 (solid curves), 
Sa0-Lgam (dotted curves), Sa0-Hgam (dot-dashed curves). 
The upper panel shows the brightness temperatures $T_{\rm B}$ as a function of orbital distance $r$
at wavelengths 0.87 mm (blue curves) and 2.9 mm (red curves) at time $t = t_{\rm snap}$.
The gas temperature profile $T$ (Equation~\eqref{eq:T}) is shown by the dashed line.
The lower panel shows the opacity index at 0.87--1.3 mm.
}
\label{fig:TB_Hgam}
\end{figure}
Figure~\ref{fig:TB_Hgam} compares the radial profiles of 
the brightness temperature and spectral slope 
at 0.87 mm and 2.9 mm from these runs with those from run Sa0.
The three snapshots are taken at different times 
but still provide similar flux densities (see Table~\ref{tab:runs}) 
because $t_{\rm snap}$ is defined as such. 
We find that the variation of $\gamma$ within the range 0.5--1.5 little affects 
the emission properties of the disk, with the variation of $T_{\rm B}$ within a factor of 2
and the variation of $\alpha_{\rm 0.87\textrm{--}1.3~mm}$ as small as $\la 25\%$ at all $r$. 
This is mainly because the steady-state radial distribution 
of the dust surface density $\Sigma_{\rm d}$ is insensitive to $\gamma$.
If we measure the aggregate size by St, the steady-state distribution of $\Sigma_{\rm d}$ 
is determined from Equation~\eqref{eq:Sigmad_approx} with either
${\rm St} = {\rm St}_{\rm drift}$ (Equation~\eqref{eq:St_drift})
or ${\rm St} = {\rm St}_{\rm frag}$ (Equation~\eqref{eq:St_frag}). 
The $\gamma$ dependence of the radial dust flux $\dot{M}_{\rm d}$ 
is weak as long as we fix $M_{\rm disk}$ and $r_{\rm c}$. 
${\rm St}_{\rm frag}$ depends on $\gamma$  only through 
$\eta \propto \gamma + 1.8$ (for $T \propto r^{-0.57}$ and $r \ll r_{\rm c}$),
and its variation is small as long as we vary $\gamma$ within the range 0.5--1.5.
For ${\rm St}_{\rm drift}$, the dependence is less obvious from Equation~\eqref{eq:St_drift},
but it turns out that the weak $\gamma$ dependences of $\eta^2\Sigma_{\rm d}$ and 
$\dot{M}_{\rm d}$ partly cancel out in this equation.

\subsection{Radial Variation of Turbulence}\label{sec:alpha}
The radial distribution of turbulence strength $\alpha_{\rm t}$ 
is another important uncertainty in our simulations. 
Our fiducial model assume $\alpha_{\rm t} \propto \sqrt{r}$, 
which gives a turbulence-driven collision velocity nearly independent of $r$
($\Delta v_{\rm t} \propto \sqrt{\alpha_{\rm t}} c_{\rm s} \propto \sqrt{\alpha_{\rm t} T}
\propto r^{-0.035}$). 
In fact, a radially constant collision velocity 
is required in our model to simultaneously reproduce dark rings at small $r$ 
and bright rings at large $r$ simultaneously.\footnote{In most of our simulation runs, the turbulence-driven velocity is the dominant component of the aggregate collision velocity. The only exception is run LLa0, in which the turbulence-driven velocity is only slightly larger than the drift-driven velocity. In this run, radial drift and turbulence nearly equally contribute to the aggregate fragmentation.}
To demonstrate this, we here consider two models in which 
$\alpha_{\rm t}$ is fixed to 0.03 and 0.1 
at all $r$ (referred as models Sa0-Lalp and Sa0-Halp, respectively).
These values correspond to the values of $\alpha_{\rm t}$ 
in the fiducial Sa0 model at $r \approx 10~\AU$ and $100~\AU$, respectively. 

\begin{figure}[t]
\centering
\includegraphics[width=8.5cm]{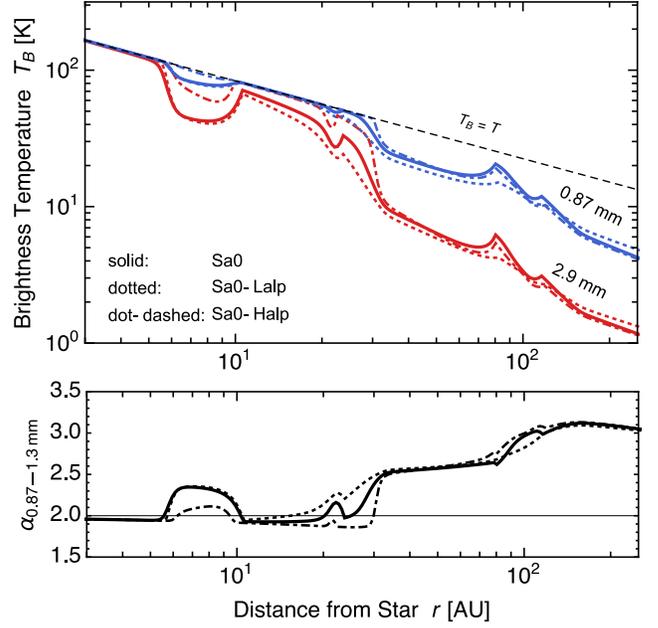}
\caption{Comparison between runs Sa0 (solid curves), 
Sa0-Lalp (dotted curves), Sa0-Halp (dot-dashed curves). 
The upper panel shows the brightness temperatures $T_{\rm B}$ as a function of orbital distance $r$
at wavelengths 0.87 mm (blue curves) and 2.9 mm (red curves) at time $t = t_{\rm snap}$.
The gas temperature profile $T$ (Equation~\eqref{eq:T}) is shown by the dashed line.
The lower panel shows the opacity index at 0.87--1.3 mm.
}
\label{fig:TB_LHalp}
\end{figure}
Figure~\ref{fig:TB_LHalp} compares the result of fiducial run Sa0
with those of runs Sa0-Lalp and Sa0-Halp.
We find that model Sa0-Lalp fails to produce an emission peak 
at $\approx 80~\AU$.
This is because turbulence is too weak to cause fragmentation of sintered aggregates 
at that location. 
By contrast, model Sa0-Halp produces only shallow dips at $r \la 30~\AU$. 
In this high-$\alpha_{\rm t}$ model, the maximum aggregate size at $r \la 30~\AU$ 
is limited by turbulent fragmentation even in the non-sintering zones. 
The suppressed dust growth causes a slowdown of radial drift, 
which acts to fill the density dips in the non-sintering zones.
Thus, a model assuming a high $\alpha_{\rm t}$ in outer regions and 
a low $\alpha_{\rm t}$ in inner regions best reproduces the observation of HL Tau. 

Theoretically, $\alpha_{\rm t}$ that is lower at smaller $r$ has been expected  
for turbulence driven by the magnetorotational instability \citep[MRI;][]{BH91}. 
Since the origin of the MRI is the coupling between the gas disk and 
magnetic fields, MRI turbulence tends to be weaker at locations where 
the ionization degree is lower. In protoplanetary disks, a lower ionization degree 
corresponds to a higher gas density and hence to a smaller orbital radius,  
because ionizing cosmic-rays or X-rays are attenuated at large column densities 
and because recombination is faster in denser gas \citep[e.g.,][]{SMUN00,B11a}. 

\subsection{Monomer Size and Dust Settling}\label{sec:a0}
\begin{figure*}
\epsscale{1.1}
\plottwo{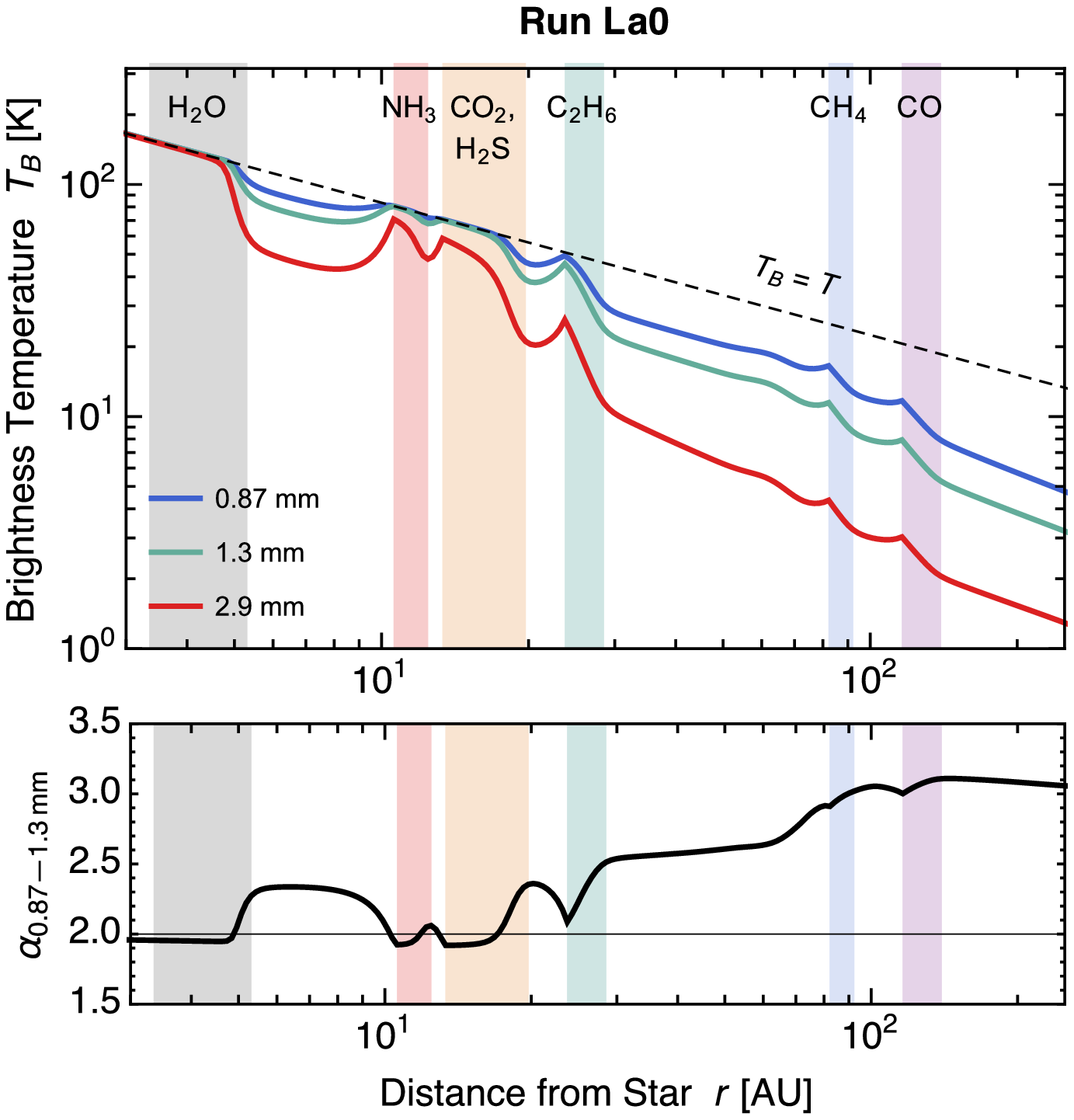}{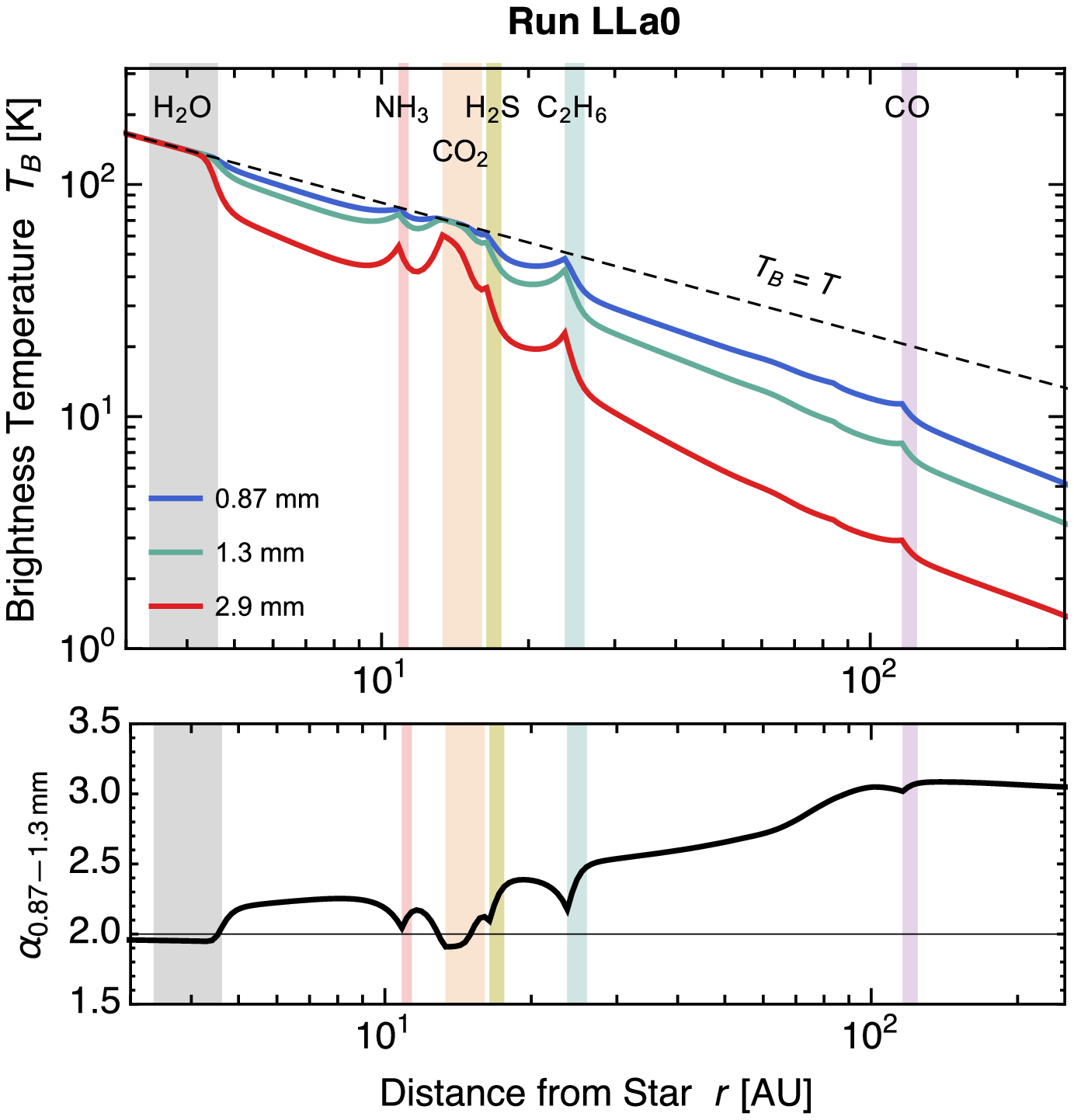}
\caption{Same as the right panels of Figure~\ref{fig:tauTB} but for models La0 (left panels) 
and LLa0 (right panels).
}
\label{fig:TB_La0}
\end{figure*}
\begin{figure*}
\centering
\includegraphics[width=18cm]{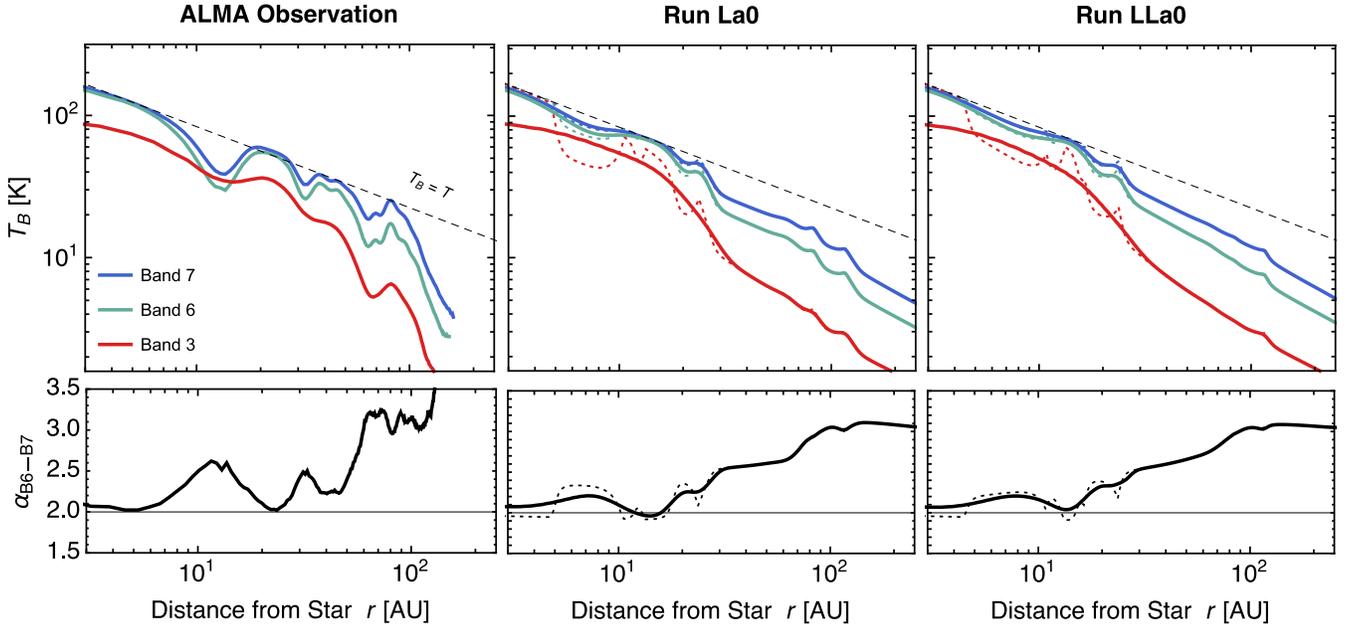}
\caption{Same as Figure~\ref{fig:TBcomp_Sa0}, but for models La0-tuned (center panels) 
and LLa0-tuned (right panels).
}
\label{fig:TBcomp_La0}
\end{figure*}
As mentioned in Section~\ref{sec:emission}, our calculations of dust thermal emission 
neglect the geometrical thickness of the dust subdisk. 
In reality, if a dust disk has an inclination of $\sim 45^\circ$ and 
a finite vertical extent $H_{\rm d}$, 
any radial structure of the disk emission is smeared out over this length scale 
in the direction of the minor axis of the disk image. 
\citet{PDM+16} point out that the scale height of large dust particles 
in the HL Tau disk must be as small as $\sim 1~{\rm AU}$ at $r =100~\AU$
in order to be consistent with the well separated morphology of the bright rings 
observed by ALMA. 
Such a small dust scale height strongly indicates that dust settling has occurred in the gas disk; 
without settling, the dust scale height would be $\sim 10~{\rm AU}$ at 100 AU. 

However, in our fiducial model, 
the settling of representative aggregates is severely prevented by turbulent diffusion.
According to Equation~\eqref{eq:Hd},
significant settling of dust particles ($H_{\rm d} \ll H_{\rm g}$) requires ${\rm St} \gg \alpha_{\rm t}$.
This condition is not satisfied in fiducial run Sa0,
because the value of ${\rm St}$ of the representative aggregates 
observed in the simulation is comparable to the value of $\alpha_{\rm t}$ assumed 
(see Figure~\ref{fig:aSigma}(c)).
In this run, $\alpha_{\rm t}$ is arranged to have a high value 
so that aggregates disrupt and pile up in the sintering zones. 
If turbulence were weak, sintered aggregates would not experience disruption 
as long as we maintain the assumption $\Delta v_{\rm frag,S} = 20~{\rm m~s^{-1}}$.
 
One way to reconcile sintering-induced ring formation with dust settling in our model
is to assume weaker turbulence {\it and} a lower fragmentation velocity. 
Within our dust model, a lower value of $\Delta v_{\rm frag}$ corresponds 
to a larger monomer size $a_0$ (see Equations~\eqref{eq:Dvfrag_NS} and \eqref{eq:Dvfrag_S}). 
However, aggregates made of larger monomers tend to be sintered less slowly
as demonstrated in Section~\ref{sec:sint}. 
To examine if there is a range of $a_0$ where the aggregates can 
experience settling and sintering simultaneously, 
we performed two simulations named La0 and LLa0. 
In run La0, we increase  $a_0$ to $1~\micron$ while lowering $\alpha_{\rm t}$  
to $10^{-3}(r/10~\AU)^{1/2}$.  
Run LLa0 is a more extreme case of $a_0 = 4~\micron$ 
and $\alpha_{\rm t} = 10^{-4}(r/10~\AU)^{1/2}$.
The other parameters are the same as in the fiducial run Sa0.  

\begin{figure}[t]
\centering
\includegraphics[width=8.5cm]{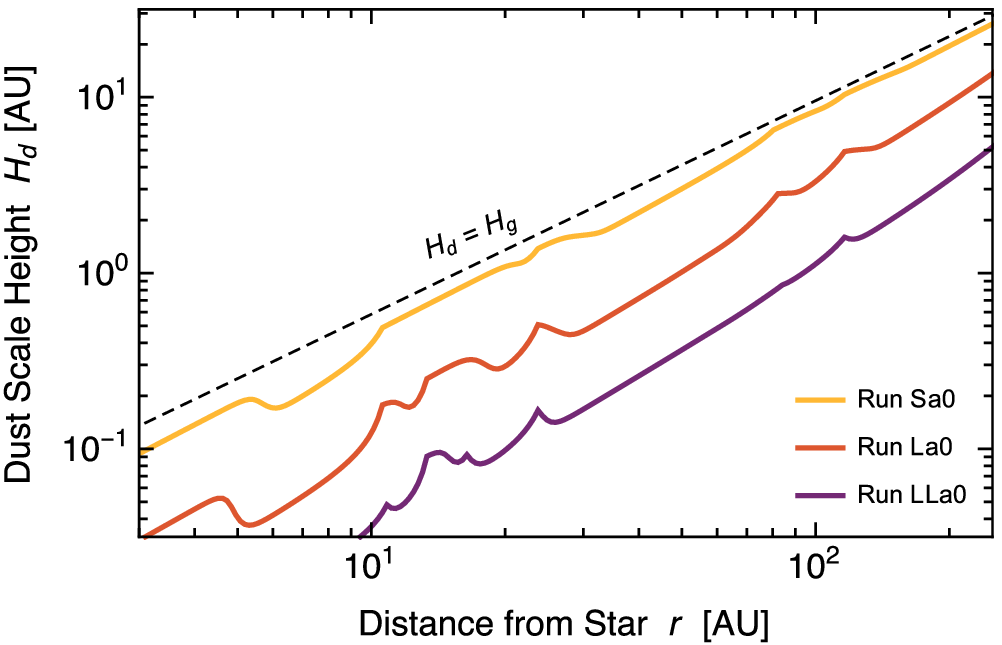}
\caption{Dust scale height $H_d$ versus orbital radius $r$ at time $t = t_{\rm snap}$
for runs Sa0 (black curve) and La0 (dashed curve).
The dashed line shows the gas scale height $H_{\rm g}$.
}
\label{fig:Hd}
\end{figure}
Figure~\ref{fig:TB_La0} shows the raw radial profiles of $T_{\rm B}$ and 0.87--1.3 mm spectral index 
at $t = t_{\rm snap}$ obtained from runs La0 and LLa0. 
The profiles after smoothing at the ALMA resolutions are shown in Figure~\ref{fig:TBcomp_La0}.
Note again that we neglect the geometric thickness of the dust disk when calculating $T_{\rm B}$.
Indeed, the dust disks in models La0 and LLa0 are geometrically thin 
unlike in model Sa0 because turbulent diffusion is inefficient in these two models.
We plot in Figure~\ref{fig:Hd} the dust scale height $H_{\rm d}$ from Equation~\eqref{eq:Hd}
as a function of $r$ for the three models.
At $r = 100~\AU$, we find $H_{\rm d} \approx 3~\AU$ for model La0 and
$H_{\rm d} \approx 1~\AU$ for model LLa0. 
Models Sa0, La0 and LLa0 give similar results for the radial distribution of $T_{\rm B}$, 
particularly in inner disk regions. 
However, model LLa0 fails to produce an emission peak at 80 AU
because the ${\rm CH_4}$ sintering zone completely disappears for $a_0 = 4~\micron$.

In summary, we find that dust settling gives a strong constraint on 
the turbulence strength and monomer grain size in the HL Tau disk. 
Dust settling and pileup occur simultaneously 
only if  $10^{-4} < \alpha_{\rm t} \la 10^{-3}$ {\it and} $1~\micron \la a_0 < 4~\micron$.
If $a_0 \ga 4~\micron$, sintering would be too slow to provide an appreciable 
emission peak at $\sim 80~\AU$. 
If $a_0 \ll 1~\micron$, sintered aggregates would be disrupted only when $\alpha_{\rm t} \gg 10^{-3}$, 
but such a strong turbulence would inhibit dust settling.

Interestingly, the above results suggest the grains constituting the aggregates in the HL Tau disk 
are considerably larger than interstellar grains ($\la 0.25~\micron$ in radius; e.g, \citealt{MRN77}).
They are also larger than grains constituting interplanetary dust particles of presumably cometary origin 
(typically 0.1--0.5 $\micron$ in diameter; e.g., \citealt{R93}).
However, such a large grain size is not excluded by the previous observations of HL Tau.
The near-infrared scattered light images of the envelope of HL Tau 
are best reproduced by models that assume the maximum particle size 
of more or less $1~\micron$ \citep{LFT+04,MOPI08}.
Since the scattered light probes the envelope's surface layer where coagulation is inefficient, 
the observed micron-sized particles might be monomers rather than aggregates.

\subsection{Sublimation Energies}\label{sec:L}
As mentioned in Section~\ref{sec:ALMA}, 
the fiducial model does not fully explain the exact locations and widths of all major HL Tau rings.
For example, the innermost dark ring predicted by model Sa0 lies somewhat closer 
to the central star than the observed one. 
If we define the position of the innermost dark ring 
as the location where $T_{\rm B}/T$ at Band 6 is maximized, 
the radius of the predicted innermost dark ring  (8 AU) is $\approx 40$\%
smaller than that of the observed one (13 AU).
Furthermore, the second innermost dark ring in model Sa0 is much narrower than that observed 
because the ${\rm C_2H_6}$ snow line lies very close to the ${\rm H_2S}$ sintering line. 
Of course, a different temperature profile would provide a different configuration of the sintering zones.
However, as long as $T(r)$ is assumed to obey a single power law,  
it is generally impossible to move some of the sintering zones while unchanging the positions of the others.
For example, a temperature profile slightly higher than Equation~\eqref{eq:T} would shift the 
${\rm C_2H_6}$ sintering zone to 40 AU, but would at the same time shift the ${\rm CH_4}$
sintering zone to 100 AU. 
Moreover, a different $T(r)$ would make our prediction for the brightness temperatures 
in the optically thick regions less good. 

However, the locations of the sintering zones depend not only on the gas temperature profile 
but also on the sublimation energies $L_j$ in $P_{{\rm ev},j}(T)$.
As mentioned in Section~\ref{sec:Pev}, there is typically a $10\%$ uncertainty in the published data of $L_j$.
In general, a 10\% uncertainty in the sublimation energy causes a
$\sim 10\%$ uncertainty  in the sublimation temperature  $T_{{\rm subl},j}$
because $P_{{\rm ev},j}$ is a function of the ratio $L_j/T$.
Assuming the temperature profile given by Equation~\eqref{eq:T}, 
we have $r_{{\rm snow},j} \propto T_{{\rm subl},j}^{-1.8}$ and hence 
$|\delta r_{{\rm snow},j}|/r_{{\rm snow},j} \approx  1.8 |\delta T_{{\rm subl},j}|/T_{{\rm subl},j}
\sim 2|\delta L_{j}|/L_{j}$, where $\delta L_{j}$, $\delta T_{{\rm subl},j}$, and 
$\delta r_{{\rm snow},j}$ denote the uncertainties of $L_{j}$, $T_{{\rm subl},j}$, and 
$r_{{\rm snow},j}$, respectively.
Consequently, a $10\%$ uncertainty in $L_j$ 
leads to a $\sim 20\%$ uncertainty in the snow line location.
Such an uncertainty can be significant because the separations 
of the observed rings are only a fraction of their radii. 

\begin{figure*}
\epsscale{1.1}
\plottwo{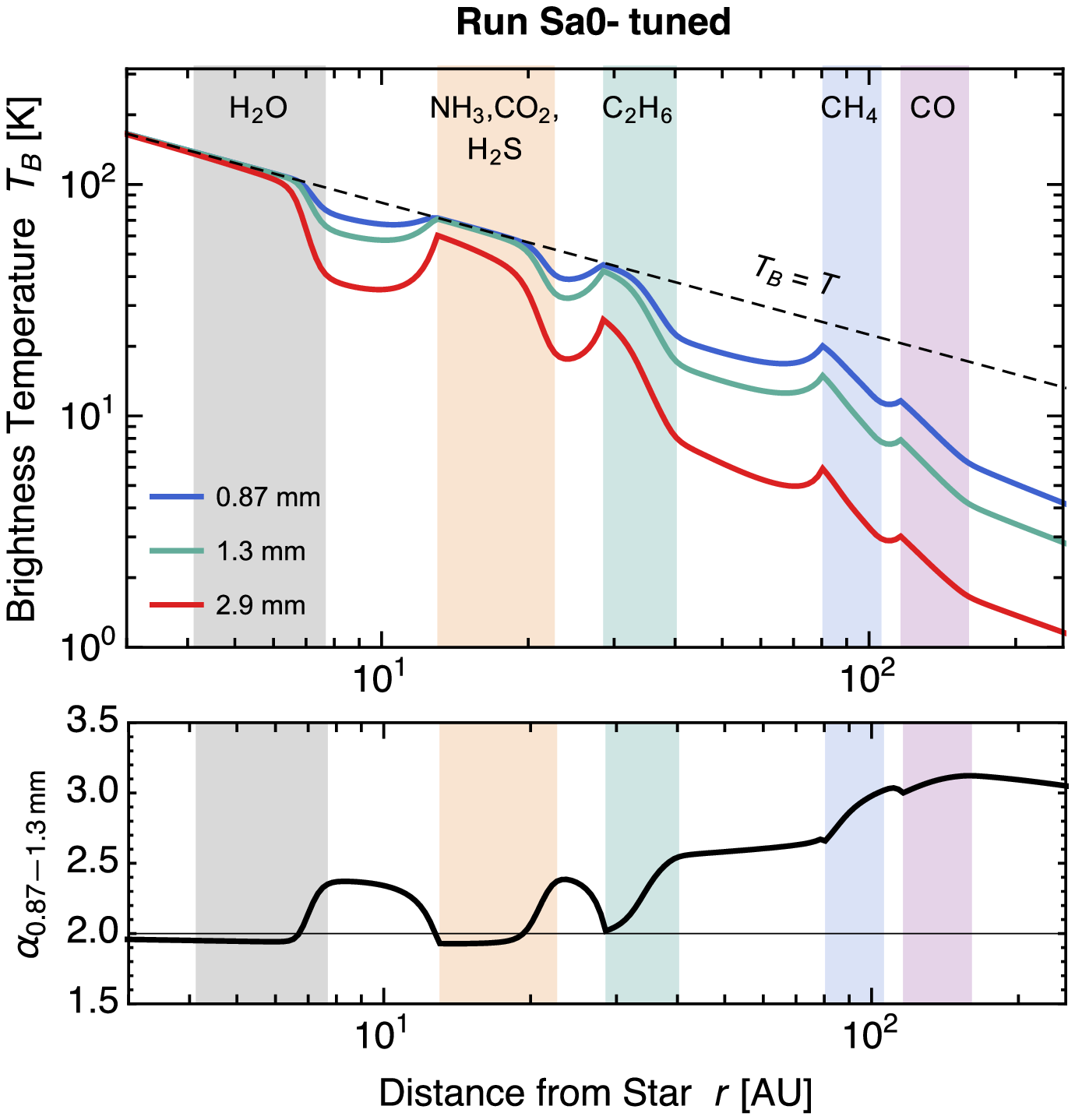}{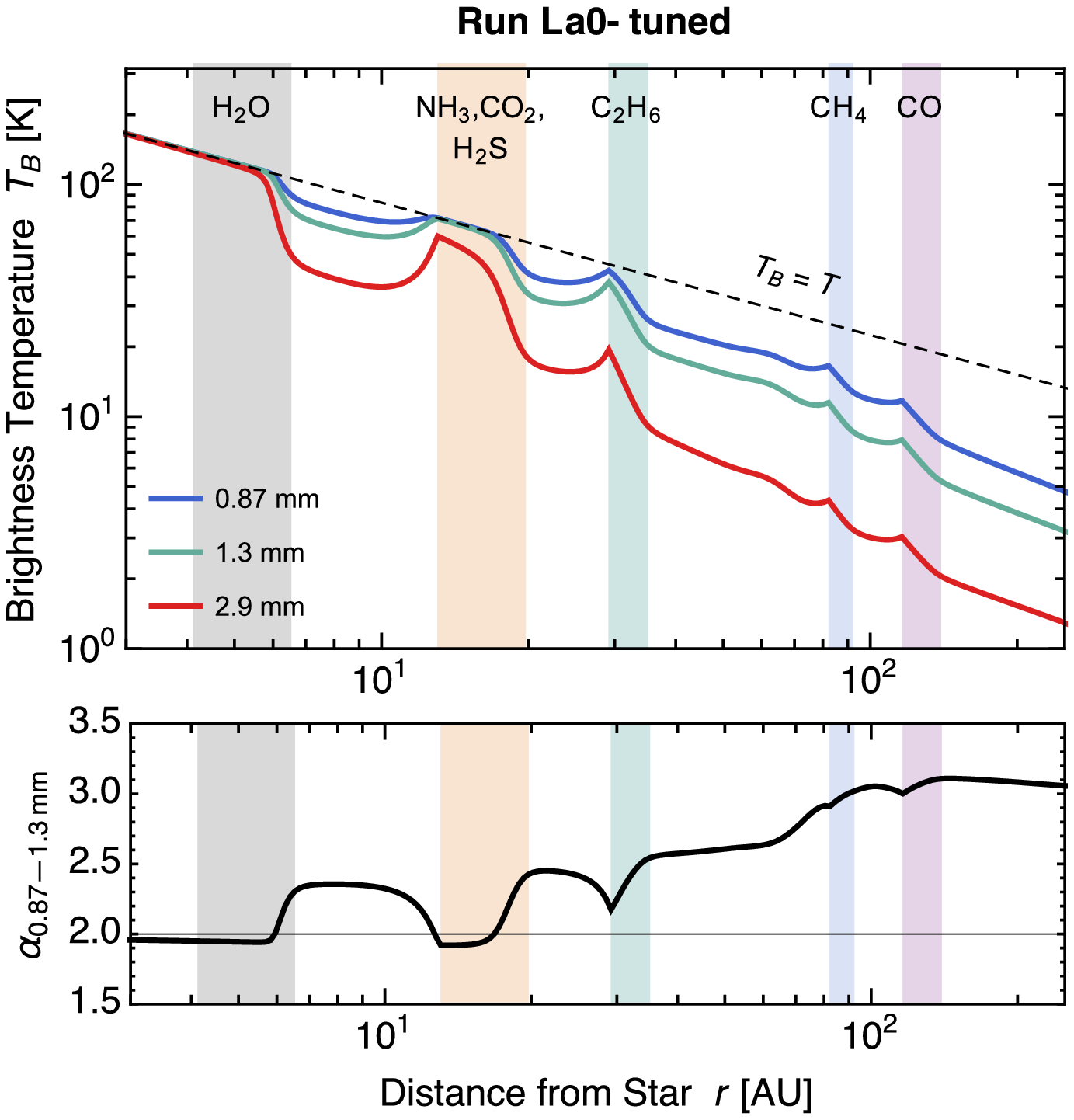}
\caption{Same as the right panels of Figure~\ref{fig:tauTB} but for runs Sa0-tuned
(left panels) and La0-tuned (right panels).
}
\label{fig:TB_tuned}
\end{figure*}
\begin{figure*}
\centering
\includegraphics[width=18cm]{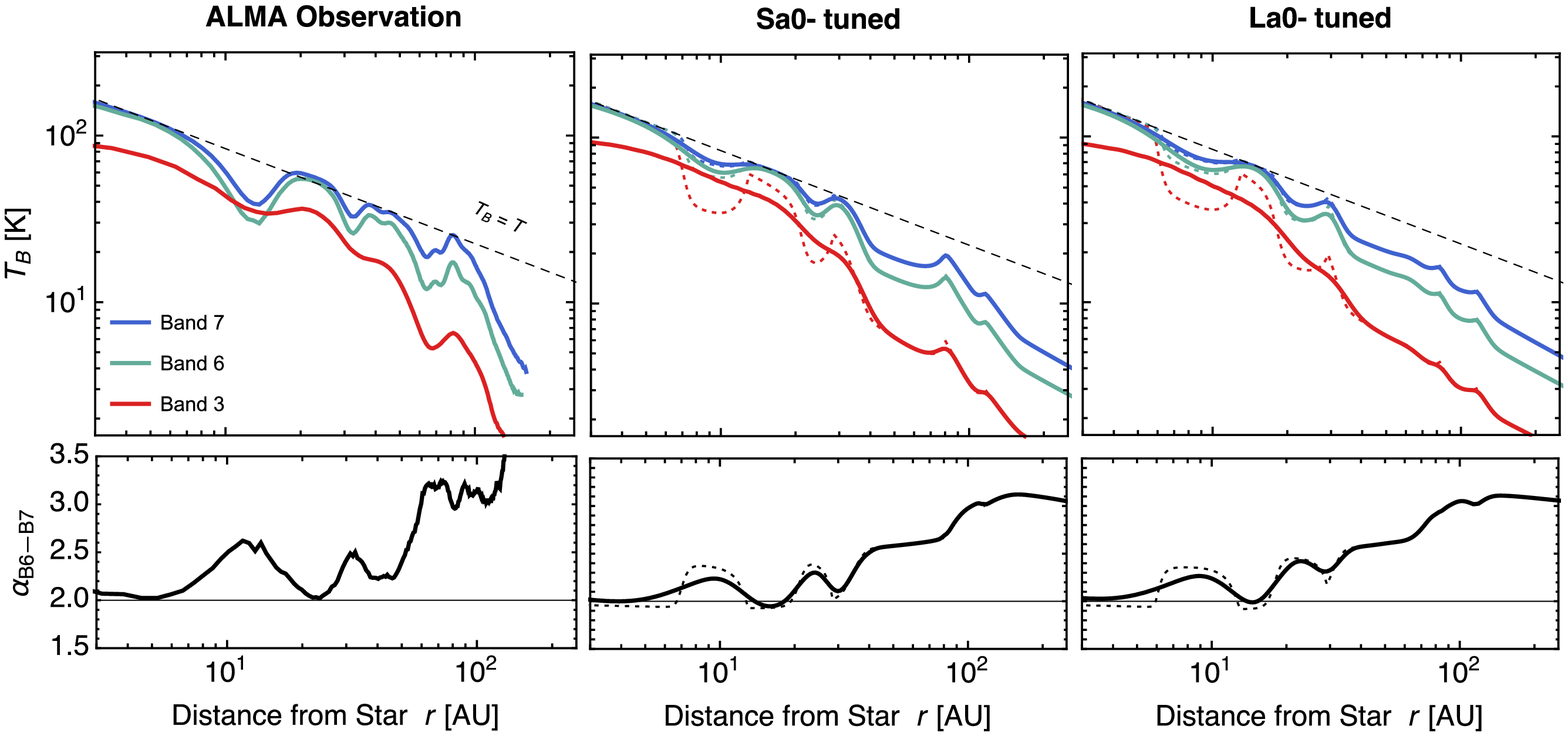}
\caption{Same as Figure~\ref{fig:TBcomp_Sa0}, but for models Sa0-tuned (center panels) 
and La0-tuned (right panels),
which adopt optimized sublimation energies for ${\rm H_2O}$, ${\rm NH_3}$, and ${\rm C_2H_6}$.
}
\label{fig:TBcomp_tuned}
\end{figure*}
To demonstrate the potential importance of uncertainties in the sublimation energies,
we performed two simulations named Sa0-tuned and La0-tuned.
The parameters adopted in these simulations are the same as those in Sa0 and La0,
respectively, except that 
the sublimation energies of ${\rm H_2O}$, ${\rm NH_3}$, and ${\rm C_2H_6}$
are lowered by 10\% from our baseline values (see column 5 of Table~\ref{tab:Pev}).
According to the estimate shown above, such modifications 
shift the sintering zones of these volatiles outward by $\sim 20\%$.
The resulting radial profiles of $T_{\rm B}$ and $\alpha_{\rm 0.87\textrm{--}1.3~mm}$ 
before smoothing are shown in Figure~\ref{fig:TB_tuned}.
In model Sa0-tuned, the three innermost sintering zones are located at 4--8 AU (${\rm H_2O}$), 
13--23 AU (${\rm NH_3}$--${\rm CO_2}$--${\rm H_2S}$), and 29--40 AU (${\rm C_2H_6}$)
instead of 3--6 AU, 11--23 AU, and 24--33 AU as in model Sa0.
Figure~\ref{fig:TBcomp_tuned} shows the radial emission profiles after smoothing, 
which we compare with the ALMA observation. 
For the sake of comparison, we define the radii of the bright and dark rings 
by the orbital radii at which $T_{\rm B}/T$ at Band 6 is locally maximized and minimized, respectively. 
Then, the radii of four innermost bright/dark rings are found to be
13, 23, 32, and 38 AU for the observation, 
9, 16, 24, and 30 AU for model Sa0-NoSint, 
and 9, 15, 24, and 29 AU for model La0-NoSint. 
Thus, model Sa0-NoSint reproduces the ring radii in the observed image 
to an accuracy of $\la 30\%$, which is about 10\% better than model Sa0.   
Model La0-NoSint is slightly less accurate with the maximum error of $35\%$, 
but is still 10\% better than the original model La0.   
Furthermore, the second innermost dark ring in these models 
is much wider than that in model Sa0.
As a consequence, the radial distribution of $\alpha_{\rm B6\textrm{--}B7}$ 
now has a clear peak structure at the position of this ring, 
and the peak value $\approx 2.3$ agrees with the observed value $\approx 2.5$
to within a relative error of 10\%. 

Thus, assuming sublimation energies of ${\rm H_2O}$, ${\rm NH_3}$, 
and ${\rm C_2H_6}$ that are only 10\% lower than the baseline values 
significantly improves our predictions for the radial configuration of the dust rings.
Interestingly, for ${\rm H_2O}$, ${\rm NH_3}$, the tuned values of the sublimation energies 
are more consistent with the results of recent experiments by \citet{MMBG14} 
(5165 K for ${\rm H_2O}$, 2965 K for ${\rm NH_3}$) than our fiducial values.
However, not every new data for sublimation energies improve our predictions. 
\citet{MMBG14} also measured the sublimation energies of ${\rm CO_2}$ and ${\rm CO}$,
and the measured values (2605 K for ${\rm CO_2}$, 890 K for ${\rm CO}$) are also lower 
than ours by about 10 \%. 
A 10 \% change in the sublimation energy of ${\rm CO_2}$ has little effect on 
the resulting ring patterns because the ${\rm CO_2}$ sintering zone 
partially overlaps with the sintering zones of ${\rm NH_3}$ and ${\rm H_2S}$.  
A 10 \% decrease in the sublimation energy of ${\rm CO}$ shifts the ${\rm CO}$ sintering zone 
to 143--197 AU, which makes the correspondence between the CO sintering zone and 
the faint 97 AU ring of HL Tau (see Section~\ref{sec:ALMA}) less good. 

\section{Discussion}\label{sec:discussion}
{
\subsection{Possible Effects of Bouncing}\label{sec:bouncing}
As mentioned in the introduction, sintered aggregates tend to bounce rather than stick 
when they collide at a velocity below the fragmentation threshold.  
For example, sintered aggregates made of 0.1~\micron-sized icy grains
bounce at $1~{\rm m~s^{-1}} \la \Delta v \la 20~{\rm m~s^{-1}}$ (\citealt{S99}; S.~Sirono, in preparation). 
This effect has been neglected in this study by simply applying Equation~\eqref{eq:Dm} to both sintered and unsintered aggregates. In principle, such a simplification causes an overestimate of the maximum size of aggregates in the sintering zones. However, this effect is expected to be minor because sintered aggregates grow only moderately even without the bouncing effect. To see this, we go back to the radial profile of the radius $a_*$ of the representative aggregates from run Sa0 already shown in Figure~\ref{fig:aSigma}(a). Since the radial inward flow of the aggregates is nearly stationary (see Figure~\ref{fig:aSigma}(d)), this figure shows how the size of individual representative aggregates evolve as they move inward. We find that $a_*$ is radially constant in the ${\rm CH_4}$ and ${\rm CO}$ sintering zones, meaning that the aggregates do not grow at all when they go through these zones. Appreciable growth of sintered aggregates occurs only in
inner regions of the ${\rm H_2O}$ and ${\rm NH_3}$--${\rm CO_2}$--${\rm H_2S}$ sintering zones.
However, even at these locations, the aggregate size increases only by a factor of less than two until 
they reach the inner edges of the sintering zones. 
Therefore, we can conclude that inclusion of bouncing collisions would 
little change the evolution of representative aggregates in the sintering zones.}

{
\subsection{Limitations of the Single Size Approach}\label{sec:distr}
Our single-size approach (Section~\ref{sec:global}) relies on 
the assumption that the mass budget of dust at each orbital radius is dominated by
a single population of aggregates having mass $m = m_*(r)$. 
This assumption might be inadequate at the boundaries 
of sintering and non-sintering zones, 
around which two populations of aggregates of different characteristic sizes 
(i.e., sintered and unsintered aggregates) can coexist.
However, this effect would only be important in the close vicinity of the boundaries
because the sintering timescale is a steep function of $r$
and because the aggregates in our simulations do not drift faster 
than they collide with each other. 

A probably more critical limitation of the single-size approach 
is that one has to assume the size distribution of fragments produced by 
the collisions of the mass-dominating aggregates.
In this study, we have avoided detailed modeling of the fragmentation process 
by assuming a simple power-law fragment size distribution (Equation~\eqref{eq:dNda}) 
independently of the collision velocity of the largest aggregates.
The assumed power-law distribution would reasonably approximate 
the true fragment size distribution when the collisions of 
the mass-dominating aggregates are highly disruptive.
In fact, however, unsintered aggregates in our simulations 
do not experience catastrophic disruption since their collision velocity
 is always below  the catastrophic disruption threshold $\Delta v_{\rm frag}$.
For example, our fiducial run Sa0 shows that $\Delta m_*/m_* \approx 0.2$--$0.4$
in the non-sintering zones, which implies that fragments would carry away only 
a few tens of percent of the total mass of two colliding mass-dominating aggregates in these zones.
Equation~\eqref{eq:dNda} might also overestimate 
the amount of fragments from sintered aggregates 
because they in reality bounce off rather than fragment at low collision velocities 
(see Section~\ref{sec:bouncing}).
Preliminary results of our aggregate collision simulations (S.~Sirono 2015, in preparation) 
show that fragments from two colliding sintered aggregates carry away only a few percent 
of their total mass even when $\Delta v \approx \Delta v_{\rm frag}$.

Therefore, it important to assess how the predictions 
from our models depend on the amount of fragments assumed. 
We here consider a fragment size distribution that is similar to Equation~\eqref{eq:dNda}
but assumes a reduced amount of fragments in the size range $a < a_*/10$, 
\beq
N_{\rm d}'(a) = \left\{ \begin{array}{ll}
C a^{-3.5}, & a_*/10 < a < a_*, \\
f_{\rm red}Ca^{-3.5}, & a_{\rm min} < a < a_*/10, \\
0, & {\rm otherwise},
\end{array}\right.
\label{eq:dNda_mod}
\eeq
where the factor $f_{\rm red}(<1)$ encapsulates the reduction of fragment production
and $C$ is again determined by the condition $\int_0^\infty m N_{\rm d}'(a) {\rm d}a = \Sigma_{\rm d}$. 
We consider two values $f_{\rm red} = 0.3$ and 0.03 based on the estimates 
for unsintered and sintered aggregates mentioned above.
 
\begin{figure*}[t]
\centering
\includegraphics[width=18cm]{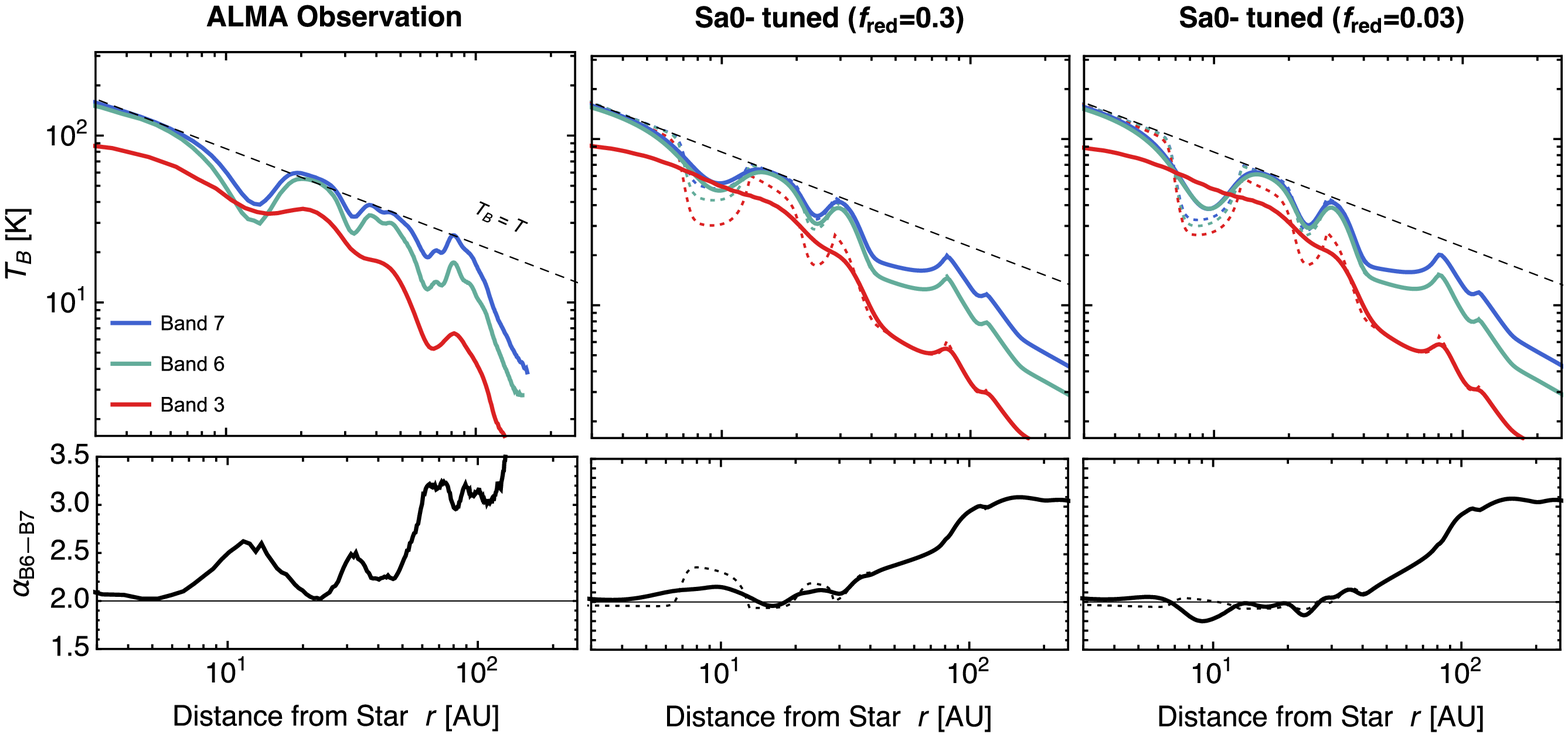}
\caption{Same as Figure~\ref{fig:TBcomp_Sa0}, but for model Sa0-tuned 
with the modified fragment size distribution given by Equation~\eqref{eq:dNda_mod}.
The center and right panels are for 
fragment reduction factor $f_{\rm red} = 0.3$ (center panels) and 0.03 (right panels).
}
\label{fig:TBcomp_fred}
\end{figure*}
We now recalculate the radial profiles of $T_{\rm B}$ and $\alpha_{\rm B6\textrm{--}B7}$
for model Sa0-tuned using Equation~\eqref{eq:dNda_mod} instead of Equation~\eqref{eq:dNda}.
The results are shown in the center and right panels of Figure~\ref{fig:TBcomp_fred}. 
These together with the result for $f_{\rm red} = 1$ (the center panels in Figure~\ref{fig:TBcomp_tuned})
show that both $T_{\rm B}$ and $\alpha_{\rm B6\textrm{--}B7}$ in the two innermost emission dips
decrease as $f_{\rm red}$ is decreased. 
In these inner regions, the radius $a_*$ of the largest (mass-dominating) aggregates is
significantly larger than millimeters, and therefore the fragments smaller than millimeters 
have a non-negligible contribution to the millimeter dust opacity. 
In this case, the opacity index $\beta_{\rm 0.87\textrm{--}1.3~mm}$ 
decreases toward zero, which is the value in the geometric optics limit, 
as the amount of the fragments decreases \citep[see, e.g.,][]{D06}.
This explains why a lower value of $f_{\rm red}$ leads to a lower value of the 
spectral slope in the optically thin inner dips where the relation 
$\alpha_{\rm 0.87\textrm{--}1.3~mm} \approx \beta_{\rm 0.87\textrm{--}1.3~mm}+2$ applies.    
In the case of $f_{\rm red}= 0.03$, the spectral slope in the innermost emission dip  falls below two 
due to the effect of the frequency-dependent angular resolution already mentioned in Section~\ref{sec:ALMA}.
We find that the results for $f_{\rm red} = 0.3$ and $0.03$ better reproduce 
the observed depths of the two innermost emission dips than the result for $f_{\rm red} = 1$.
However, these low-$f_{\rm red}$ models yield poorer agreement with 
the observed value of $\alpha_{\rm B6\textrm{--}B7}$ in these emission dips.
Varying the value of $f_{\rm red}$ within the range $0.03$--1 
has no effect on the predictions for $T_{\rm B}$ and $\alpha_{\rm B6\textrm{--}B7}$
outside the two innermost emission dips.
Taken together, we cannot judge at this point which value of $f_{\rm red}$ best reproduces 
the observed appearance of HL Tau. 
In any case, the effects of assuming lower values of $f_{\rm red}$ are not significant
as long as $f_{\rm red}$ is higher than 0.3, the value we expect for unsintered aggregates in the dark rings.
}

{
\subsection{Evolution of the Monomer Size Distribution and Possible Planetesimal Formation near the ${\rm H_2O}$ Sintering Zone}\label{sec:ripening}
We have assumed that vapor transport within an aggregate, which drives sintering,
does not alter the size of monomer grains. 
In reality, this is not true because the growth of necks must be compensated 
by the shrinkage of the bodies of monomers. 
Furthermore, if the monomers within an aggregate are not uniform in size,
a small number of large monomers may grow, 
the phenomenon known as Ostwald ripening (see, e.g., \citealt{LP81}),
while smaller ones may completely evaporate leaving silicate grains \citep{S11a,KS11}.\footnote{In general, vapor tends to be transported from solid surfaces of a high positive curvature (e.g., bumps) to surfaces of a high negative curvature (e.g., dips) 
so that the total energy of the solid associated with surface tension is minimized (see, e.g., \citealt{B07}).
Sintering (neck growth) is simply driven by the high negative curvature of the necks.
Ostwald ripening is the phenomenon in which 
large monomers having a (positive) surface curvature lower than
the average surface curvature within the aggregate grow 
at the expense of small monomers having a curvature higher than the average.  
} 
The evolution of the monomer size distribution is negligible in the sintering zones of 
minor volatiles like ${\rm NH_3}$ and ${\rm CH_4}$ (as noted in the introduction, sintering can still occur because the neck volume is much smaller than the monomer volume).
However, this is not the case for water ice since it constitutes more than half of the monomer volume.

As pointed out by \citet{KS11}, the evolution of the monomer size distribution due 
to water vapor transport would result in a decrease in the sticking efficiency of the icy aggregates. 
\citet{KS11} showed that water vapor transport within an aggregate produces 
a small number of large ice-rich monomers and a large number of small bare silicate monomers.
Aggregates made of the two populations of monomers would be fragile, 
like sintered aggregates made of equally ice-coated grains, 
because the total binding energy ($\propto$ number of contacts) of the former
would be determined by weak silicate--silicate contacts.
These aggregates would experience catastrophic disruption and pile up 
near the ${\rm H_2O}$ snow line in a similar way to sintered aggregates.
 
In addition, the evaporation of small monomers could directly result in fragmentation of the icy aggregates. \citet{S11a} demonstrated that an aggregate made of initially submicron-sized ${\rm H_2O}$ monomers breaks up into large detached monomers of radii $\sim 1$--10~${\rm \mu m}$ (see also \citealt{KS11}).
This spontaneous breakup would also produce a pileup of dust near the ${\rm H_2O}$ snow line, 
but the surface density contrast provided by this effect  
that could be much more significant than that by than sintering-induced fragmentation because the micron-sized monomers are much more strongly coupled to the disk gas than aggregates. \citet{S11a} points out the dust surface density in the ${\rm H_2O}$ sintering zone can become high enough to trigger the formation of icy planetesimals via the gravitational instability \citep{GW73,S98,YS02}.

Thus, the evolution of the monomer size distribution due to vapor transport 
would further promote the fragmentation and accumulation of aggregates in the ${\rm H_2O}$ sintering zone. 
This may not affect the observational appearance of the HL Tau disk 
because the ${\rm H_2O}$ sintering zone is already optically thick even without the monomer size evolution.
However, this could have a significant effect on planet(esimal) formation near the ${\rm H_2O}$ snow line.
}

{
\subsection{Possible Effects of Porosity Evolution}\label{sec:porosity}
We have neglected the evolution of the porosity of icy aggregates by 
assuming the fixed aggregate internal density of $0.24~{\rm g~cm^{-3}}$. 
In reality, the porosity of aggregates (much smaller than planetesimals) 
can vary greatly through collisions \citep[e.g.,][]{OST07,SWT08,OTS09}, 
compression by gas drag \citep{KTOW13a,KTOW13b}, 
and condensation and sintering of ${\rm H_2O}$ ice (again, minor volatile ices occupy
only a small fraction of grain volume and are therefore negligible here).
In the absence of the compaction due to condensation and sintering, 
the internal density of icy aggregates in protoplanetary disks can decline to as low as 
$10^{-4}$--$10^{-3}~{\rm g~cm^{-3}}$ \citep{OTKW12,KTOW13a}. 

The porosity evolution has two important effects on the evolution of icy aggregates. 
First, fluffy aggregates tend to grow to the size regime 
where the gas drag onto the aggregates is determined by Stokes' law.
This particularly occurs in inner regions of protoplanetary disks 
where the mean free path of the disk gas molecules is short. 
In the Stokes drag regime, the growth limit given by Equation~\eqref{eq:St_drift} 
no longer applies because that limit assumes Epstein's drag law. 
\citet{OTKW12} found that fluffy aggregates that are subject to Stokes' drag law 
grow fast enough to overcome the radial drift barrier.
In a massive protoplanetary disk as considered in this study, 
icy aggregates break through the drift barrier and form icy planetesimals 
at $r \la 10~{\rm AU}$ (see the bottom right panel of \citealt{KTOW13b}).
Second, the millimeter absorption opacity of large porous aggregates 
is similar to that of small compact aggregates because $\kappa_{\rm d,\nu}$ 
can be approximated as a function of the product $\rho_{\rm int}a$ \citep{KOTN14}.

In this study, all these important effects have been neglected
by fixing the aggregate porosity to a relatively low value. 
Rapid coagulation beyond the Epstein drag regime could 
strongly reduce the optical depth in the non-sintering zone lying at $r \sim 10~{\rm AU}$,
where the radial drift barrier determines the maximum size of the relatively compact aggregates
that we assumed in this study. 
This effect should be studied in detail in future work. 

On the other hand, the above two effects are less important at $r \ga 10~{\rm AU}$, 
where even the highly fluffy aggregates would grow within the Epstein drag regime.
To show this, we first note that neither ${\rm St}_{\rm drift}$ (Equation~\eqref{eq:St_drift})
nor ${\rm St}_{\rm frag}$ (Equation~\eqref{eq:St_frag}) depends on $\rho_{\rm int}$
(note again that Epstein's law is used in the derivation of Equation~\eqref{eq:St_drift}).
This means that porosity evolution has no effect on the maximum attainable Stokes number at each $r$,
and therefore does not change the radial distribution of $\Sigma_{\rm d}$ since 
the radial advection velocity $v_r$ is a function of ${\rm St}$ (see Equation~\eqref{eq:vr}). 
Porosity evolution does not change the dust absorption opacity $\kappa_{\rm d,\nu}$ either
since the product $\rho_{\rm int} a_*$ is proportional to ${\rm St}$ in the Epstein regime (see Equation~\eqref{eq:St}). 
This is a natural consequence of the fact that both the stopping time in the Epstein regime
and the absorption opacity are determined by the mass-to-area ratio 
$m/\pi a^2  (\sim \rho_{\rm int}a)$ of the aggregates.
Taken together, the radial distribution of the absorption optical depth 
$\tau_{\nu} \propto \kappa_{\rm d,\nu}\Sigma_{\rm d}$ is independent of 
the aggregate porosity as long as the aggregates grow within the Epstein drag regime. 

Nevertheless, porosity evolution might have some important observational consequence in the outer disk region.
The millimeter continuum emission from the HL Tau disk is known to be polarized in the direction roughly 
perpendicular to the disk major axis at a level of 0.9\% in the average degree of polarization \citep{SLK+14}.
\citet{YLLS16} and \citet{KMM+15} point out that self-scattering of thermal emission by dust rings 
\citep{KMMTD15} can explain the polarization observation. 
However, this mechanism produces a degree of polarization of $\approx 1\%$ 
only when the size of the dust particles falls within a particular range. 
For example, if the particles are compact and their size distribution obeys our Equation~\eqref{eq:dNda},
the self-scattering scenario is consistent with the observation only when 
the maximum particle size $a_{\rm max}$ ($= a_*$ in this study) is $\sim 0.1~{\rm mm}$
\citep{YLLS16,KMM+15}, which is considerably smaller than 
the size of mass-dominating aggregates predicted in this study.
In the compact particle model, a particle size larger than a millimeter is ruled out because 
the scattered emission would then be too weakly polarized \citep[see also][]{KMMTD15}.
By contrast, large but fluffy aggregates are known to produce a highly polarization scattered light 
like constituent monomers do \citep{MRWDM16}, and therefore might be able to explain 
the polarization observation of HL Tau.
}

\subsection{Possible Effects of Condensation Growth}
We have neglected the condensation growth of ice particles near the snow lines. 
In fact, a recent study by \cite{RJ13} has shown that condensation growth can be effective 
at locations slightly outside the ${\rm H_2O}$ snow line. 
This effect might change our predictions for dust evolution near ${\rm H_2O}$ snow line, 
because condensation dominantly takes place on the smallest particles, i.e., {monomers}.  
If condensation dominates over sintering-induced fragmentation and bouncing, 
the regions slightly outside the snow lines would be seen as ${\it dark}$ rings 
at millimeter wavelengths as already pointed out by \citet{ZBB15}.
However, it is unclear whether condensation growth becomes important 
for volatiles less abundant than ${\rm H_2O}$, like ${\rm CO}$ and ${\rm NH_3}$. 
To address this open question, we will incorporate condensation growth into our global dust evolution simulations in future work.

\subsection{Sintering-induced Ring Formation in Other Objects?}
Perhaps the strongest prediction from our model, 
as well as from the model of \citet{ZBB15},
is that the multiple dust rings may not be peculiar to the HL Tau disk. 
In principle, sintering-induced ring formation operates as long as 
the disk is not too depleted of dust (or not too old), 
the monomers are sufficiently small to cause rapid sintering, 
and turbulence is strong enough to cause disruption of sintered aggregates.
If these conditions are satisfied, axisymmetric dust rings emerge 
slightly outside the snow lines of relatively major volatiles.
An intriguing object in this context is TW Hya.
For this system, the CO snow line has been indirectly detected 
at an orbital distance of $\sim 30$ AU \citep{QOW+13}, and furthermore, 
near-infrared scattered light images of the disk suggest the presence of 
two axisymmetric dust gaps at $\sim 80$ AU \citep{DJW+13} and $\sim 20$ AU \citep{AMK+15,RKMD15}.
{A latest ALMA observation has shown that there is also a gap in millimeter dust emission 
in the vicinity of the 20 AU scattered light gap \citep{NTK+16}.}
{The gaps at 80 AU and 20 AU} could be associated with non-sintering zones {exterior} to and 
{interior} to the CO sintering zone, respectively.  
However, it is not obvious whether our ring formation mechanism directly applies to TW Hya
because this object is known to be much older than HL Tau (at least 3 Myr; \citealt{VS11}). 
This question should also be addressed in future work.

\section{Conclusions}\label{sec:summary}
Motivated by the recent ALMA observations of the HL Tau disk, 
we have explored the possibility that sintering of icy dust aggregates 
might lead to the formation of multiple dust rings in a protoplanetary disk.
Sintering is a particle fusion process that occurs 
when the temperature is slightly below the melting point. 
Sintered aggregates are generally harder but less sticky than unsintered aggregates.
Therefore, if dust aggregates in a protoplanetary disk consist of various materials, 
their growth can be suppressed at different orbits corresponding to the sublimation fronts
of different materials.
This possibility was originally pointed out by \citet{S99,S11b}, 
and here we have for the first time studied its effects on the global evolution of dust in a protoplanetary disk. 

Following \citet{S11b}, we have regard aggregates as sintered if their sintering timescale 
is shorter than their collision timescale (Section~\ref{sec:sint}). 
This criterion defines the ``sintering zones'' in which one of the volatile species included in the aggregates 
causes sintering. The temperature profile of the HL Tau disk has been modeled based on 
the millimeter intensity maps provided by the ALMA observations \citep{ALMA+15}
together with the assumption that the central emission peak and inner bright rings 
are optically thick at 1 mm (Section~\ref{sec:temp},  Figure~\ref{fig:T}).
Based on the aggregate collision simulations by \citet{SU14}, 
we have assumed that sintered aggregates have a maximum sticking velocity 
that is 60\% lower than that for unsintered aggregates (Equations~\eqref{eq:Dvfrag_NS} and \eqref{eq:Dvfrag_S}; 
Figure~\ref{fig:coll}(a)).
For both sintered and unsintered aggregates, we have regarded 
collisions at velocities higher than the threshold 
disruptive (Equation~\eqref{eq:Dm}; Figure~\ref{fig:coll}(b)).

Using the aggregate sintering model described above, we have simulated 
the global evolution of dust in the HL Tau disk for various sets of model parameters.
As a first step toward more comprehensive modeling, we  focused 
on the evolution of mass-dominating (largest) aggregates, 
and assumed a power-law size distribution for smaller aggregates when 
we convert the simulation data into radial profiles of millimeter dust emission.
Key parameters in our model are the strength of turbulence ($\alpha_{\rm t}$), 
the size of monomers that constitute the aggregates ($a_0$), and the sublimation energies of the volatiles ($L_j$).
The monomer size is relevant here because it controls the timescale of sintering 
(Equation~\eqref{eq:tsint}) and the fragmentation strength of dust aggregates 
(Equations~\eqref{eq:Dvfrag_NS} and \eqref{eq:Dvfrag_S}).

Our key findings are summarized as follows.

\begin{enumerate}
\item
Because dust is gradually lost to the central star owing to radial drift, 
the total dust mass in the disk decreases with time (Section~\ref{sec:time}). 
For the total disk mass of $0.2~M_\sun$ and the initial dust-to-gas mass ratio of 0.01, 
our HL Tau disk models best reproduce the millimeter flux densities from
the ALMA observations when the disk age in the models is chosen to be 0.1--0.5 Myr.
This is consistent with the general belief that HL Tau is a young ($\la$ 1 Myr) protoplanetary disk. 

\item
Dust aggregates pile up in the sintering zones due to the combined effect of 
radial drift and sintering-induced fragmentation (Section~\ref{sec:aSigma}).
In general, aggregates grow locally until either rapid radial drift or 
fragmentation starts to halt their growth.
After that, the aggregates start to drift toward the central star 
at a velocity proportional to the maximum size.
Sintered aggregates have a lower maximum size and hence a lower inward drift velocity 
than unsintered aggregates simply because the former tend to disrupt more easily upon collision.
For this reason, aggregates tend to pile up in the sintering zones.

\item
At millimeter wavelengths, the sintering zones are seen as {\it bright} rings 
because the dust surface density in the sintering zones is higher than 
in the non-sintering zones (Section~\ref{sec:tauTB}). 
In particular, at the wavelengths of  0.87 mm and 1.3 mm,
the three innermost sintering zones (which correspond 
to ${\rm H_2O}$, ${\rm NH_3}$--${\rm CO_2}$--${\rm H_2S}$, 
and ${\rm C_2H_6}$, respectively) are optically thick, producing a millimeter spectral index of $\approx 2$.
The predicted spectral index and brightness temperatures are consistent 
with those of the central emission peak and two innermost bright rings
of the HL Tau disk (Section~\ref{sec:ALMA}).
Our model also predicts an optically thin emission peak at $\approx 80$ AU,
which is associated with the ${\rm CH_4}$ sintering zone,  
and two optically thin dark rings of a spectral slope of 2.3--2.5 at $\la 40~{\rm AU}$,
which are associated with the two innermost non-sintering zones.
These are all consistent with the ALMA observation.  
 
\item
The sintering-induced ring patterns diminish as the disk becomes depleted of dust (Section~\ref{sec:lifetime}).
As the dust-to-gas mass ratio decreases, aggregates collide with each other less frequently,
making their maximum size more severely limited by radial drift rather than fragmentation.
The sintering-induced rings disappear once radial drift dominates over fragmentation even in the sintering zones.
For a disk having the total mass of $0.2~M_\sun$, the characteristic radius of 150 AU, 
and the initial dust-to-gas mass ratio of 0.01, we find that the sintering-induced rings decay in 2 Myr.
The ring patterns of HL Tau might be a sign of its youth. 

\item
Models that assume lower turbulence parameter $\alpha_{\rm t}$
toward the central star best reproduce the multiple ring structure of HL Tau (Section~\ref{sec:alpha}). 
If $\alpha_{\rm t}$ were radially constant, 
turbulence-driven collision velocity $\propto \sqrt{\alpha_{\rm t}T}$
would increase with decreasing radial distance $r$.
In this case, either unsintered aggregates would fragment at small $r$, 
or sintered aggregates would fragment at large $r$. 
The former case does not reproduce dark rings at small $r$, 
while the latter case does not reproduce bright rings at large $r$. 
The radial dependence of $\alpha_{\rm t}$ suggested by our model
is qualitatively consistent with the predictions from magnetohydrodynamical turbulence models. 

\item
The vertical extent the observed dust rings places a strong constraint on the turbulence 
strength and monomer size assumed in our model (Section~\ref{sec:a0}).
When $a_0 \gg 1~\micron$, sintering would be too slow to induce dust ring formation.   
When $a_0 \ll 1~\micron$, disruption of sintered aggregates 
would require turbulence that is too strong to allow dust settling to the midplane.
If the macroscopic dust particles in the HL Tau disk must already have 
settled as suggested by previous studies \citep{KLM11,PDM+16}, 
disk turbulence must be moderately weak ($10^{-4} < \alpha_{\rm t} \la 10^{-3}$) 
{\it and} monomers must be micron-sized ($1~\micron \la a_0 < 4~\micron$).
The predicted monomer size might be consistent with the 
near-infrared observations of HL Tau that suggesting 
the presence of micron-sized grains on the surface of its circumstellar envelope \citep{LFT+04,MOPI08}. 

\item
The exact locations and widths of the dust rings predicted by our model 
are subject to uncertainties in the vapor pressure data (Section~\ref{sec:L}).  
In general, a 10\% uncertainty in the sublimation energy of a volatile species 
causes a $\sim 20\%$ uncertainty in the predicted location of the volatile's sintering zone.
We find that reducing the sublimation energies of ${\rm H_2O}$, ${\rm NH_3}$, and 
${\rm C_2H_6}$ by only 10\% from our fiducial values 
significantly improves our predictions for the observational appearance of
the ring structures in an inner region of the HL Tau disk.
The models using the tuned sublimation energies reproduce 
the radial positions of the HL Tau's inner rings to an accuracy of $\la 30\%$.

\end{enumerate}

\acknowledgments
The authors thank Neal Turner for discussions on the temperature distribution of the HL Tau disk, 
and Akimasa Kataoka for useful comments on the modeling of aggregate porosity. 
We also thank Takashi Tsukagoshi, Tetsuya Hama, Misato Fukagawa, Hideko Nomura, and Mario Flock 
for comments and discussions, and the anonymous referee for his/her prompt and constructive comments.
This work is supported by Grants-in-Aid for Scientific Research 
(\#23103004, 23103005, 25400447, 26287101, 15H02065) from MEXT of Japan and 
by  the Astrobiology Center Project of National Institutes of Natural Sciences (NINS) (Grant Number AB271020).  
This paper makes use of the following ALMA data: ADS/JAO.ALMA\#2011.0.00015.SV.
ALMA is a partnership of ESO (representing its member states), NSF (USA)
and NINS (Japan), together with NRC (Canada), NSC and ASIAA (Taiwan), and
KASI (Republic of Korea), in cooperation with the Republic of Chile.
The Joint ALMA Observatory is operated by ESO, AUI/NRAO and NAOJ.

\bibliographystyle{apj}
\bibliography{myrefs_160203}

\def\SortNoop#1{}
\begin{thebibliography}{}
\expandafter\ifx\csname natexlab\endcsname\relax\def\natexlab#1{#1}\fi

\bibitem[{{Adachi} {et~al.}(1976){Adachi}, {Hayashi}, \& {Nakazawa}}]{AHN76}
{Adachi}, I., {Hayashi}, C., \& {Nakazawa}, K. 1976, Progress of Theoretical
  Physics, 56, 1756

\bibitem[{{Akiyama} {et~al.}(2015){Akiyama}, {Muto}, {Kusakabe}, {Kataoka},
  {Hashimoto}, {Tsukagoshi}, {Kwon}, {Kudo}, {Kandori}, {Grady}, {Takami},
  {Janson}, {Kuzuhara}, {Henning}, {Sitko}, {Carson}, {Mayama}, {Currie},
  {Thalmann}, {Wisniewski}, {Momose}, {Ohashi}, {Abe}, {Brandner}, {Brandt},
  {Egner}, {Feldt}, {Goto}, {Guyon}, {Hayano}, {Hayashi}, {Hayashi}, {Hodapp},
  {Ishi}, {Iye}, {Knapp}, {Matsuo}, {Mcelwain}, {Miyama}, {Morino},
  {Moro-Martin}, {Nishimura}, {Pyo}, {Serabyn}, {Suenaga}, {Suto}, {Suzuki},
  {Takahashi}, {Takato}, {Terada}, {Tomono}, {Turner}, {Watanabe}, {Yamada},
  {Takami}, {Usuda}, \& {Tamura}}]{AMK+15}
{Akiyama}, E., {Muto}, T., {Kusakabe}, N., {et~al.} 2015, \apjl, 802, L17

\bibitem[{{ALMA Partnership} {et~al.}(2015){ALMA Partnership}, {Brogan},
  {P{\'e}rez}, {Hunter}, {Dent}, {Hales}, {Hills}, {Corder}, {Fomalont},
  {Vlahakis}, {Asaki}, {Barkats}, {Hirota}, {Hodge}, {Impellizzeri}, {Kneissl},
  {Liuzzo}, {Lucas}, {Marcelino}, {Matsushita}, {Nakanishi}, {Phillips},
  {Richards}, {Toledo}, {Aladro}, {Broguiere}, {Cortes}, {Cortes}, {Espada},
  {Galarza}, {Garcia-Appadoo}, {Guzman-Ramirez}, {Humphreys}, {Jung}, {Kameno},
  {Laing}, {Leon}, {Marconi}, {Mignano}, {Nikolic}, {Nyman}, {Radiszcz},
  {Remijan}, {Rod{\'o}n}, {Sawada}, {Takahashi}, {Tilanus}, {Vila Vilaro},
  {Watson}, {Wiklind}, {Akiyama}, {Chapillon}, {de Gregorio-Monsalvo}, {Di
  Francesco}, {Gueth}, {Kawamura}, {Lee}, {Nguyen Luong}, {Mangum}, {Pietu},
  {Sanhueza}, {Saigo}, {Takakuwa}, {Ubach}, {van Kempen}, {Wootten},
  {Castro-Carrizo}, {Francke}, {Gallardo}, {Garcia}, {Gonzalez}, {Hill},
  {Kaminski}, {Kurono}, {Liu}, {Lopez}, {Morales}, {Plarre}, {Schieven},
  {Testi}, {Videla}, {Villard}, {Andreani}, {Hibbard}, \&
  {Tatematsu}}]{ALMA+15}
{ALMA Partnership}, {Brogan}, C.~L., {P{\'e}rez}, L.~M., {et~al.} 2015, \apjl,
  808, L3

\bibitem[{{Andrews} {et~al.}(2009){Andrews}, {Wilner}, {Hughes}, {Qi}, \&
  {Dullemond}}]{AWH+09}
{Andrews}, S.~M., {Wilner}, D.~J., {Hughes}, A.~M., {Qi}, C., \& {Dullemond},
  C.~P. 2009, \apj, 700, 1502

\bibitem[{{Bai}(2011)}]{B11a}
{Bai}, X.-N. 2011, \apj, 739, 50

\bibitem[{{Balbus} \& {Hawley}(1991)}]{BH91}
{Balbus}, S.~A., \& {Hawley}, J.~F. 1991, \apj, 376, 214

\bibitem[{{Bauer} {et~al.}(1997){Bauer}, {Finocchi}, {Duschl}, {Gail}, \&
  {Schloeder}}]{BFDGS97}
{Bauer}, I., {Finocchi}, F., {Duschl}, W.~J., {Gail}, H.-P., \& {Schloeder},
  J.~P. 1997, \aap, 317, 273

\bibitem[{{Beckwith} {et~al.}(1990){Beckwith}, {Sargent}, {Chini}, \&
  {Guesten}}]{BSCG90}
{Beckwith}, S.~V.~W., {Sargent}, A.~I., {Chini}, R.~S., \& {Guesten}, R. 1990,
  \aj, 99, 924

\bibitem[{{Birnstiel} {et~al.}(2010){Birnstiel}, {Dullemond}, \&
  {Brauer}}]{BDB10}
{Birnstiel}, T., {Dullemond}, C.~P., \& {Brauer}, F. 2010, \aap, 513, A79

\bibitem[{{Birnstiel} {et~al.}(2012){Birnstiel}, {Klahr}, \&
  {Ercolano}}]{BKE12}
{Birnstiel}, T., {Klahr}, H., \& {Ercolano}, B. 2012, \aap, 539, A148

\bibitem[{{Birnstiel} {et~al.}(2011){Birnstiel}, {Ormel}, \&
  {Dullemond}}]{BOD11}
{Birnstiel}, T., {Ormel}, C.~W., \& {Dullemond}, C.~P. 2011, \aap, 525, A11

\bibitem[{{Blackford}(2007)}]{B07}
{Blackford}, J.~R. 2007, Journal of Physics D Applied Physics, 40, 355

\bibitem[{{Brauer} {et~al.}(2008){Brauer}, {Dullemond}, \& {Henning}}]{BDH08}
{Brauer}, F., {Dullemond}, C.~P., \& {Henning}, T. 2008, \aap, 480, 859

\bibitem[{{Chokshi} {et~al.}(1993){Chokshi}, {Tielens}, \&
  {Hollenbach}}]{CTH93}
{Chokshi}, A., {Tielens}, A.~G.~G.~M., \& {Hollenbach}, D. 1993, \apj, 407, 806

\bibitem[{{Collings} {et~al.}(2004){Collings}, {Anderson}, {Chen}, {Dever},
  {Viti}, {Williams}, \& {McCoustra}}]{CAC+04}
{Collings}, M.~P., {Anderson}, M.~A., {Chen}, R., {et~al.} 2004, \mnras, 354,
  1133

\bibitem[{{Collings} {et~al.}(2003){Collings}, {Dever}, {Fraser}, {McCoustra},
  \& {Williams}}]{CDF+03}
{Collings}, M.~P., {Dever}, J.~W., {Fraser}, H.~J., {McCoustra}, M.~R.~S., \&
  {Williams}, D.~A. 2003, \apj, 583, 1058

\bibitem[{{Debes} {et~al.}(2013){Debes}, {Jang-Condell}, {Weinberger},
  {Roberge}, \& {Schneider}}]{DJW+13}
{Debes}, J.~H., {Jang-Condell}, H., {Weinberger}, A.~J., {Roberge}, A., \&
  {Schneider}, G. 2013, \apj, 771, 45

\bibitem[{{Dipierro} {et~al.}(2015){Dipierro}, {Price}, {Laibe}, {Hirsh},
  {Cerioli}, \& {Lodato}}]{DPL+15}
{Dipierro}, G., {Price}, D., {Laibe}, G., {et~al.} 2015, \mnras, 453, L73

\bibitem[{{Dohnanyi}(1969)}]{D69}
{Dohnanyi}, J.~S. 1969, \jgr, 74, 2531

\bibitem[{{Dominik} \& {Tielens}(1997)}]{DT97}
{Dominik}, C., \& {Tielens}, A.~G.~G.~M. 1997, \apj, 480, 647

\bibitem[{{Dong} {et~al.}(2015){Dong}, {Zhu}, \& {Whitney}}]{DZW15}
{Dong}, R., {Zhu}, Z., \& {Whitney}, B. 2015, \apj, 809, 93

\bibitem[{{Draine}(2003)}]{D03b}
{Draine}, B.~T. 2003, \apj, 598, 1026

\bibitem[{{Draine}(2006)}]{D06}
---. 2006, \apj, 636, 1114

\bibitem[{{Dubrulle} {et~al.}(1995){Dubrulle}, {Morfill}, \& {Sterzik}}]{DMS95}
{Dubrulle}, B., {Morfill}, G., \& {Sterzik}, M. 1995, \icarus, 114, 237

\bibitem[{{Dzyurkevich} {et~al.}(2010){Dzyurkevich}, {Flock}, {Turner},
  {Klahr}, \& {Henning}}]{DFTKH10}
{Dzyurkevich}, N., {Flock}, M., {Turner}, N.~J., {Klahr}, H., \& {Henning}, T.
  2010, \aap, 515, A70

\bibitem[{{Estrada} {et~al.}(2016){Estrada}, {Cuzzi}, \& {Morgan}}]{ECM16}
{Estrada}, P.~R., {Cuzzi}, J.~N., \& {Morgan}, D.~A. 2016, \apj, in press,
  arXiv:1506.01420

\bibitem[{{Ferrier}(1994)}]{F94}
{Ferrier}, B.~S. 1994, Journal of the Atmospheric Sciences, 51, 249

\bibitem[{{Flock} {et~al.}(2015){Flock}, {Ruge}, {Dzyurkevich}, {Henning},
  {Klahr}, \& {Wolf}}]{FRD+15}
{Flock}, M., {Ruge}, J.~P., {Dzyurkevich}, N., {et~al.} 2015, \aap, 574, A68

\bibitem[{{Fouchet} {et~al.}(2010){Fouchet}, {Gonzalez}, \& {Maddison}}]{FGM10}
{Fouchet}, L., {Gonzalez}, J.-F., \& {Maddison}, S.~T. 2010, \aap, 518, A16

\bibitem[{{Garaud}(2007)}]{G07}
{Garaud}, P. 2007, \apj, 671, 2091

\bibitem[{{Goldreich} \& {Ward}(1973)}]{GW73}
{Goldreich}, P., \& {Ward}, W.~R. 1973, \apj, 183, 1051

\bibitem[{{Gonzalez} {et~al.}(2015){Gonzalez}, {Laibe}, {Maddison}, {Pinte}, \&
  {M{\'e}nard}}]{GLM+15}
{Gonzalez}, J.-F., {Laibe}, G., {Maddison}, S.~T., {Pinte}, C., \&
  {M{\'e}nard}, F. 2015, \mnras, 454, L36

\bibitem[{{Gonzalez} {et~al.}(2012){Gonzalez}, {Pinte}, {Maddison},
  {M{\'e}nard}, \& {Fouchet}}]{GPMMF12}
{Gonzalez}, J.-F., {Pinte}, C., {Maddison}, S.~T., {M{\'e}nard}, F., \&
  {Fouchet}, L. 2012, \aap, 547, A58

\bibitem[{{Guilloteau} {et~al.}(2011){Guilloteau}, {Dutrey}, {Pi{\'e}tu}, \&
  {Boehler}}]{GDPB11}
{Guilloteau}, S., {Dutrey}, A., {Pi{\'e}tu}, V., \& {Boehler}, Y. 2011, \aap,
  529, A105

\bibitem[{{Hartmann} {et~al.}(1998){Hartmann}, {Calvet}, {Gullbring}, \&
  {D'Alessio}}]{HCGD98}
{Hartmann}, L., {Calvet}, N., {Gullbring}, E., \& {D'Alessio}, P. 1998, \apj,
  495, 385

\bibitem[{{Hayashi} {et~al.}(1993){Hayashi}, {Ohashi}, \& {Miyama}}]{HOM93}
{Hayashi}, M., {Ohashi}, N., \& {Miyama}, S.~M. 1993, \apjl, 418, L71

\bibitem[{Haynes(2014)}]{CRC14}
Haynes, W.~M. 2014, CRC Handbook of Chemistry and Physics, 95th edn. (Boca
  Raton, FL: CRC press)

\bibitem[{{Johansen} {et~al.}(2009){Johansen}, {Youdin}, \& {Klahr}}]{JYK09}
{Johansen}, A., {Youdin}, A., \& {Klahr}, H. 2009, \apj, 697, 1269

\bibitem[{{Johnson} {et~al.}(1971){Johnson}, {Kendall}, \& {Roberts}}]{JKR71}
{Johnson}, K.~L., {Kendall}, K., \& {Roberts}, A.~D. 1971, Royal Society of
  London Proceedings Series A, 324, 301

\bibitem[{{Kanagawa} {et~al.}(2015){Kanagawa}, {Muto}, {Tanaka}, {Tanigawa},
  {Takeuchi}, {Tsukagoshi}, \& {Momose}}]{KMT+15}
{Kanagawa}, K.~D., {Muto}, T., {Tanaka}, H., {et~al.} 2015, \apjl, 806, L15

\bibitem[{{Kataoka} {et~al.}(2015{\natexlab{a}}){Kataoka}, {Muto}, {Momose},
  {Tsukagoshi}, \& {Dullemond}}]{KMMTD15}
{Kataoka}, A., {Muto}, T., {Momose}, M., {Tsukagoshi}, T., \& {Dullemond},
  C.~P. 2015{\natexlab{a}}, ArXiv e-prints, arXiv:1507.08902

\bibitem[{{Kataoka} {et~al.}(2014){Kataoka}, {Okuzumi}, {Tanaka}, \&
  {Nomura}}]{KOTN14}
{Kataoka}, A., {Okuzumi}, S., {Tanaka}, H., \& {Nomura}, H. 2014, \aap, 568,
  A42

\bibitem[{{Kataoka} {et~al.}(2013{\natexlab{a}}){Kataoka}, {Tanaka}, {Okuzumi},
  \& {Wada}}]{KTOW13b}
{Kataoka}, A., {Tanaka}, H., {Okuzumi}, S., \& {Wada}, K. 2013{\natexlab{a}},
  \aap, 557, L4

\bibitem[{{Kataoka} {et~al.}(2013{\natexlab{b}}){Kataoka}, {Tanaka}, {Okuzumi},
  \& {Wada}}]{KTOW13a}
---. 2013{\natexlab{b}}, \aap, 554, A4

\bibitem[{{Kataoka} {et~al.}(2015{\natexlab{b}}){Kataoka}, {Muto}, {Momose},
  {Tsukagoshi}, {Fukagawa}, {Shibai}, {Hanawa}, {Murakawa}, \&
  {Dullemond}}]{KMM+15}
{Kataoka}, A., {Muto}, T., {Momose}, M., {et~al.} 2015{\natexlab{b}}, \apj,
  809, 78

\bibitem[{{Kitamura} {et~al.}(2002){Kitamura}, {Momose}, {Yokogawa}, {Kawabe},
  {Tamura}, \& {Ida}}]{KMY+02}
{Kitamura}, Y., {Momose}, M., {Yokogawa}, S., {et~al.} 2002, \apj, 581, 357

\bibitem[{{Kornet} {et~al.}(2001){Kornet}, {Stepinski}, \&
  {R{\'o}{\.z}yczka}}]{KSR01}
{Kornet}, K., {Stepinski}, T.~F., \& {R{\'o}{\.z}yczka}, M. 2001, \aap, 378,
  180

\bibitem[{{Kretke} \& {Lin}(2007)}]{KL07}
{Kretke}, K.~A., \& {Lin}, D.~N.~C. 2007, \apjl, 664, L55

\bibitem[{{Krijt} {et~al.}(2015){Krijt}, {Ormel}, {Dominik}, \&
  {Tielens}}]{KODT15}
{Krijt}, S., {Ormel}, C.~W., {Dominik}, C., \& {Tielens}, A.~G.~G.~M. 2015,
  \aap, 574, A83

\bibitem[{{Krijt} {et~al.}(2016){Krijt}, {Ormel}, {Dominik}, \&
  {Tielens}}]{KODT16}
---. 2016, \aap, 586, A20

\bibitem[{{Kuroiwa} \& {Sirono}(2011)}]{KS11}
{Kuroiwa}, T., \& {Sirono}, S.-i. 2011, \apj, 739, 18

\bibitem[{{Kwon} {et~al.}(2011){Kwon}, {Looney}, \& {Mundy}}]{KLM11}
{Kwon}, W., {Looney}, L.~W., \& {Mundy}, L.~G. 2011, \apj, 741, 3

\bibitem[{{Kwon} {et~al.}(2015){Kwon}, {Looney}, {Mundy}, \& {Welch}}]{KLMW15}
{Kwon}, W., {Looney}, L.~W., {Mundy}, L.~G., \& {Welch}, W.~J. 2015, \apj, 808,
  102

\bibitem[{{Laibe}(2014)}]{L14}
{Laibe}, G. 2014, \mnras, 437, 3037

\bibitem[{{Lambrechts} \& {Johansen}(2014)}]{LJ14}
{Lambrechts}, M., \& {Johansen}, A. 2014, \aap, 572, A107

\bibitem[{{Lifshitz} \& {Pitaevskii}(1981)}]{LP81}
{Lifshitz}, E.~M., \& {Pitaevskii}, L.~P. 1981, {Physical Kinetics} (Oxford:
  Pergamon)

\bibitem[{{Lor{\'e}n-Aguilar} \& {Bate}(2015)}]{LB15}
{Lor{\'e}n-Aguilar}, P., \& {Bate}, M.~R. 2015, \mnras, 453, L78

\bibitem[{{Lucas} {et~al.}(2004){Lucas}, {Fukagawa}, {Tamura}, {Beckford},
  {Itoh}, {Murakawa}, {Suto}, {Hayashi}, {Oasa}, {Naoi}, {Doi}, {Ebizuka}, \&
  {Kaifu}}]{LFT+04}
{Lucas}, P.~W., {Fukagawa}, M., {Tamura}, M., {et~al.} 2004, \mnras, 352, 1347

\bibitem[{{Luna} {et~al.}(2014){Luna}, {Satorre}, {Santonja}, \&
  {Domingo}}]{LSSD14}
{Luna}, R., {Satorre}, M.~{\'A}., {Santonja}, C., \& {Domingo}, M. 2014, \aap,
  566, A27

\bibitem[{{Mart{\'{\i}}n-Dom{\'e}nech}
  {et~al.}(2014){Mart{\'{\i}}n-Dom{\'e}nech}, {Mu{\~n}oz Caro}, {Bueno}, \&
  {Goesmann}}]{MMBG14}
{Mart{\'{\i}}n-Dom{\'e}nech}, R., {Mu{\~n}oz Caro}, G.~M., {Bueno}, J., \&
  {Goesmann}, F. 2014, \aap, 564, A8

\bibitem[{{Mathis} {et~al.}(1977){Mathis}, {Rumpl}, \& {Nordsieck}}]{MRN77}
{Mathis}, J.~S., {Rumpl}, W., \& {Nordsieck}, K.~H. 1977, \apj, 217, 425

\bibitem[{{Men'shchikov} {et~al.}(1999){Men'shchikov}, {Henning}, \&
  {Fischer}}]{MHF99}
{Men'shchikov}, A.~B., {Henning}, T., \& {Fischer}, O. 1999, \apj, 519, 257

\bibitem[{Meyer(1977)}]{M77}
Meyer, B. 1977, Sulfur, energy, and environment (Elsevier Scientific Pub. Co.)

\bibitem[{{Min} {et~al.}(2016){Min}, {Rab}, {Woitke}, {Dominik}, \&
  {M{\'e}nard}}]{MRWDM16}
{Min}, M., {Rab}, C., {Woitke}, P., {Dominik}, C., \& {M{\'e}nard}, F. 2016,
  \aap, 585, A13

\bibitem[{{Moses} {et~al.}(1992){Moses}, {Allen}, \& {Yung}}]{MAY92}
{Moses}, J.~I., {Allen}, M., \& {Yung}, Y.~L. 1992, \icarus, 99, 318

\bibitem[{{Mumma} \& {Charnley}(2011)}]{MC11}
{Mumma}, M.~J., \& {Charnley}, S.~B. 2011, \araa, 49, 471

\bibitem[{{Mundt} {et~al.}(1988){Mundt}, {Buehrke}, \& {Ray}}]{MBR88}
{Mundt}, R., {Buehrke}, T., \& {Ray}, T.~P. 1988, \apjl, 333, L69

\bibitem[{{Murakawa} {et~al.}(2008){Murakawa}, {Oya}, {Pyo}, \&
  {Ishii}}]{MOPI08}
{Murakawa}, K., {Oya}, S., {Pyo}, T.-S., \& {Ishii}, M. 2008, \aap, 492, 731

\bibitem[{{Nomura} {et~al.}(2015){Nomura}, {Tsukagoshi}, {Kawabe}, {Ishimoto},
  {Okuzumi}, {Muto}, {Kanagawa}, {Ida}, {Walsh}, {Millar}, \& {Bai}}]{NTK+16}
{Nomura}, H., {Tsukagoshi}, T., {Kawabe}, R., {et~al.} 2015, ArXiv e-prints,
  arXiv:1512.05440

\bibitem[{{Okuzumi} \& {Hirose}(2012)}]{OH12}
{Okuzumi}, S., \& {Hirose}, S. 2012, \apjl, 753, L8

\bibitem[{{Okuzumi} {et~al.}(2012){Okuzumi}, {Tanaka}, {Kobayashi}, \&
  {Wada}}]{OTKW12}
{Okuzumi}, S., {Tanaka}, H., {Kobayashi}, H., \& {Wada}, K. 2012, \apj, 752,
  106

\bibitem[{{Okuzumi} {et~al.}(2009){Okuzumi}, {Tanaka}, \& {Sakagami}}]{OTS09}
{Okuzumi}, S., {Tanaka}, H., \& {Sakagami}, M.-a. 2009, \apj, 707, 1247

\bibitem[{{Ormel}(2014)}]{O14}
{Ormel}, C.~W. 2014, \apjl, 789, L18

\bibitem[{{Ormel} \& {Cuzzi}(2007)}]{OC07}
{Ormel}, C.~W., \& {Cuzzi}, J.~N. 2007, \aap, 466, 413

\bibitem[{{Ormel} {et~al.}(2007){Ormel}, {Spaans}, \& {Tielens}}]{OST07}
{Ormel}, C.~W., {Spaans}, M., \& {Tielens}, A.~G.~G.~M. 2007, \aap, 461, 215

\bibitem[{{Paardekooper} \& {Mellema}(2004)}]{PM04}
{Paardekooper}, S.-J., \& {Mellema}, G. 2004, \aap, 425, L9

\bibitem[{{Paardekooper} \& {Mellema}(2006)}]{PM06}
---. 2006, \aap, 453, 1129

\bibitem[{{Pinilla} {et~al.}(2012){Pinilla}, {Benisty}, \& {Birnstiel}}]{PBB12}
{Pinilla}, P., {Benisty}, M., \& {Birnstiel}, T. 2012, \aap, 545, A81

\bibitem[{{Pinte} {et~al.}(2016){Pinte}, {Dent}, {M{\'e}nard}, {Hales}, {Hill},
  {Cortes}, \& {de Gregorio-Monsalvo}}]{PDM+16}
{Pinte}, C., {Dent}, W.~R.~F., {M{\'e}nard}, F., {et~al.} 2016, \apj, 816, 25

\bibitem[{{Pollack} {et~al.}(1994){Pollack}, {Hollenbach}, {Beckwith},
  {Simonelli}, {Roush}, \& {Fong}}]{PHB+94}
{Pollack}, J.~B., {Hollenbach}, D., {Beckwith}, S., {et~al.} 1994, \apj, 421,
  615

\bibitem[{{Poppe}(2003)}]{P03}
{Poppe}, T. 2003, \icarus, 164, 139

\bibitem[{{Qi} {et~al.}(2013){Qi}, {{\"O}berg}, {Wilner}, {D'Alessio},
  {Bergin}, {Andrews}, {Blake}, {Hogerheijde}, \& {van Dishoeck}}]{QOW+13}
{Qi}, C., {{\"O}berg}, K.~I., {Wilner}, D.~J., {et~al.} 2013, Science, 341, 630

\bibitem[{{Rapson} {et~al.}(2015){Rapson}, {Kastner}, {Millar-Blanchaer}, \&
  {Dong}}]{RKMD15}
{Rapson}, V.~A., {Kastner}, J.~H., {Millar-Blanchaer}, M.~A., \& {Dong}, R.
  2015, \apjl, 815, L26

\bibitem[{{Rietmeijer}(1993)}]{R93}
{Rietmeijer}, F.~J.~M. 1993, Earth and Planetary Science Letters, 117, 609

\bibitem[{{Ros} \& {Johansen}(2013)}]{RJ13}
{Ros}, K., \& {Johansen}, A. 2013, \aap, 552, A137

\bibitem[{{Saito} \& {Sirono}(2011)}]{SS11}
{Saito}, E., \& {Sirono}, S.-i. 2011, \apj, 728, 20

\bibitem[{{Sano} {et~al.}(2000){Sano}, {Miyama}, {Umebayashi}, \&
  {Nakano}}]{SMUN00}
{Sano}, T., {Miyama}, S.~M., {Umebayashi}, T., \& {Nakano}, T. 2000, \apj, 543,
  486

\bibitem[{{Sargent} \& {Beckwith}(1991)}]{SB91}
{Sargent}, A.~I., \& {Beckwith}, S.~V.~W. 1991, \apjl, 382, L31

\bibitem[{{Sato} {et~al.}(2016){Sato}, {Okuzumi}, \& {Ida}}]{SOI16}
{Sato}, T., {Okuzumi}, S., \& {Ida}, S. 2016, \aap, in press, arXiv:1512.02414

\bibitem[{{Seizinger} {et~al.}(2013){Seizinger}, {Krijt}, \& {Kley}}]{SKK13}
{Seizinger}, A., {Krijt}, S., \& {Kley}, W. 2013, \aap, 560, A45

\bibitem[{{Sekiya}(1998)}]{S98}
{Sekiya}, M. 1998, \icarus, 133, 298

\bibitem[{{Sirono}(1999)}]{S99}
{Sirono}, S. 1999, \aap, 347, 720

\bibitem[{{Sirono}(2011{\natexlab{a}})}]{S11a}
{Sirono}, S.-i. 2011{\natexlab{a}}, \apjl, 733, L41

\bibitem[{{Sirono}(2011{\natexlab{b}})}]{S11b}
---. 2011{\natexlab{b}}, \apj, 735, 131

\bibitem[{{Sirono} \& {Ueno}(2014)}]{SU14}
{Sirono}, S.-i., \& {Ueno}, H. 2014, in COSPAR Meeting, Vol.~40, 40th COSPAR
  Scientific Assembly, 3110

\bibitem[{{Stephens} {et~al.}(2014){Stephens}, {Looney}, {Kwon},
  {Fern{\'a}ndez-L{\'o}pez}, {Hughes}, {Mundy}, {Crutcher}, {Li}, \&
  {Rao}}]{SLK+14}
{Stephens}, I.~W., {Looney}, L.~W., {Kwon}, W., {et~al.} 2014, \nat, 514, 597

\bibitem[{{Suyama} {et~al.}(2008){Suyama}, {Wada}, \& {Tanaka}}]{SWT08}
{Suyama}, T., {Wada}, K., \& {Tanaka}, H. 2008, \apj, 684, 1310

\bibitem[{Swinkels \& Ashby(1981)}]{SA81}
Swinkels, F., \& Ashby, M. 1981, Acta Metallurgica, 29, 259

\bibitem[{{Takahashi} \& {Inutsuka}(2014)}]{TI14}
{Takahashi}, S.~Z., \& {Inutsuka}, S.-i. 2014, \apj, 794, 55

\bibitem[{{Takeuchi} \& {Lin}(2005)}]{TL05}
{Takeuchi}, T., \& {Lin}, D.~N.~C. 2005, \apj, 623, 482

\bibitem[{{Tanaka} {et~al.}(1996){Tanaka}, {Inaba}, \& {Nakazawa}}]{TIN96}
{Tanaka}, H., {Inaba}, S., \& {Nakazawa}, K. 1996, \icarus, 123, 450

\bibitem[{{Uribe} {et~al.}(2011){Uribe}, {Klahr}, {Flock}, \&
  {Henning}}]{UKFH11}
{Uribe}, A.~L., {Klahr}, H., {Flock}, M., \& {Henning}, T. 2011, \apj, 736, 85

\bibitem[{{Vacca} \& {Sandell}(2011)}]{VS11}
{Vacca}, W.~D., \& {Sandell}, G. 2011, \apj, 732, 8

\bibitem[{{Wada} {et~al.}(2013){Wada}, {Tanaka}, {Okuzumi}, {Kobayashi},
  {Suyama}, {Kimura}, \& {Yamamoto}}]{WTO+13}
{Wada}, K., {Tanaka}, H., {Okuzumi}, S., {et~al.} 2013, \aap, 559, A62

\bibitem[{{Wada} {et~al.}(2008){Wada}, {Tanaka}, {Suyama}, {Kimura}, \&
  {Yamamoto}}]{WTS+08}
{Wada}, K., {Tanaka}, H., {Suyama}, T., {Kimura}, H., \& {Yamamoto}, T. 2008,
  \apj, 677, 1296

\bibitem[{{Wada} {et~al.}(2009){Wada}, {Tanaka}, {Suyama}, {Kimura}, \&
  {Yamamoto}}]{WTS+09}
---. 2009, \apj, 702, 1490

\bibitem[{{Warren}(1984)}]{W84}
{Warren}, S.~G. 1984, \ao, 23, 1206

\bibitem[{{Weidenschilling}(1977)}]{W77a}
{Weidenschilling}, S.~J. 1977, \mnras, 180, 57

\bibitem[{{Whipple}(1972)}]{W72}
{Whipple}, F.~L. 1972, in From Plasma to Planet, ed. A.~{Elvius}, 211

\bibitem[{{White} \& {Hillenbrand}(2004)}]{WH04}
{White}, R.~J., \& {Hillenbrand}, L.~A. 2004, \apj, 616, 998

\bibitem[{{Yamamoto} {et~al.}(1983){Yamamoto}, {Nakagawa}, \& {Fukui}}]{YNF83}
{Yamamoto}, T., {Nakagawa}, N., \& {Fukui}, Y. 1983, \aap, 122, 171

\bibitem[{{Yang} {et~al.}(2015){Yang}, {Li}, {Looney}, \& {Stephens}}]{YLLS16}
{Yang}, H., {Li}, Z.-Y., {Looney}, L., \& {Stephens}, I. 2015, \mnras, in
  press, arXiv:1507.08353

\bibitem[{{Youdin}(2011)}]{Y11}
{Youdin}, A.~N. 2011, \apj, 731, 99

\bibitem[{{Youdin} \& {Lithwick}(2007)}]{YL07}
{Youdin}, A.~N., \& {Lithwick}, Y. 2007, \icarus, 192, 588

\bibitem[{{Youdin} \& {Shu}(2002)}]{YS02}
{Youdin}, A.~N., \& {Shu}, F.~H. 2002, \apj, 580, 494

\bibitem[{{Zhang} {et~al.}(2015){Zhang}, {Blake}, \& {Bergin}}]{ZBB15}
{Zhang}, K., {Blake}, G.~A., \& {Bergin}, E.~A. 2015, \apjl, 806, L7

\bibitem[{{Zhu} {et~al.}(2012){Zhu}, {Nelson}, {Dong}, {Espaillat}, \&
  {Hartmann}}]{ZNDEH12}
{Zhu}, Z., {Nelson}, R.~P., {Dong}, R., {Espaillat}, C., \& {Hartmann}, L.
  2012, \apj, 755, 6

\bibitem[{{Zubko} {et~al.}(1996){Zubko}, {Mennella}, {Colangeli}, \&
  {Bussoletti}}]{ZMCB96}
{Zubko}, V.~G., {Mennella}, V., {Colangeli}, L., \& {Bussoletti}, E. 1996,
  \mnras, 282, 1321

\end{thebibliography}

\end{document}